\documentclass[a4paper,11pt]{article}
\usepackage[toc,page]{appendix}
\usepackage{graphicx}
\usepackage{colortbl}
\usepackage{amsmath}
\usepackage{subfigure}
\usepackage{mathtools}
\usepackage[utf8]{inputenc}
\usepackage{float}
\usepackage[normalem]{ulem}
\usepackage{epstopdf}
\usepackage{amsfonts,amsmath,amssymb}
\usepackage{hyperref}

\usepackage{epsfig,multicol,bbm}
\usepackage{color}
\usepackage[dvipsnames]{xcolor}

\definecolor{darkblue}{rgb}{0.0, 0.0, 0.55}
\definecolor{grey}{rgb}{0.57, 0.64, 0.69}
\definecolor{lightbrown}{rgb}{0.71, 0.4, 0.11}
\newcommand{\tcb}{\textcolor{blue}}

\newcommand{\be}{\begin{equation}}
\newcommand{\ee}{\end{equation}}

 \newcommand{\rbm}[1]{{\color{red}\bf [Robb: #1]}}

\newcommand\fverb{\setbox\pippobox=\hbox\bgroup\verb}
\newcommand\fverbit{\egroup\item[\fbox{\unhbox\pippobox}]}

\newbox\pippobox

\begin{document}
\title{\bf Analytically Approximation Solution to Higher Derivative Gravity}
\author{S. N. Sajadi\thanks{Electronic address: naseh.sajadi@gmail.com}\,,\, Robert B. Mann\thanks{Electronic address: rbmann@uwaterloo.ca}\,,\, N. Riazi\thanks{Electronic address: n\_riazi@sbu.ac.ir}\,,\, Saeed Fakhry\thanks{Electronic address: s$_{-}$fakhry@sbu.ac.ir}
\\
\small Department of Physics, Shahid Beheshti University, G.C., Evin, Tehran 19839,  Iran \\
\small Physics Department and Biruni Observatory, College of Sciences, Shiraz University, Shiraz 71454, Iran\\
\small Department of Physics and Astronomy, University of Waterloo, Waterloo, Ontario, Canada N2L 3G1}
\maketitle
\begin{abstract}
We obtain  analytical approximate black hole solutions for higher derivative gravity in the presence of Maxwell electromagnetic source.  We construct near horizon and asymptotic solutions and then  use these to obtain an approximate analytic solution using a continued fraction method   to get a complete solution. We compute the thermodynamic quantities and check the first law and Smarr formula. Finally, we investigate the null and time-like geodesics of this black hole.   
\end{abstract}

\section{Introduction}

Einstein's general relativity is a remarkably successful theory of gravity. By predicting and describing new fundamental phenomena such as black holes, gravitational waves and cosmic expansion, it has become a cornerstone of modern theoretical physics and astronomy. Recently, some of its predictions have been confirmed by the direct detection of black hole and its shadow and gravitational waves from binary black hole merger. 

Despite such enormous successes, it has some limitations. As a classical field theory, it does not take quantum effects into account. In order to understand them, with an ultimate vision to unify general relativity with quantum theory, it is necessary to go beyond general relativity. In effective field theories, Einstein gravity is extended by higher-order terms in curvature that represent quantum corrections that emerge from high energy regimes. If one adds all possible quadratic curvature invariants to the usual Einstein-Hilbert action one obtains a theory free of ultraviolet divergences \cite{tHooft:1974toh}. However the linearized equations of motion of these theories allow unphysical ghost-like modes \cite{Stelle:1976gc}. Although such ghosts in general violate unitarity, and thus the probabilistic interpretation of quantum theory, there are some arguments \cite{Smilga:2013vba} indicating that this is not a severe problem. In this case there are some theories that lead to ghosts \cite{Smilga:2004cy}. The details of this renormalization have been discussed elsewhere \cite{Stelle:1976gc} and will not be considered here. 

We consider here the search for electric charged black hole solutions in Einstein Quadratic Gravity, which is
general relativity extended by  quadratic curvature invariants in the action.   Black holes are fundamental objects in a theory of gravity,  providing powerful probes for studying subtle aspects of a theory of gravity.
 In addition to the Schwarzschild solution, there is another spherically symmetric asymptotically flat non Schwarzschild black hole solution within the same theory that  admits positive and negative values for the  black hole mass  \cite{Lu:2015cqa}. The search for new electrically charged black hole solutions, has been carried out numerically \cite{Lin:2016jjl}. However numerical solutions do not give a clear picture of the metric dependence on physical parameters of the system.  We therefore seek  an analytic method to obtain a highly accurate analytic approximate solution to the field equations.

 To this end we employ a continued fraction expansion ansatz \cite{Rezzolla:2014mua}. This ansatz is designed so that the coefficients in the continued fraction are fixed by behaviour of the metric near the event horizon, while the pre-factors are introduced to match the asymptotic behaviour at infinity. This way an accurate analytic expression approximating the metric  can be obtained for the whole space outside the event horizon, and not only near the black hole or far from it. The continued fraction approximation is not only useful for the present analysis, but also, proves useful in concerning quasi-normal modes \cite{Leaver:1985ax}. With the continued fraction solution in hand, we study the properties of the black hole solution. Specifically we analyze the motion of particles around the black hole, constraining the coupling with solar system tests, and investigating the properties of its shadow. We exhibit constraints on  the coupling constant $\alpha$ of the theory by using Shapiro time delay. We find that the theory can be compatible with solar system tests whilst maintaining relatively large values of the coupling. Furthermore, we find that the radius of the innermost stable circular orbit around the black hole and the angular momentum of a test body at this radius increase with increasing $\alpha$ as compared to their corresponding values in general relativity.

Our paper is organized as follows. In the next section we review the near horizon and asymptotic solutions. Then,   using a continued fraction expansion,  we obtain an approximate analytic solution from the first law thermodynamics and Smarr formula. In Sec. \ref{sectwo} we study some properties of the black holes and investigate the orbit of particles around it. we constrain the coupling constant by using Shapiro test in solar system. Finally, in sec. \ref{con} we present our conclusions.

\section{Basic equations}

The most general Lagrangian with electromagnetic field and the cosmological constant $ \Lambda $ can
be written as 
\begin{equation}\label{eq1}
L=\gamma(R-2\Lambda)-\alpha C_{a b c d}C^{a b c d}+\beta R^{2}-k F_{a b}F^{a b}
\end{equation}
where $ F_{a b} $ is the electromagnetic tensor and $ C_{a b c d} $ is the Weyl tensor, $ \alpha $, $ \beta $, $ \gamma $ and $ \kappa $ are coupling constants.  Since the trace of the equations of motion in the absence of a cosmological constant
vanishes, the term proportional to $\beta$ does not contribute to the solution.  Henceforth we set
$\beta=0$, and  for simplicity we also set $ \gamma=\kappa=1 $ \cite{Lu:2015cqa}.  

The field equations are then given by
\begin{equation}
E_{a b}=R_{a b}-\dfrac{1}{2}g_{a b}R-\Lambda g_{a b}-4\alpha B_{a b}-2T_{a b}=0 \hspace{0.5cm},\hspace{0.5cm}\nabla_{a}F^{a b}=0
\end{equation}
where $ T_{a b} $ and $ B_{a b} $ are
\begin{equation}
B_{a b}=\left(\nabla^{m}\nabla^{n}+\dfrac{1}{2}R^{m n} \right)C_{a m b n}\hspace{0.5cm},\hspace{.5cm}T_{a b}=F_{d a}F^{d}{}_{b}-\dfrac{1}{4}g_{a b} F_{d e}F^{d e}.
\end{equation}
We consider following static, spherical symmetric metric
\begin{equation}\label{metform}
dS^{2}=-h(r)dt^{2}+\dfrac{dr^{2}}{f(r)}+r^{2}\left(d\theta^{2}+\dfrac{\sin^{2}(\sqrt{k}\theta)}{k}d\phi^{2}\right).
\end{equation}
By inserting the metric into the field equations   we obtain the differential equations for $ f(r) $ and $h(r)$, with
\begin{equation}\label{eq5}
R+4\Lambda=-2fhr^{2}h^{''}+fr^{2}h^{'}{}^{2}-rhh^{'}(rf^{'}+4f)+4h^{2}(2\Lambda r^{2}-rf^{'}+k-f)=0,
\end{equation} 
from taking the trace (where the prime denotes the $r$ derivative), with
\begin{eqnarray}\label{eq6}
&&E_{r r}-2T_{r r}=\nonumber \\
&&\left( -6\,\alpha\,h^{2}f{r}^{3}h^{'} +12 {r}^{2}\alpha
h^{3}f \right)f^{''}+\left(12 \alpha h^{3}r k-3 \alpha h f{r}^{3}h^{'}{}^{2}-12 \alpha
h^{3}f r-6 \alpha h^{2}f {r}^{2}h^{'}+24 \alpha h^{3}{r}^{3}\Lambda \right)f^{'}\nonumber \\
&&-9 \alpha h^{3}{r}^{2}f^{'}{}^{2}+\left(16 \alpha h^{3}\Lambda {r}^{2}+6 h^{3}{r}^{2}+6 {r}^{3}h^{2}h^{'}+24\alpha k h^{3}+6 {q}^{2} h^{2}+16 \alpha {r}^{3} h^{2}h^{'} \Lambda \right)f \nonumber \\
&&+\left(-9 \alpha h{r}^{2}h^{'}{}^{2}-24 \alpha h^{3}+3 \alpha {r}^{3}h^{'}{}^{3}\right)f^{2}+ \left(( -16 \alpha {\Lambda}^{2}-6 \Lambda
) {r}^{2} -6\,k-16\,\alpha\,k\Lambda \right)r^{2}h^{3}=0
\end{eqnarray}
being the only other non-redundant field equation.
Expanding the functions $h(r)$ and $f(r)$ around the event horizon $ r_{+} $  
%\rbm{We can absorb $c$ into the time coordinate, so what is the point of including $c$ in the expansion of $h(r)$?}\rbm{I don't see why.  I could redefine $t\to t/\sqrt{c}$ and eliminate $c$.  Could you please explain  how this works?}\textcolor{blue}{Basically, we have included $c$ to show the freedom of rescale of time coordinate, but, as you can see we have put $ c=1 $ in the following in our plots.}
\begin{align}\label{eq7}
h(r) &= h_{1}(r-r_{+})+h_{2}(r-r_{+})^{2}+h_{3}(r-r_{+})^{3}+... \\
f(r)  &= f_{1}(r-r_{+})+f_{2}(r-r_{+})^{2}+f_{3}(r-r_{+})^{3}+...
\label{eq8}
\end{align}
and then inserting these expressions into equations (\ref{eq5}) and (\ref{eq6}), we find
\begin{equation}\label{eq9}
h_{2}=\dfrac{r_{+}h_{1}\Lambda^{2}}{3f_{1}^{2}}+\left(\dfrac{5h_{1}}{3f_{1}}+\dfrac{kh_{1}}{3r_{+}f_{1}^{2}}+\dfrac{r_{+}h_{1}}{8\alpha f_{1}^{2}}\right)\Lambda +\dfrac{kh_{1}-2f_{1}h_{1}r_{+}}{f_{1}r_{+}^{2}}+\dfrac{ k h_{1}r_{+}^{2}- f_{1}h_{1}r_{+}^{3}-q^{2}f_{1}}{8\alpha f_{1}^{2}r_{+}^{3}},
\end{equation}
and 
\begin{equation}\label{eq10}
f_{2}=-\dfrac{r_{+}\Lambda^{2}}{f_{1}}+\left(3-\dfrac{k}{r_{+}f_{1}}-\dfrac{3r_{+}}{8\alpha f_{1}}\right)\Lambda +\dfrac{k-2f_{1}r_{+}}{r_{+}^{2}}-\dfrac{3( k h_{1}r_{+}^{2}-f_{1}h_{1}r_{+}^{3}-q^{2}f_{1})}{8\alpha f_{1}h_{1}r_{+}^{3}}
\end{equation}
where $r_+$ and $f_1$ are undetermined constants of integration.  The parameter $h_1$ can be absorbed into the definition of the $t$ coordinate in the near-horizon expansion; however 
doing so would have implications for the large-$r$ and continued fraction solutions we shall obtain, and so we have retained it in the above.
For $q=0$ this solution reduces to the near-horizon solution obtained previously \cite{Lu:2015cqa} for the theory given in \eqref{eq1}.  We also note the existence of an alternate solution whose near-horizon expansion has a well-defined small-$\alpha$ limit -- we discuss this in Appendix B.

All higher-order coefficients in \eqref{eq7} and \eqref{eq8}
are determined in terms of these quantities and $r_+$ (and the coupling parameters); we  provide expressions for  the higher-order coefficients in the Appendix.

To obtain a solution for $h(r)$ and $f(r)$  at large $r$, we write
\begin{equation}\label{linearex}
h(r)=1+\varepsilon \mathcal{H}(r)+\mathcal{O}(\varepsilon^{2}),\hspace{0.5cm}f(r)=1+\varepsilon \mathcal{F}(r)+\mathcal{O}(\varepsilon^{2})
\end{equation}
where $\varepsilon\ll 1$. Substituting Eqs. (\ref{linearex}) into Eqs. (\ref{eq5}) and (\ref{eq6}) the field equations become
\begin{align}
2\alpha r^{2}\mathcal{F}^{''}+r^{3}\mathcal{H}^{'}+r^{2}\mathcal{F}-4\alpha \mathcal{F}+q^2 &=0,\label{eqF}\\
\dfrac{1}{2}r^{2}\mathcal{H}^{''}+r \mathcal{H}^{'}+r \mathcal{F}^{'}+\mathcal{F} &=0,\label{eqH}
\end{align}
to order $\varepsilon$. Solving for $\mathcal{H}^{'}$ from Eq. (\ref{eqF})
\begin{equation}\label{eqdH}
\mathcal{H}^{'}=-\dfrac{2\alpha}{r}\mathcal{F}^{''}-\left(1-\dfrac{4\alpha}{r^2}\right)\dfrac{\mathcal{F}}{r}-\dfrac{q^2}{r^3}
\end{equation}
and  inserted this into  (\ref{eqH}) yields
\begin{equation}\label{eqnonhom}
-2\alpha r \mathcal{F}^{'''}-2\alpha \mathcal{F}^{''}+\left(r+\dfrac{4\alpha}{r}\right)\mathcal{F}^{'}+\left(1-\dfrac{4\alpha}{r^2}\right)\mathcal{F}=-\dfrac{q^2}{r^2},
\end{equation}
The corresponding homogenous equation is
\begin{equation}
-2\alpha r \mathcal{F}_{g}^{'''}-2\alpha \mathcal{F}_{g}^{''}+\left(r+\dfrac{4\alpha}{r}\right)\mathcal{F}_{g}^{'}+\left(1-\dfrac{4\alpha}{r^2}\right)\mathcal{F}_{g}=0
\end{equation}
whose solution is
\begin{equation}
\mathcal{F}_{g}=\dfrac{C_{1}}{r}+\dfrac{C_{2}(\sqrt{2\alpha}+r)e^{-\dfrac{r}{\sqrt{2\alpha}}}}{r}-\dfrac{C_{3}(-\sqrt{2\alpha}+r)e^{\dfrac{r}{\sqrt{2\alpha}}}}{r}.
\label{Yuk}
\end{equation}
This solution consists of a growing mode and a decaying mode. Asymptotic flatness demands that we set $ C_{3}=0$ \cite{Bonanno:2019rsq}  and the constant $C_{1}$ can be interpreted as the black hole’s mass ($C_{1}=-2M$). 
The second term decays exponentially and one can therefore be neglected. 

The particular solution to \eqref{eqnonhom} can be obtained using the  ansatz
\begin{equation}
\mathcal{F}_{s}=\sum_{n=2}\dfrac{F_{n}}{r^{n}},
\end{equation}
from which we find
\begin{equation}
\mathcal{F}_{s}=\dfrac{{{q}^{2} }}{r^{2}}+\dfrac{8\alpha{q}^{2}}{r^{4}}+\dfrac{288\alpha^2 q^{2}}{r^{6}}+\dfrac{23040\alpha^3 q^{2}}{r^{8}}+...
\label{eq12a}
\end{equation}
yielding
\begin{equation}\label{eqf11}
f(r)=1+\varepsilon(\mathcal{F}_{g}+\mathcal{F}_{s})=1-\dfrac{2M}{r}+\dfrac{{{q}^{2} }}{r^{2}}+\dfrac{8\alpha{q}^{2}}{r^{4}}+\dfrac{288\alpha^2 q^{2}}{r^{6}}+\dfrac{23040\alpha^3 q^{2}}{r^{8}}+...
\end{equation}
as the solution for $f(r)$, neglecting the exponentially decaying terms. Inserting this  into (\ref{eqdH}), we get
\begin{equation}\label{eqh12}
h(r)=1-\dfrac{2M}{r}+\dfrac{q^{2}-2\alpha}{r^{2}}+\dfrac{4\alpha{q}^{2}}{r^{4}}+\dfrac{96\alpha^2 q^{2}}{r^{6}}+\dfrac{5760\alpha^3 q^2}{r^8}+... 
\end{equation}
where we have set\footnote{We note that our asymptotic expansions (\ref{eqf11},\ref{eqh12}) do not agree with those obtained previously \cite{Lin:2016jjl} for
the charged case.} $\varepsilon=1$.

We wish to obtain an approximate analytic solution  (for $k=1$) that is valid near the horizon and at large $r$.  To this end we employ a continued fraction expansion \cite{Rez}, and write
\begin{equation}\label{eq17}
h(r)=xA(x),\hspace{0.5cm}\dfrac{h(r)}{f(r)}=B^{2}(x),
\end{equation}
with
\begin{align}
A(x) &=1-\epsilon(1-x)+(a_{0}-\epsilon)(1-x)^{2}+\tilde{A}(x)(1-x)^{3}
\label{Ax}
\\
B(x) &=1+b_{0}(1-x)+\tilde{B}(x)(1-x)^{2}
\label{Bx}
\end{align} 
where
\begin{equation}
x = 1- \frac{r_+}{r} \qquad 
\tilde{A}(x)=\dfrac{a_{1}}{1+\dfrac{a_{2}x}{1+\dfrac{a_{3}x}{1+\dfrac{a_{4}x}{1+...}}}}
\qquad 
\tilde{B}(x)=\dfrac{b_{1}}{1+\dfrac{b_{2}x}{1+\dfrac{b_{3}x}{1+\dfrac{b_{4}x}{1+...}}}}
\label{cfrac}
\end{equation}
where we truncate the continued fraction at order $4$.
By expanding (\ref{eq17}) near the horizon ($ x\to 0 $) and 
the asymptotic  region ($ x\to 1 $)  we obtain  
\be
\epsilon=-\dfrac{H_{1}}{r_{+}}-1,  \qquad b_{0}=0,  \qquad a_{0}=\dfrac{q^{2}-2\alpha}{r_{+}^{2}}
\ee
for the lowest order expansion coefficients, with the remaining
$a_i$ and $b_i$ given in terms of $(r_+, q, h_1, f_1)$; we provide these expressions in the  Appendix.  
 
The resultant expressions are somewhat cumbersome to deal with, so henceforth we
set $f_{1} = h_{1}$ for the sake of simplicity.  This yields a more restricted set of solutions that still
capture the basic physics of the higher-curvature effects.  Note that this restriction does not imply
that $ f(r) = h(r) $.

\begin{figure}[H]
\centering
\subfigure[]{
 \includegraphics[width=0.3\columnwidth]{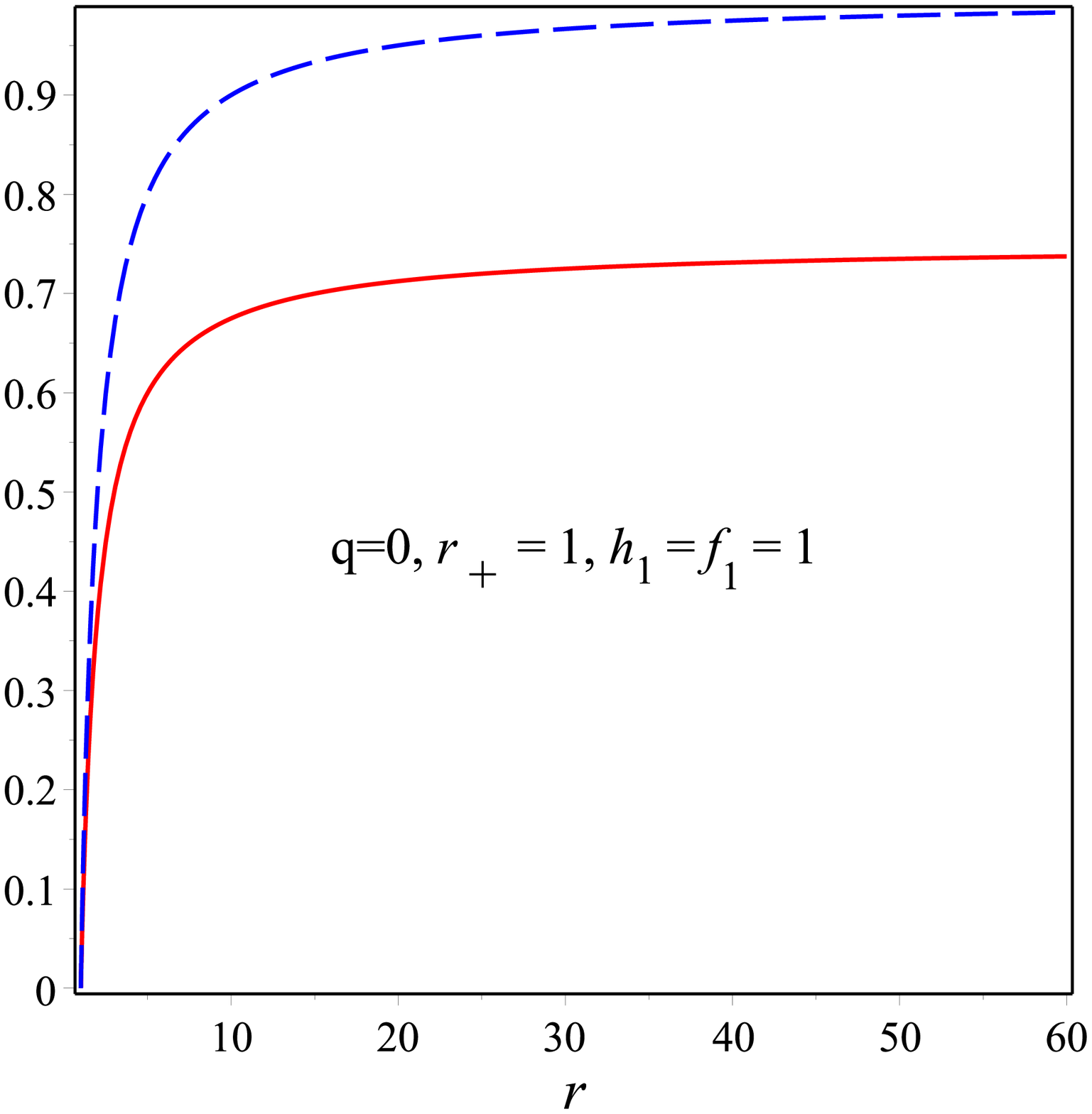}
 \label{fig1q1}
 }
 \subfigure[near horizon]
 {
 \includegraphics[width=0.3\columnwidth]{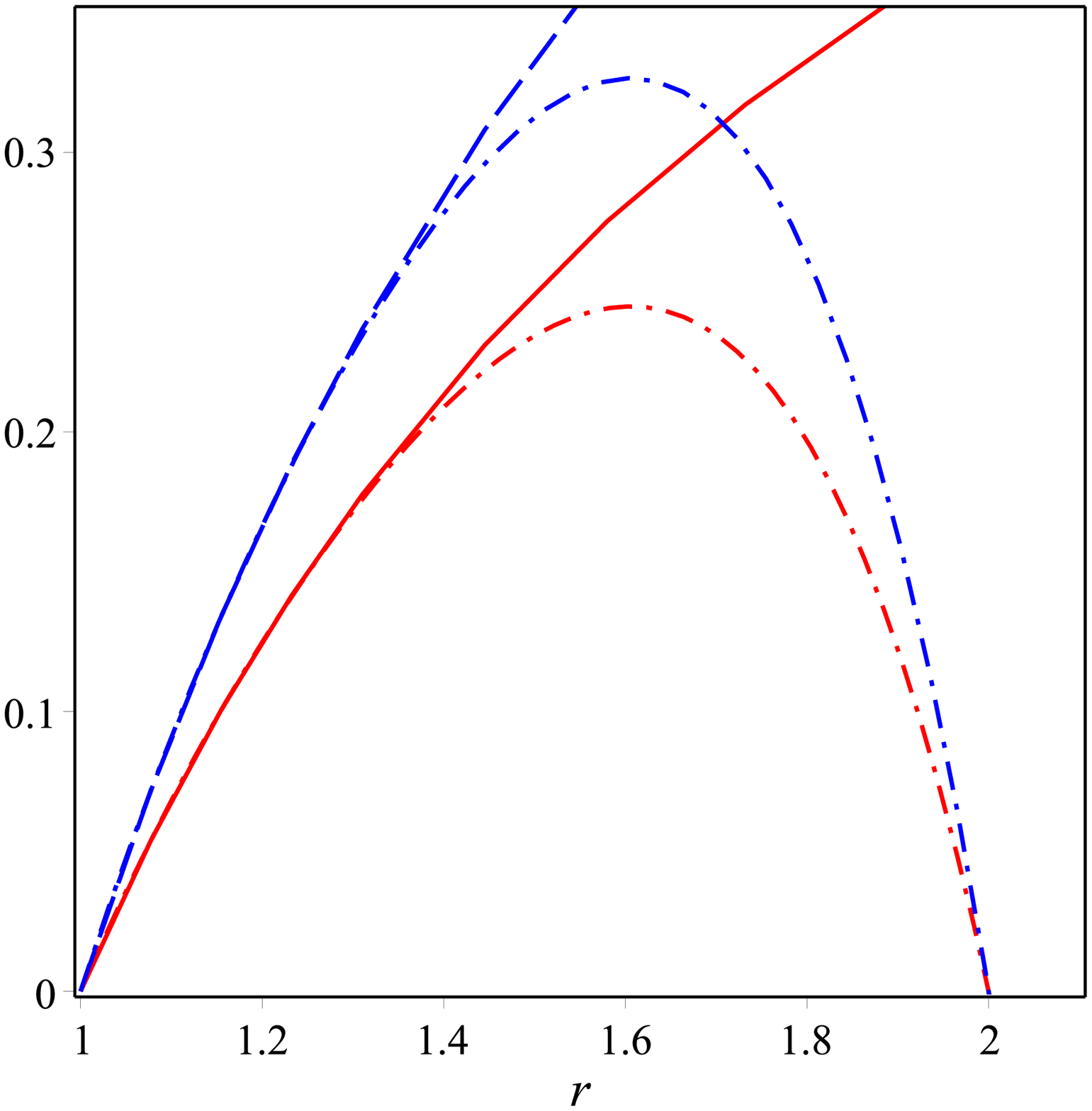}
 \label{fig1q2}
 }
 \subfigure[asymptotic]
 {
 \includegraphics[width=0.3\columnwidth]{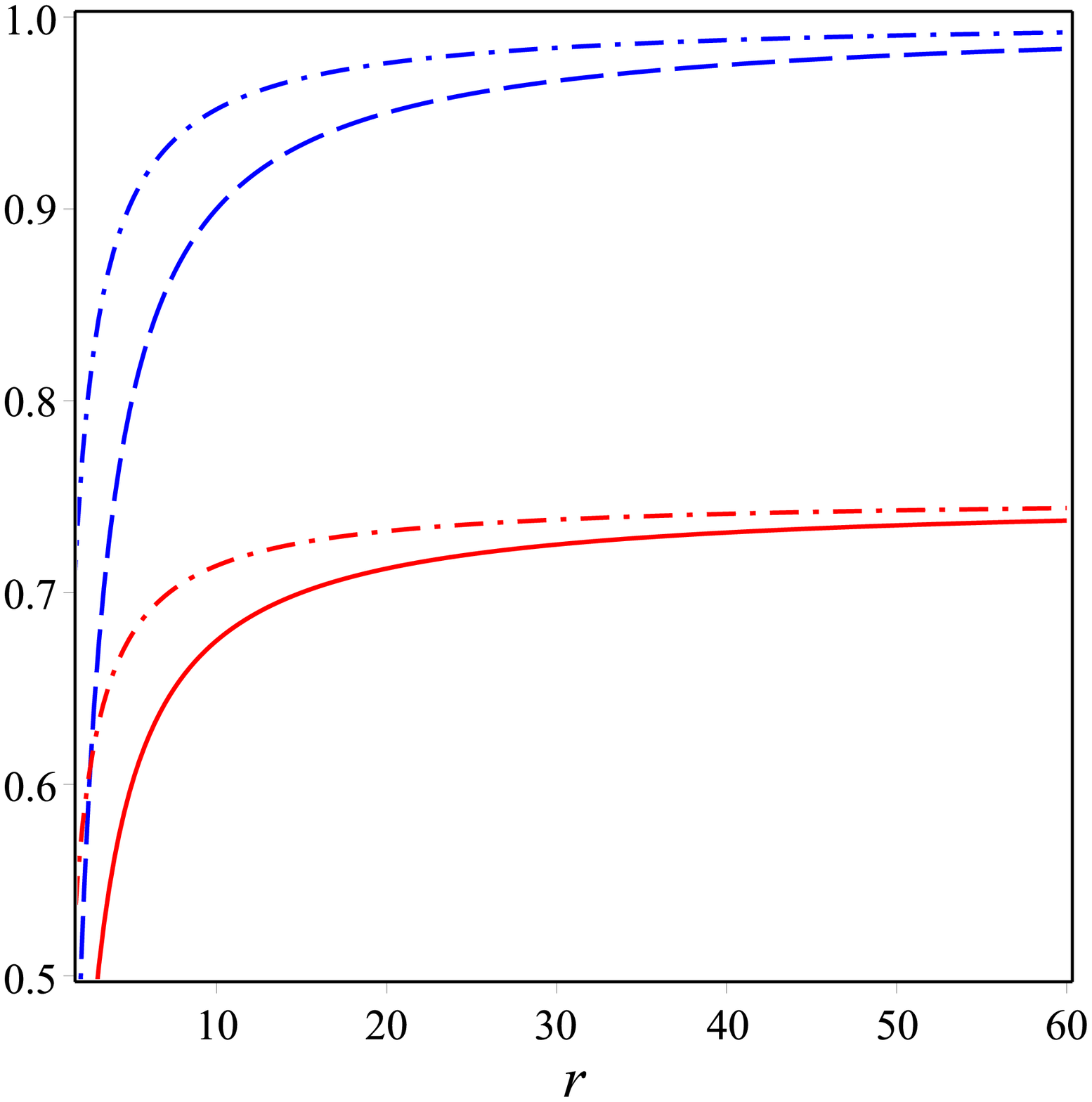}
 \label{fig1q3}
 }
 \subfigure[]
 {
 \includegraphics[width=0.3\columnwidth]{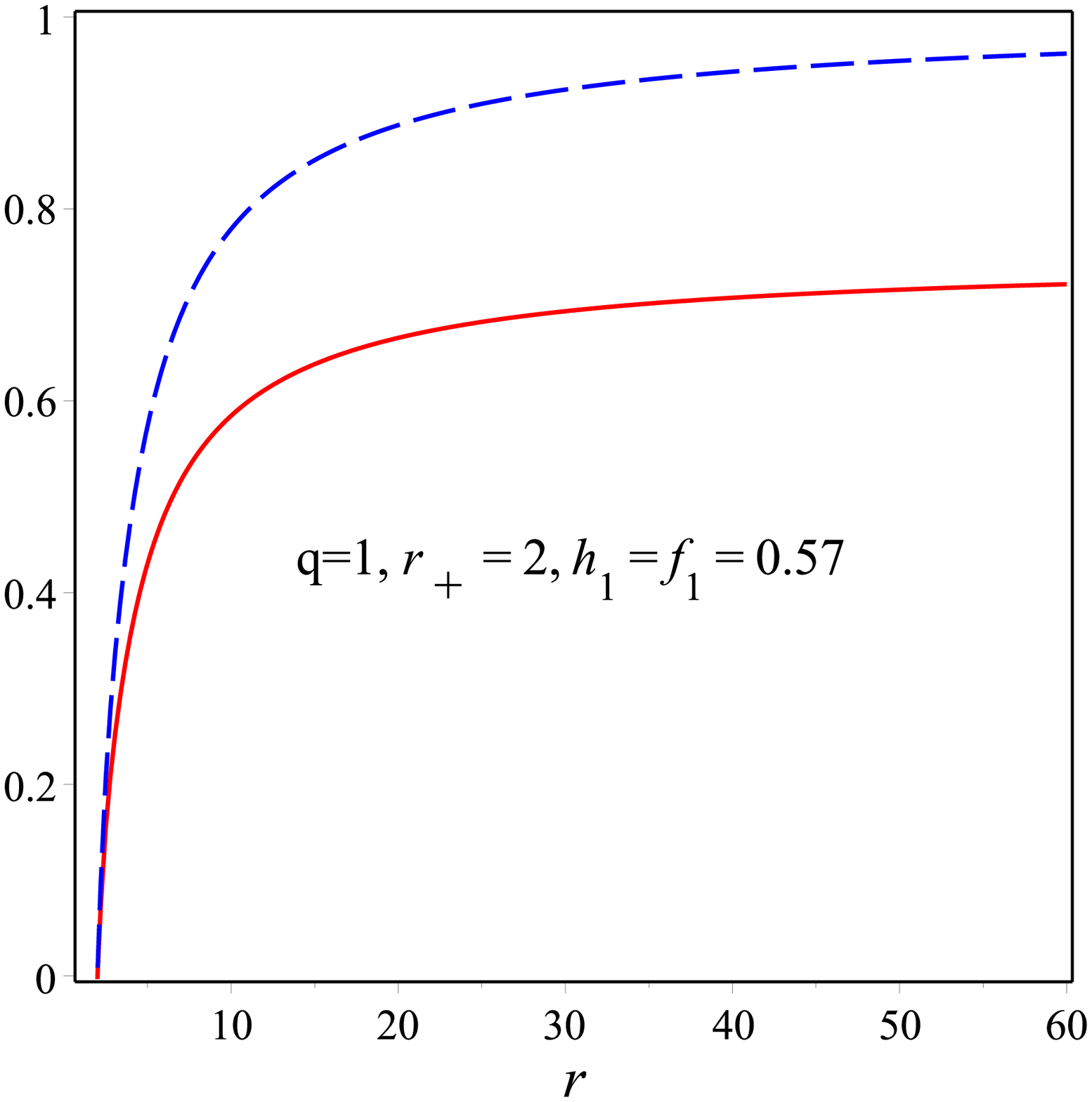}
 \label{fig1q4}
 }
 \subfigure[near horizon]
 {
 \includegraphics[width=0.3\columnwidth]{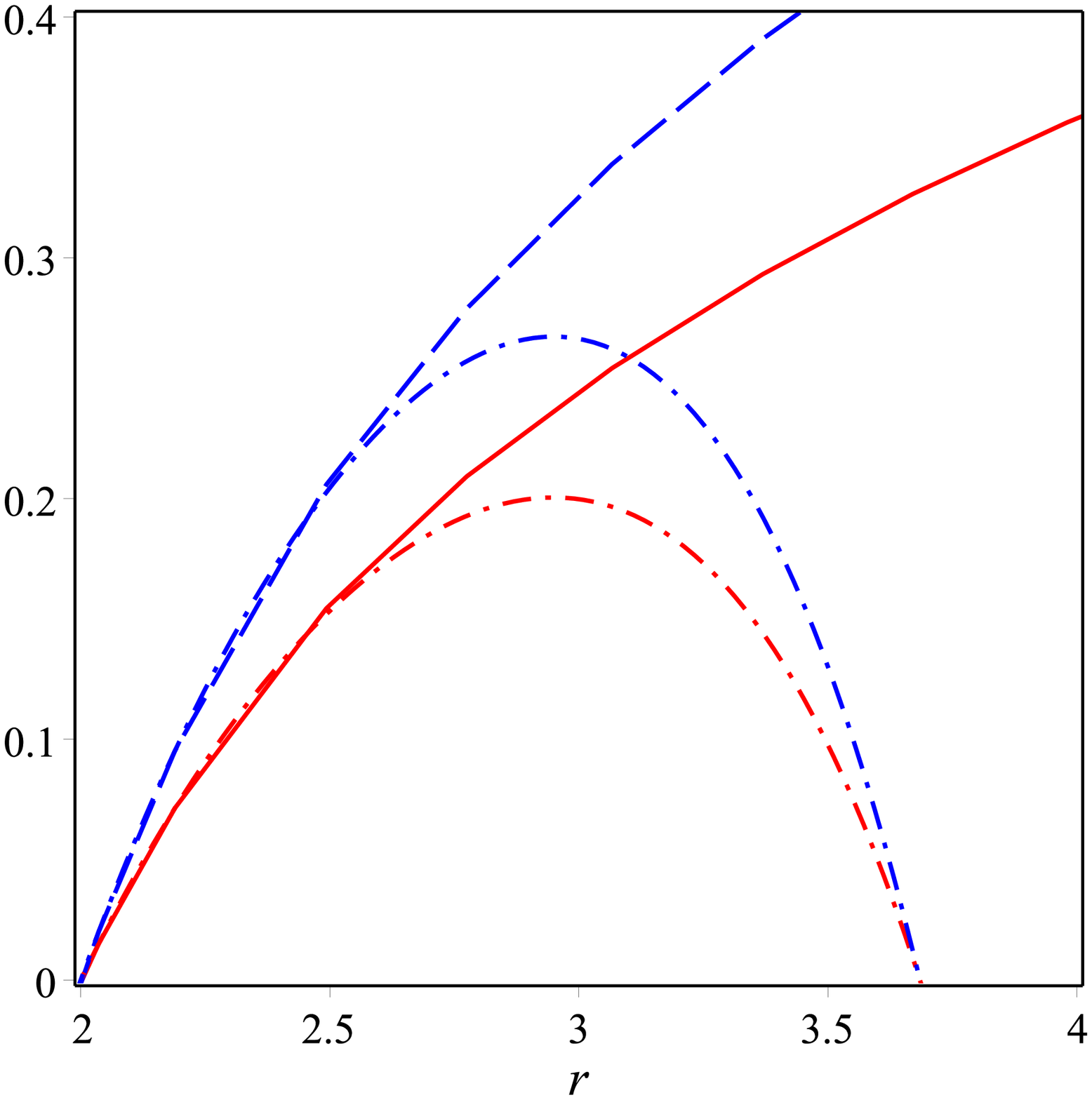}
 \label{fig1q5}
 }
 \subfigure[asymptotic]
 {
 \includegraphics[width=0.3\columnwidth]{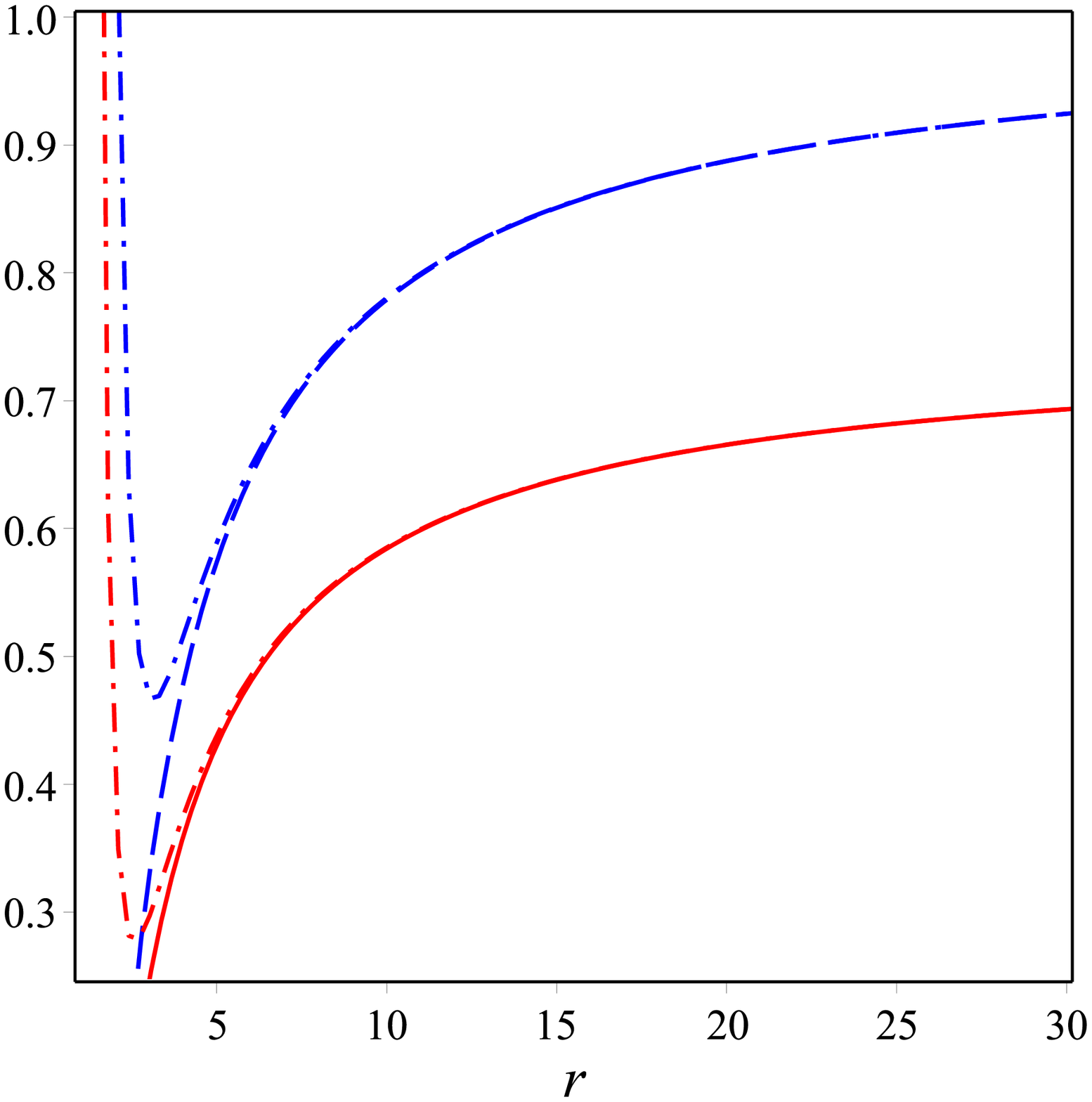}
 \label{fig1q6}
 }
 \subfigure[]
 {
 \includegraphics[width=0.3\columnwidth]{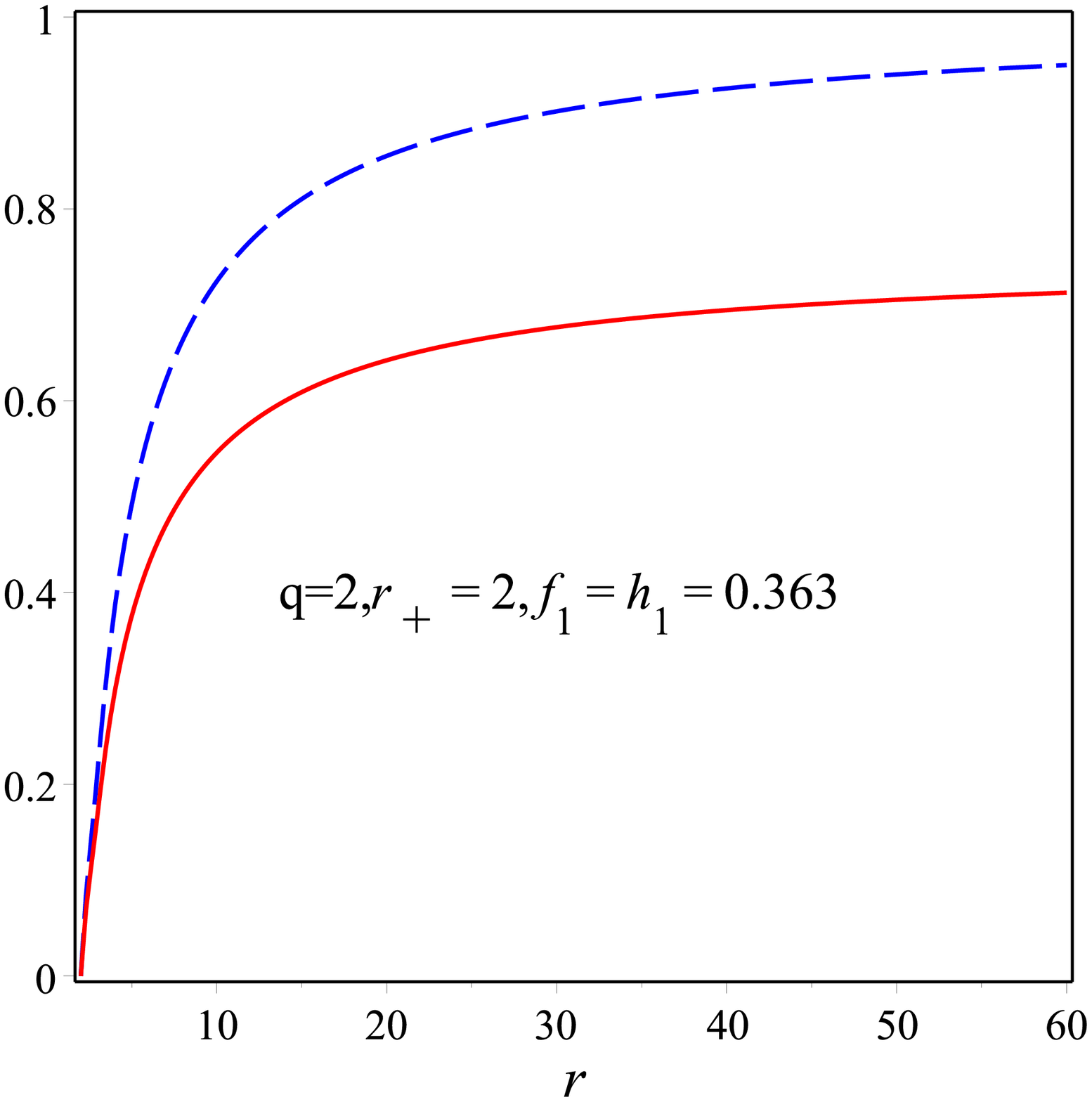}
 \label{fig1q7}
 }
 \subfigure[near horizon]
 {
 \includegraphics[width=0.3\columnwidth]{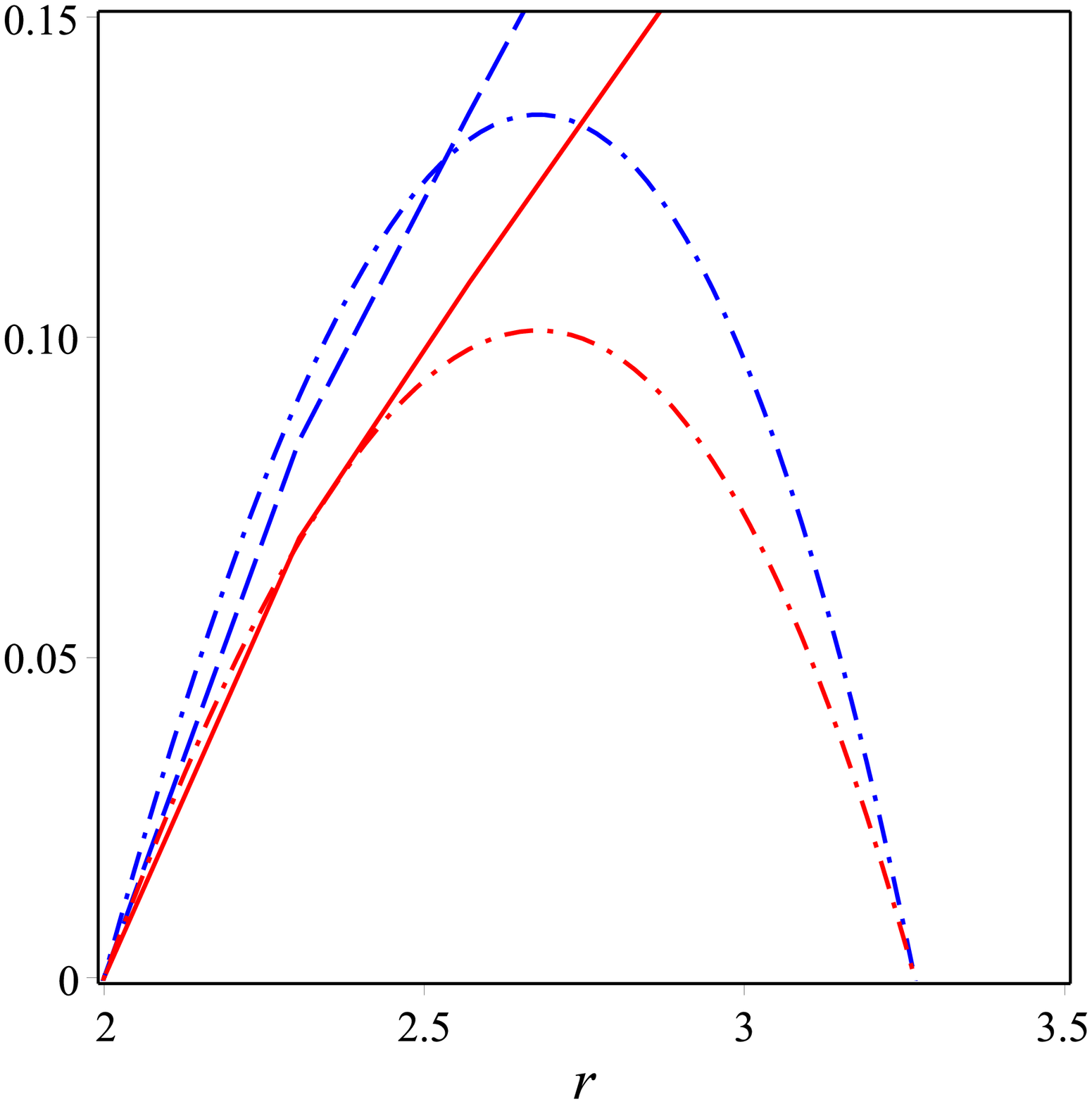}
 \label{fig1q8}
 }
 \subfigure[asymptotic]
 {
 \includegraphics[width=0.3\columnwidth]{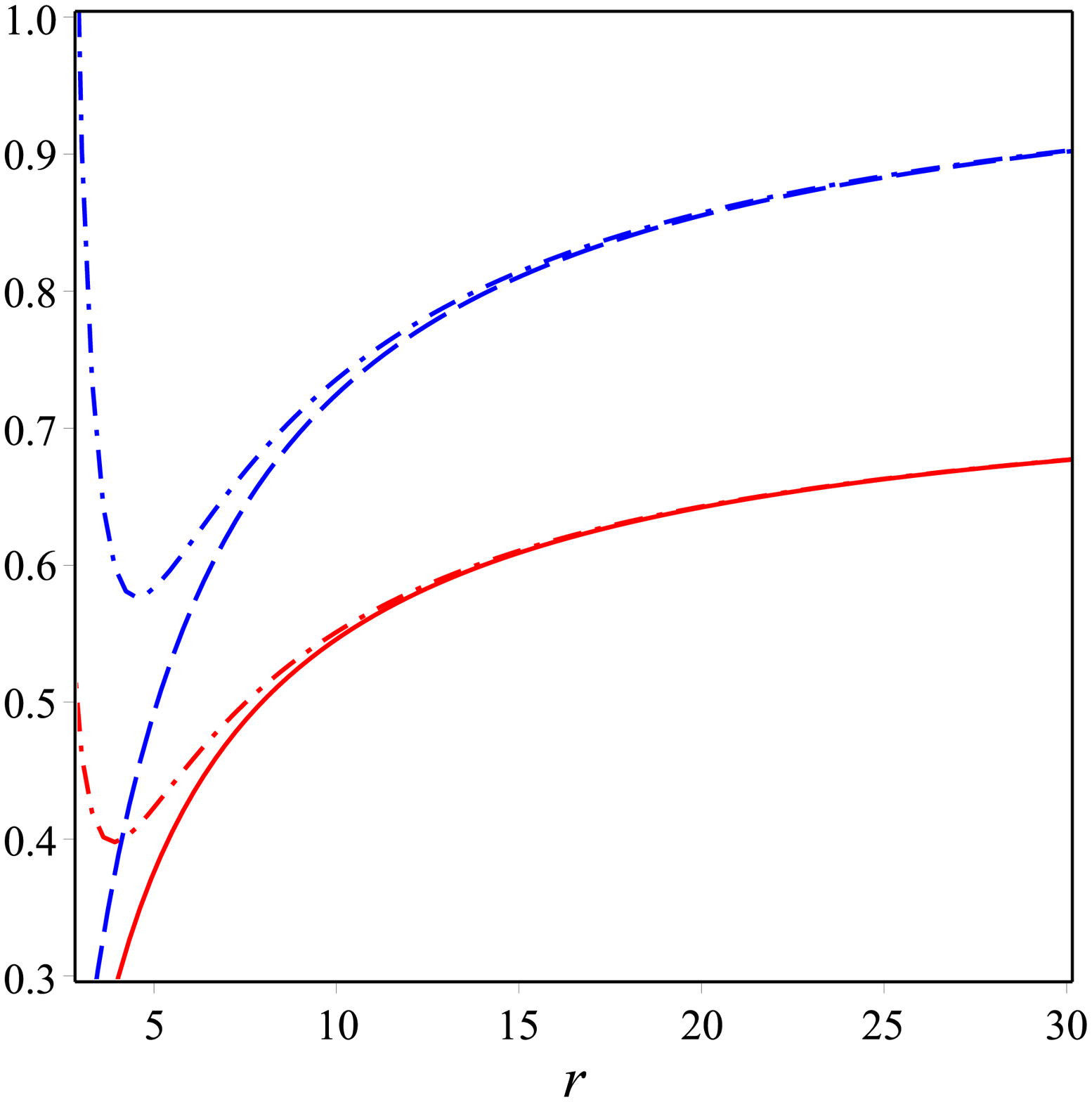}
 \label{fig1q9}
 }
\caption{The behavior of $f(r)  $ (\textcolor{blue}{ blue dashed line}) and $ 0.75h(r) $ (\textcolor{red}{ red solid line}) in terms of $ r $ for $ k=1, \alpha=0.5$, for the first group of solutions.
The top row is  $q=0$, the middle row $q=1$, and the bottom row $q=2$.  The left column
is the full continued fraction solution, the middle column compares this solution to the near-horizon approximation (dot-dashed lines), and that right column compares this solution to the large-$r$ approximation (dot-dashed lines). All dimensionful quantities are in units of $M$.
}
\label{fig1q}
\end{figure}

\begin{figure}[H]
\centering
\subfigure[]{
 \includegraphics[width=0.3\columnwidth]{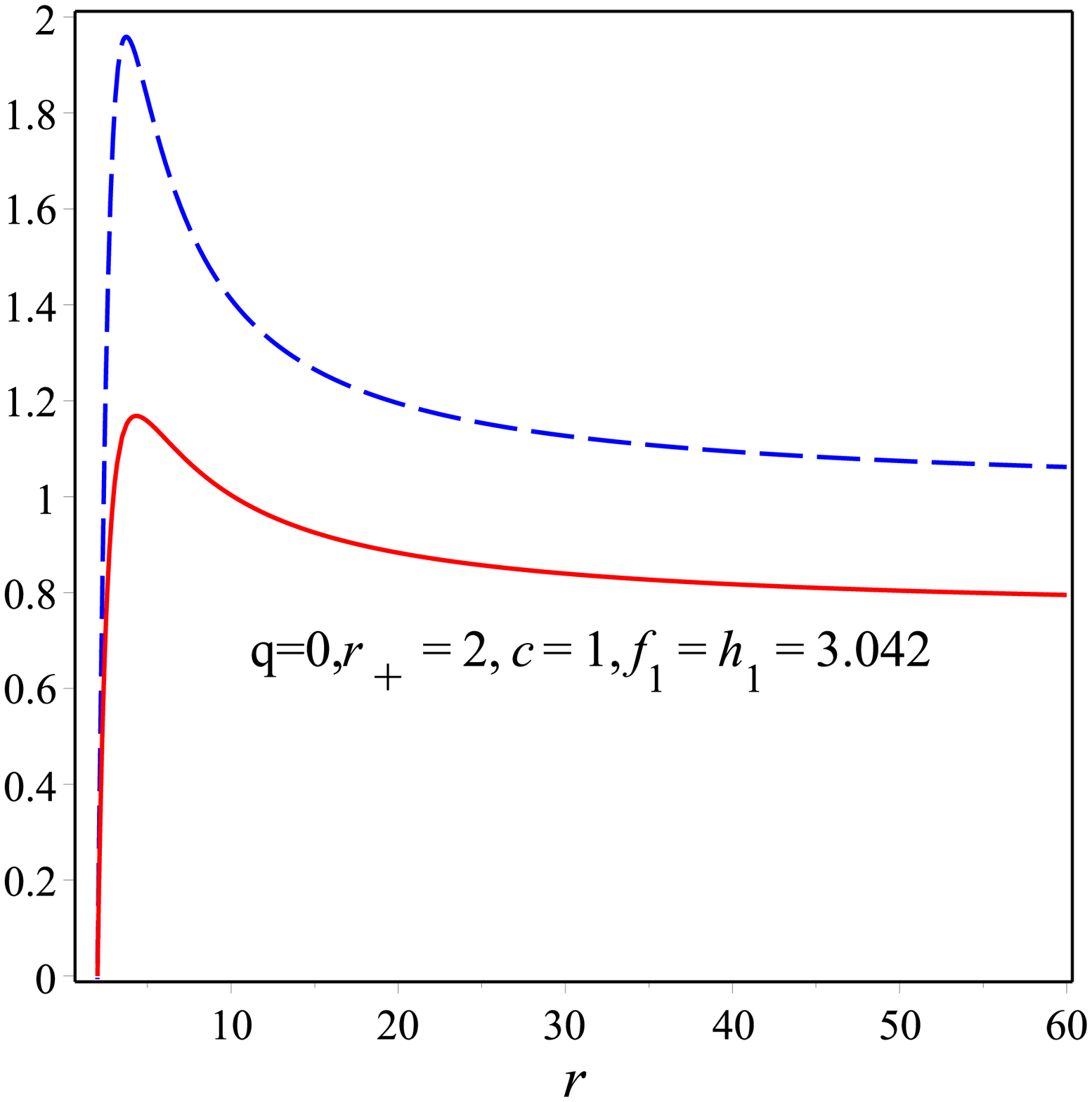}
 \label{fig2q7}
 }
 \subfigure[near horizon]
 {
 \includegraphics[width=0.3\columnwidth]{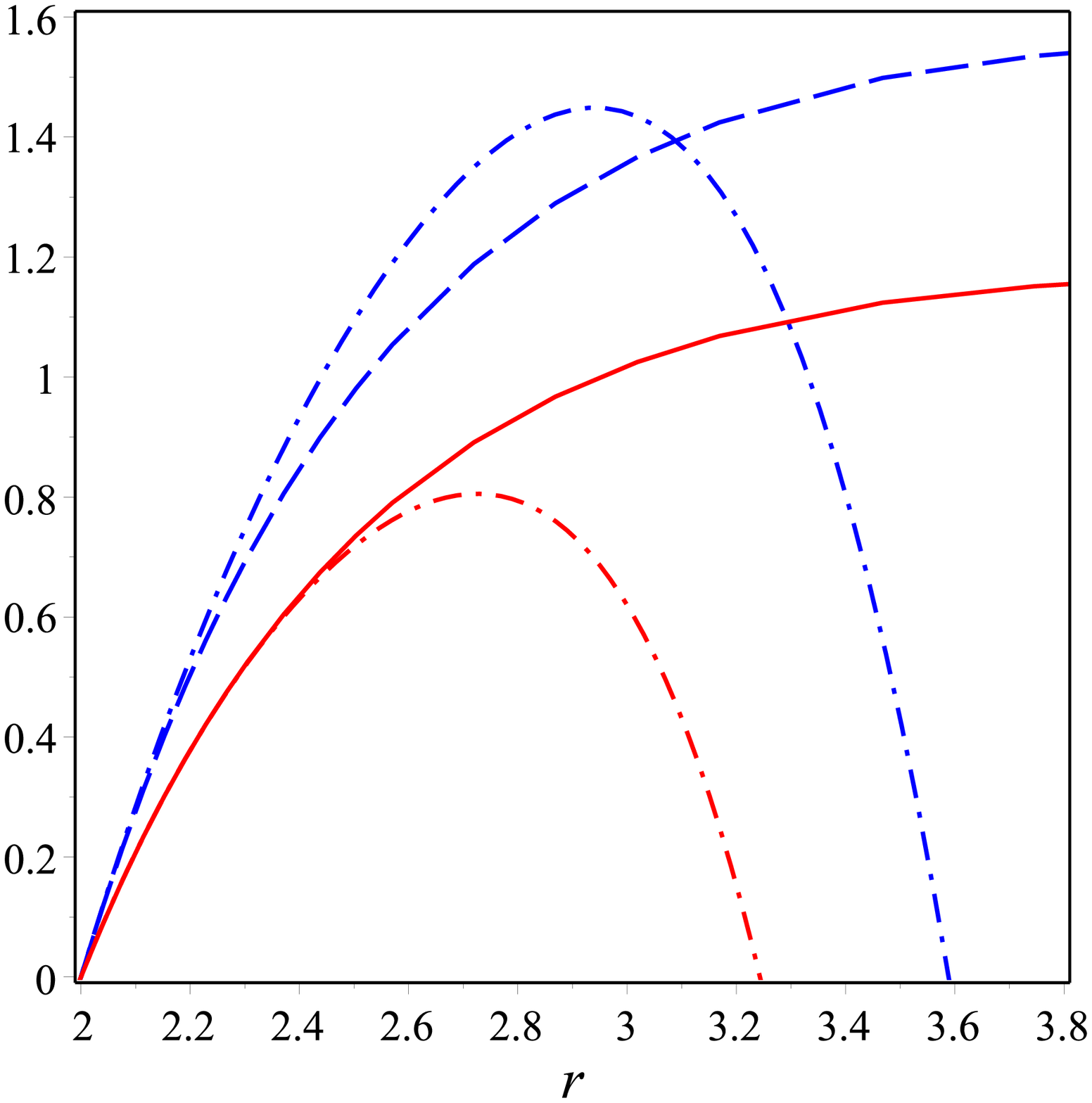}
 \label{fig2q8}
 }
 \subfigure[asymptotic]
 {
 \includegraphics[width=0.3\columnwidth]{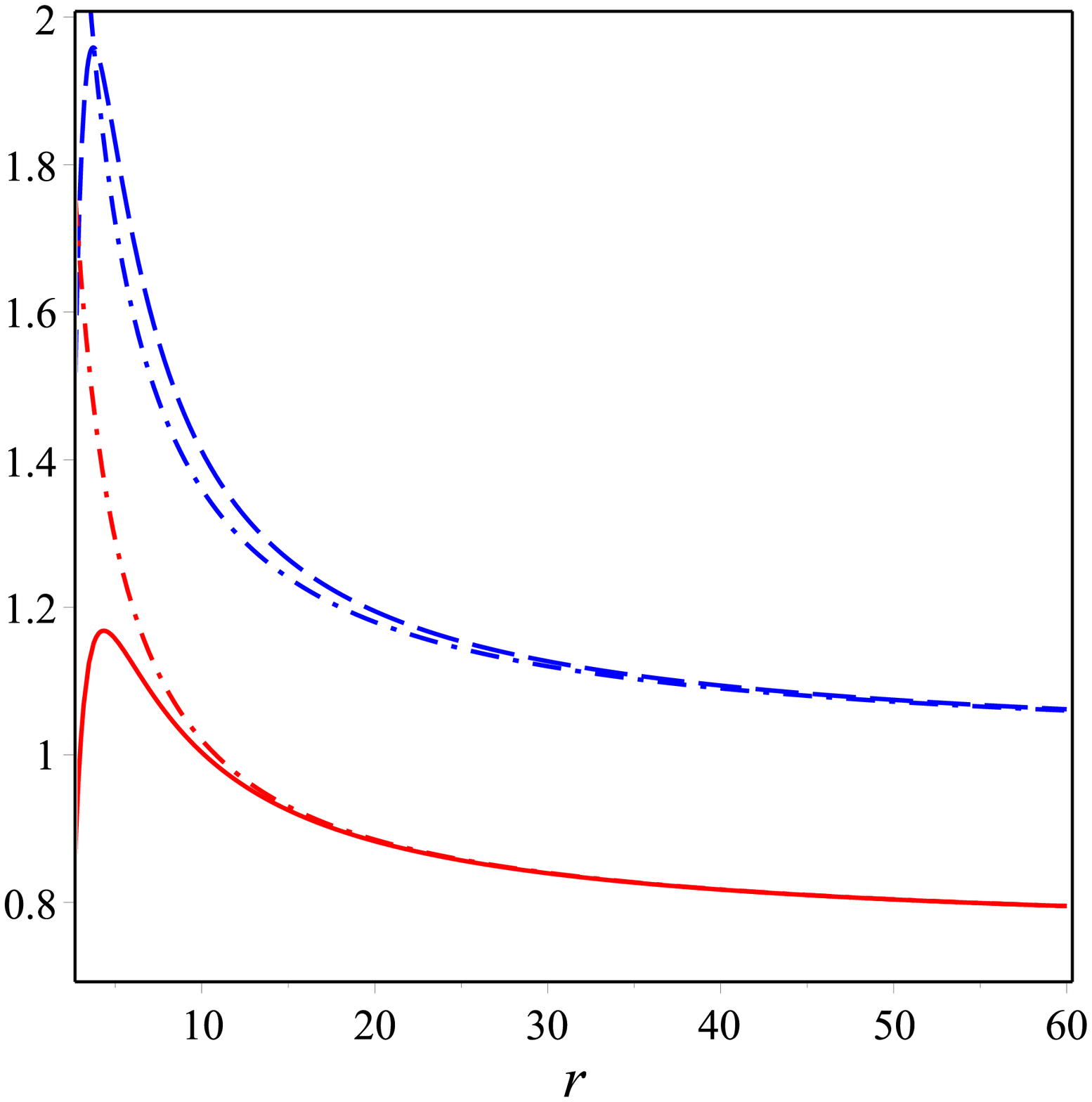}
 \label{fig2q9}
 }
 %\subfigure[]
% {
 %\includegraphics[width=0.3\columnwidth]{hhff}
 %%\label{fig2q10}
% }
 %\subfigure[near horizon]
% {
 %\includegraphics[width=0.3\columnwidth]{hhff2}
% \label{fig2q11}
% }
% \subfigure[asymptotic]
 %{
% \includegraphics[width=0.3\columnwidth]{hhff1}
% \label{fig2q12}
 %}
 %\subfigure[]
 %{
 %\includegraphics[width=0.3\columnwidth]{ffhh}
% \label{fig:FDMsavingCW2}
% }
 %\subfigure[near horizon]
 %{
 %\includegraphics[width=0.3\columnwidth]{ffhh1}
 %\label{fig:FDMsavingCW2}
 %}
% \subfigure[asymptotic]
 %{
 %\includegraphics[width=0.3\columnwidth]{ffhh2}
 %\label{fig:FDMsavingCW2}
% }
 \subfigure[]
 {
 \includegraphics[width=0.3\columnwidth]{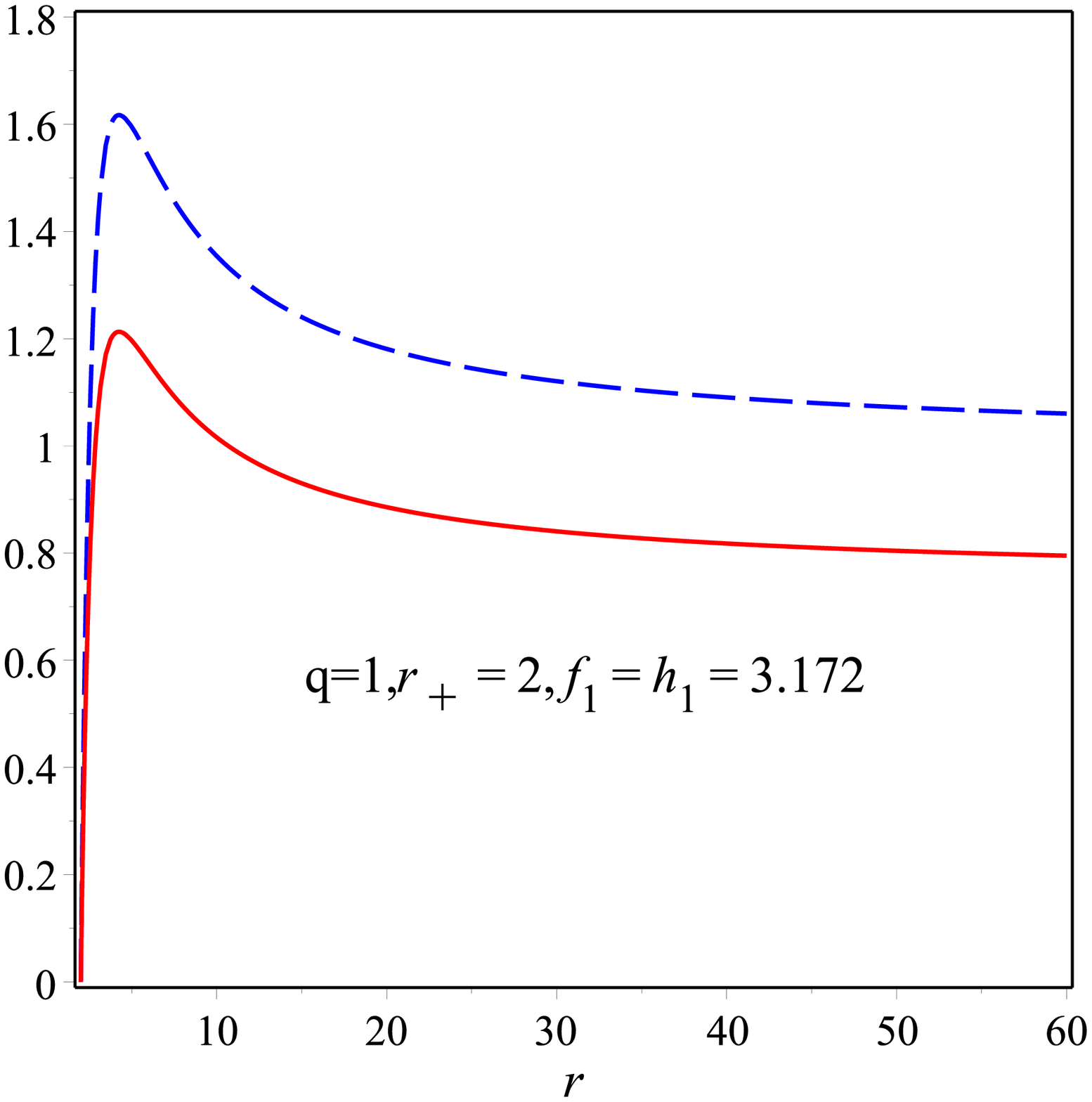}
 \label{fig2q1}
 }
 \subfigure[near horizon]
 {
 \includegraphics[width=0.3\columnwidth]{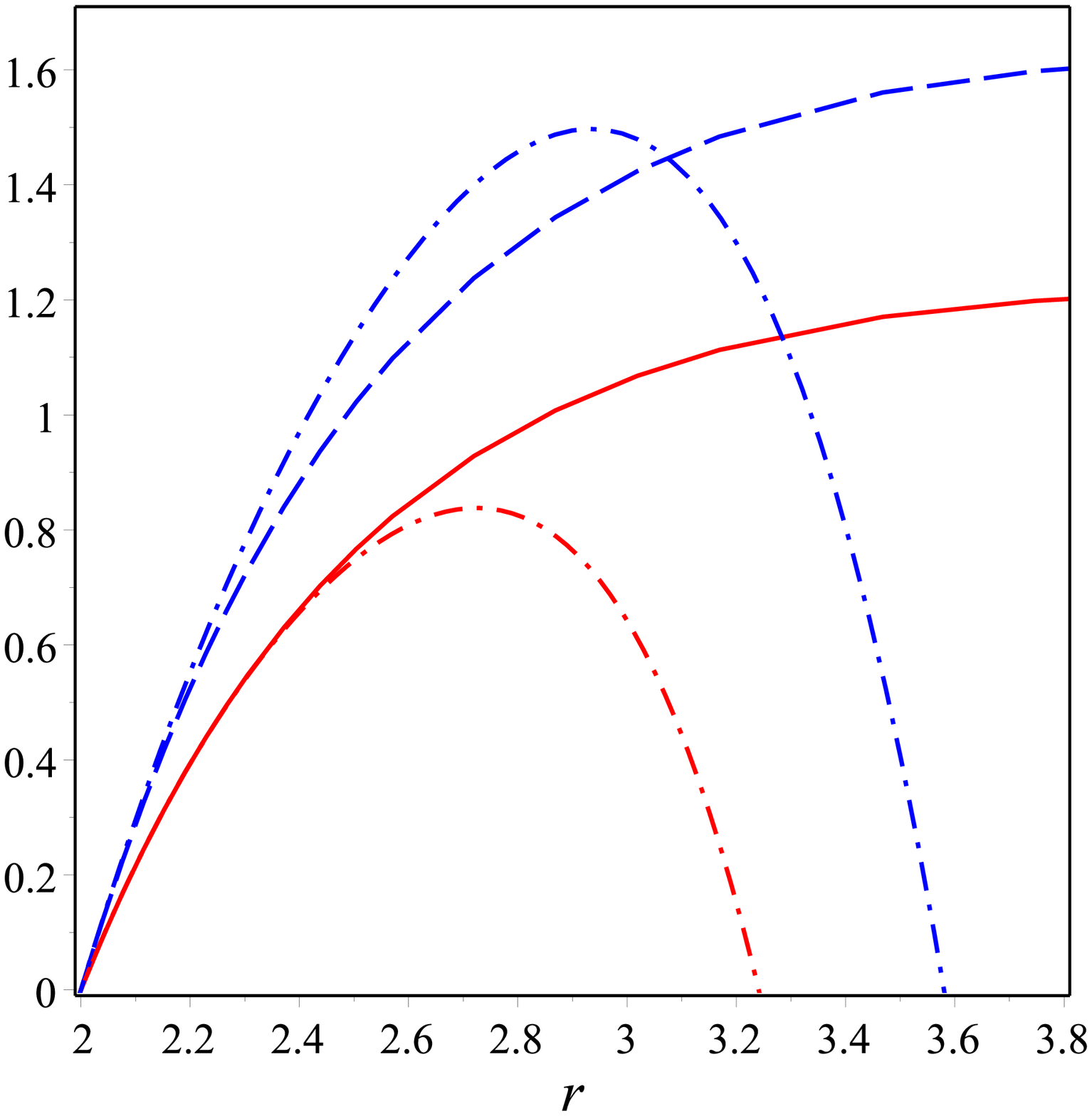}
 \label{fig2q2}
 }
 \subfigure[asymptotic]
 {
 \includegraphics[width=0.3\columnwidth]{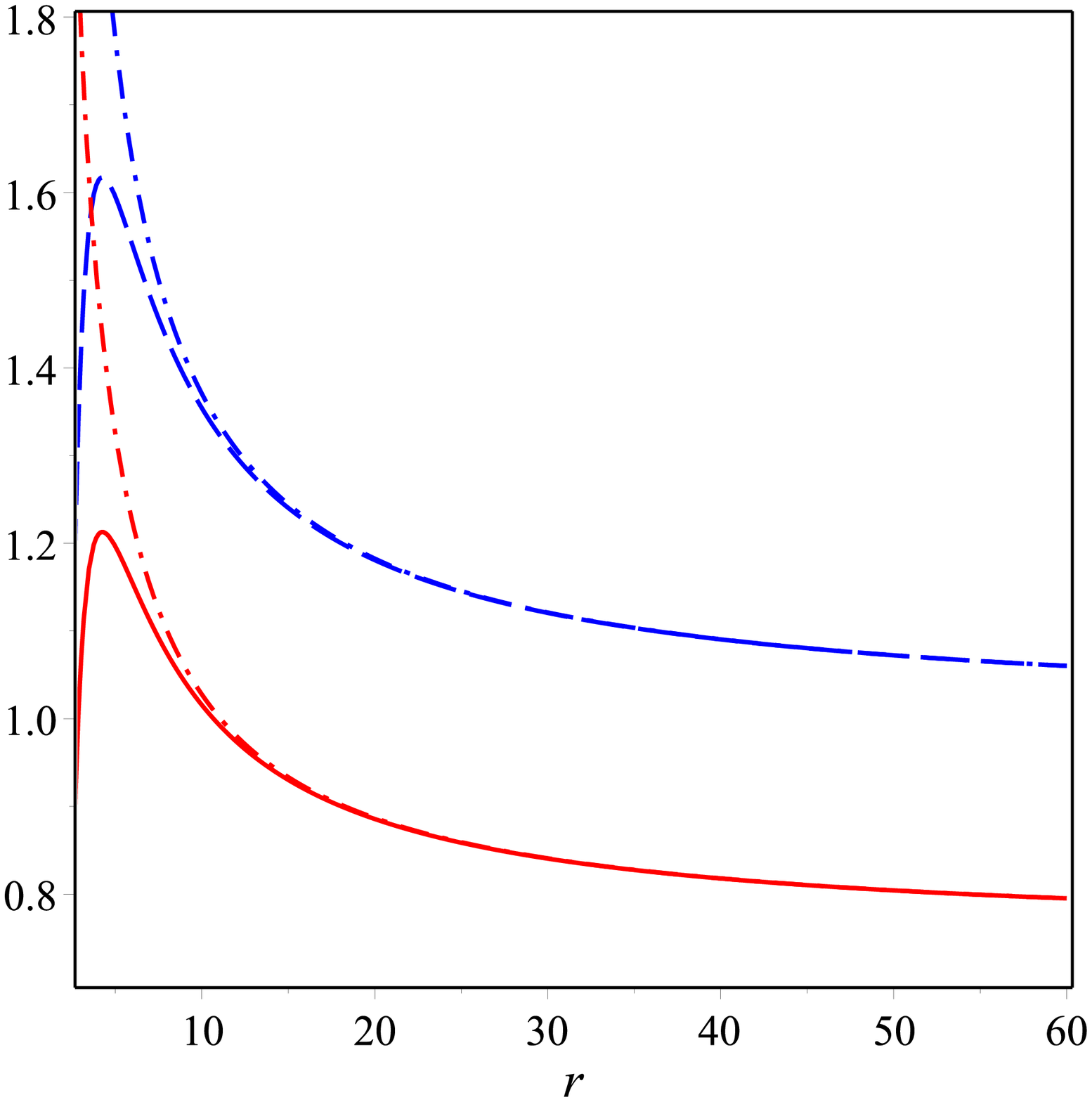}
 \label{fig2q3}
 }
 \subfigure[]
 {
 \includegraphics[width=0.3\columnwidth]{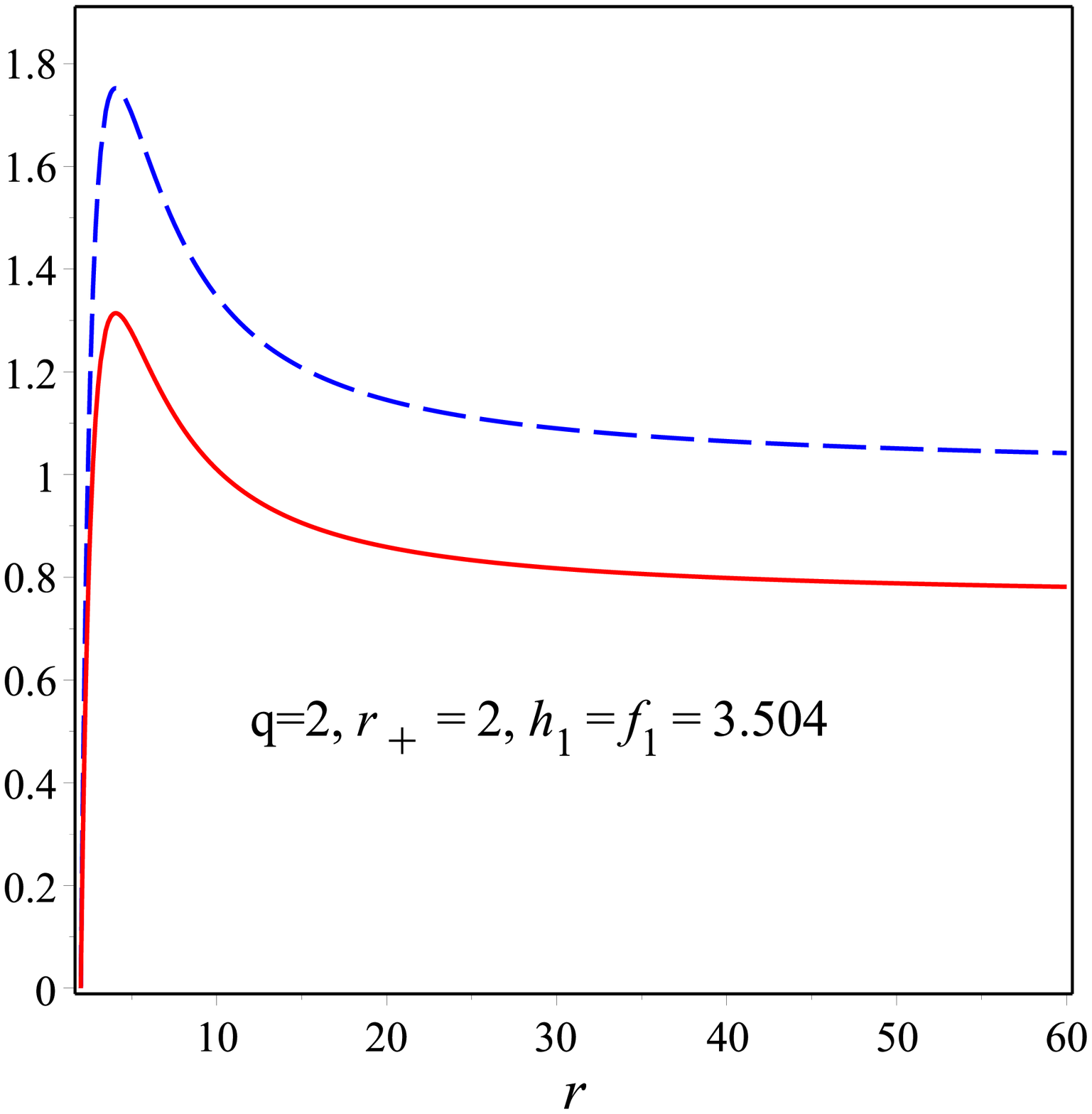}
 \label{fig2q4}
 }
 \subfigure[near horizon]
 {
 \includegraphics[width=0.3\columnwidth]{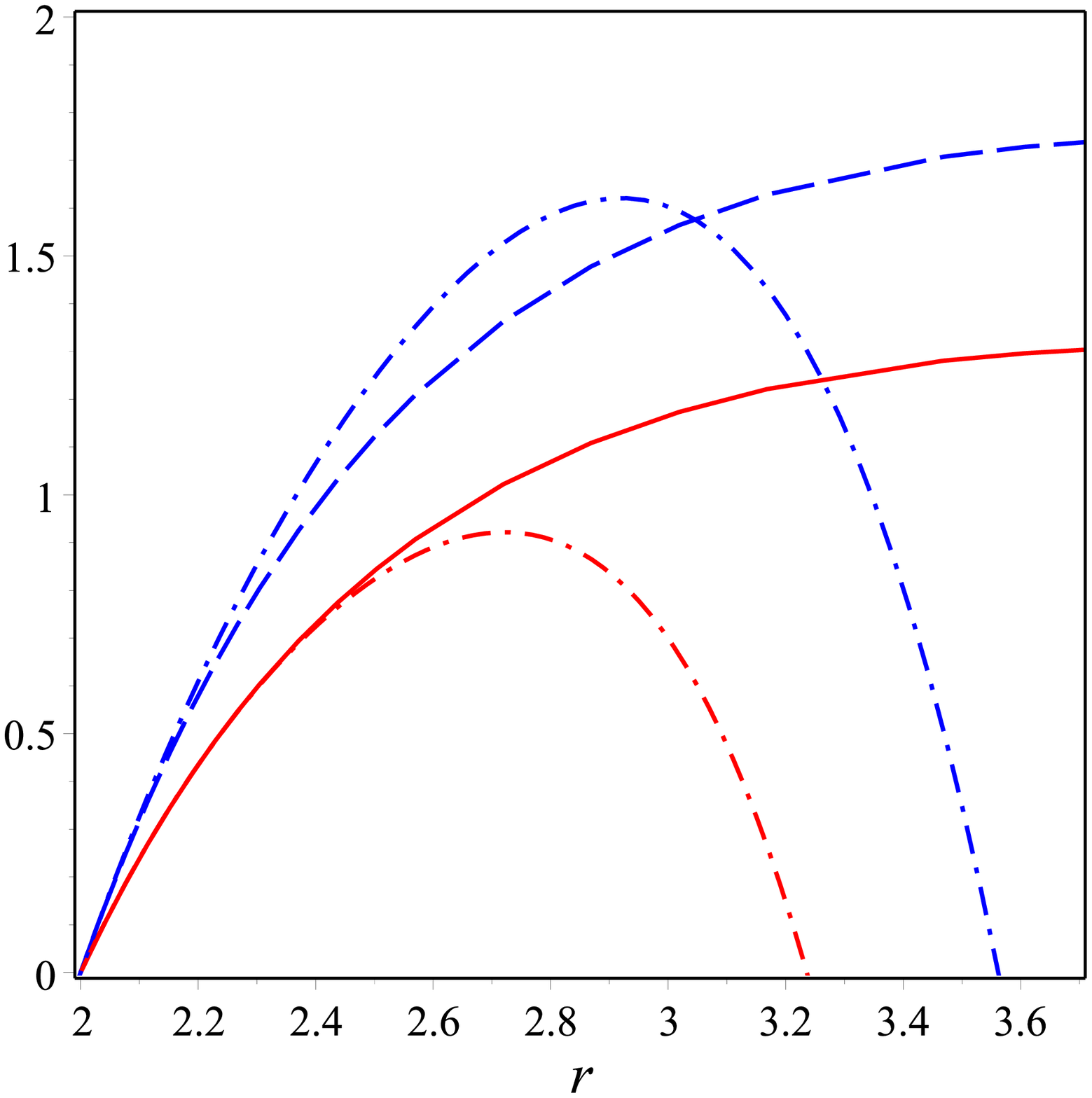}
 \label{fig2q5}
 }
 \subfigure[asymptotic]
 {
 \includegraphics[width=0.3\columnwidth]{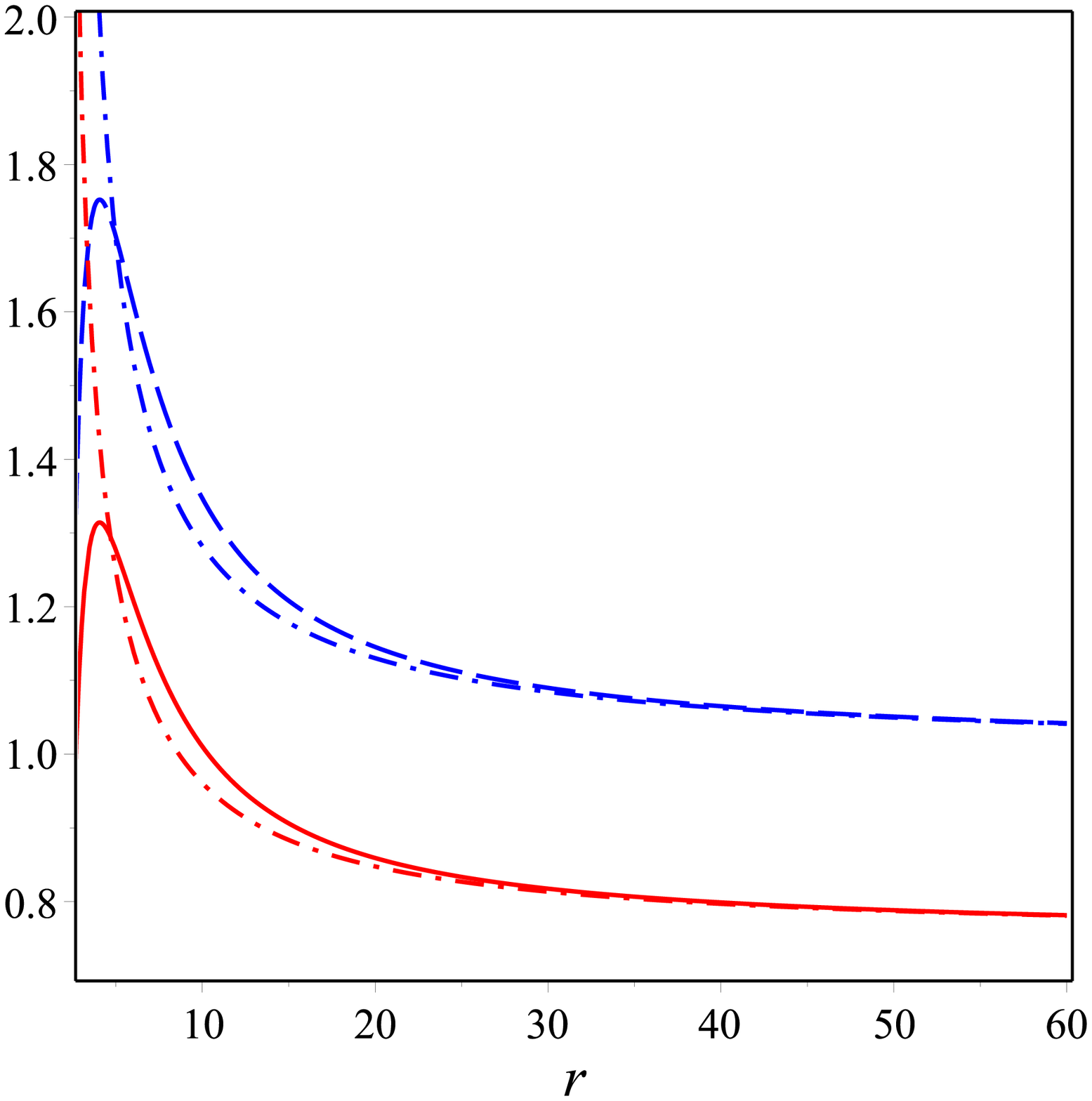}
 \label{fig2q6}
 }
\caption{The behavior of $f(r)  $ (\textcolor{blue}{ blue dashed line}) and $ 0.75h(r) $ (\textcolor{red}{ red solid line}) in terms of $ r $ for $ k=1, \alpha=0.5$, for the second group of solutions.
The top row is  $q=0$, the middle row $q=1$, and the bottom row $q=2$.  The left column
is the full continued fraction solution, the middle column compares this solution to the near-horizon approximation (dot-dashed lines), and that right column compares this solution to the large-$r$ approximation (dot-dashed lines).  All dimensionful quantities are in units of $M$.}
\label{fig2q}
\end{figure}

The result is an approximate analytic solution for both metric functions everywhere outside the horizon.   In Figures (\ref{fig1q})-(\ref{fig2q}) we present the solutions for $ f(r) $ and $ h(r) $, depicting the full continued
fraction solution (\ref{eq17}) along with its comparison to the near-horizon and large-$r$ 
series expansions, the latter given by dot-dashed lines.  We see that
the continued  fraction expansion converges to both of these other approximations.

We find two groups of solutions.  One group reduces to the Reissner-Nordstrom 
as $\alpha\to 0$ --   these are charged generalizations of the  uncharged case
studied previously \cite{Lu:2015cqa}.  This group of solutions is shown in
Figure~\ref{fig1q} for three different values of $q$.
The metric functions are increasing functions in $ r\geq r_{+} $.

Figure~\ref{fig2q} illustrates the second group of solutions for the same values of $q$. These solutions are physically distinct from the first group, having a peak outside of the event horizon.  This peak is
related to a negative mass \cite{Lu:2015cqa} for the black hole. For a static space time 
we have a timelike Killing vector 
$ \xi=\partial_{t} $ everywhere outside the horizon and so  we obtain
\begin{align}\label{eq20}
T &=\dfrac{1}{4\pi}\left. \sqrt{\dfrac{f(r_{+})}{h(r_{+})}}h^{'}(r)\right\vert_{r_{+}}
= \dfrac{f_{1}}{4\pi} = {\dfrac {(1-2\epsilon+a_{1}+a_{0})}{{4\pi r_{+}} \left( 1+{ b_{1}} \right) }}
 =\dfrac{ (1+\delta(r_{+},q))}{4\pi r_{+}} 
\end{align}
for the temperature $T$, where we have defined 
$ f_{1}=\dfrac{1+\delta}{r_{+}} $. We find in \eqref{b5} that $b_{1}$ has two values, which are $b_{1}=b^{-}_{1}=0$ and $b_{1}=b^+_{1}=-2$
for $f_1 = h_1$. We shall only consider the first of these, as the second one leads to negative temperature.

Extreme charged black hole solutions   exist 
if  $f_{1} = 0$, implying that  $a_1 = 2\epsilon-1-a_{0} $ and so $h_{1}=0$.  We then must also have
 $q^2 = r_+^2$  in order that the remaining 
 parameters in (\ref{eq7}) and (\ref{eq8}) are finite.  The mass for this branch of solutions is not always positive \cite{Lu:2015cqa}.
%Tile $ a_{3} $, metric function is as follow:
%\begin{equation}\label{eqap}
%h_{app}(r)\approx\left(1-{\dfrac {2{M}}{r}}+{\dfrac {{q}^{2}}{{r}^{2}}}-{\dfrac {{r_{+}
%} \left( {q}^{2}-2{ M}{r_{+}}+{{ r_{+}}}^{2} \right) }{{r}^
%{3}}}\right)+{\dfrac { 16\left( - \left( 1+\delta \right) {{r_{+}}}^{2}-4
%{M}{r_{+}}+{q}^{2}+3{{r_{+}}}^{2} \right) ^{2} \left( 1+
%\delta \right) }{{ r_{+}}{r}^{3} \left( 2 \left( 1+\delta \right) 
%{{r_{+}}}^{2}+{q}^{2}-2{{r_{+}}}^{2} \right) }\alpha} 
%\end{equation} 
%\begin{equation}
%B^{2}\approx1+O(\alpha^{2})
%\end{equation}

%\begin{figure}[H]\hspace{0.4cm}
%\centering
%\subfigure[]{\includegraphics[width=0.4\columnwidth]{trp3}}
%\subfigure[]{\includegraphics[width=0.4\columnwidth]{trp4}}
%\caption{ Plots of Temprature in terms of $r_{+}$ for $\alpha=0.5,k=1,\Lambda=0$ and $q=1, \delta=1.358, \delta_{1}=0.1,c=1$ (\textcolor{green}{green dashed line}), $ q=1, \delta=5.344,c=1,\delta_{1}=5.342 $ (\textcolor{yellow}{ yellow solid line}), $ q=2,\delta=2.52,\delta_{1}=2.51,c=1 $(\textcolor{black}{ black space dash line}), $ q=2,\delta_{1}=\delta=6.08,c=1 $(\textcolor{blue}{ blue dot dash line} )(left) and $ q=0, \delta_{1}=\delta=0,c=1 $ (\textcolor{red}{ red solid line}), $ q=1, \delta=0.144,\delta_{1}=-0.81,c=1 $(\textcolor{green}{green dashed line}), $ q=3,\delta=\delta_{1}=-0.504,c=1 $(black dot dash line)(right)} \label{fign}
%\end{figure}
We compute the entropy as follows \cite{Wald1,Wald2,Lu:2015psa}
\begin{align} 
S&=-2\pi\int_{Horizon}d^{2}x\sqrt{\eta}\dfrac{\delta L}{\delta R_{a b c d}}\epsilon_{a b}\epsilon_{c d}=\dfrac{A}{4}\left[1+4\alpha\left( \dfrac{1}{r_{+}^{2}}-\dfrac{f^{'}(r_+)}{r_{+}}\right)\right]
=\dfrac{A}{4}\left(1+\dfrac{4\alpha}{r_{+}^{2}}(1-r_{+}f_{1}) \right) \nonumber
\\
&=\dfrac{A}{4}\left(1-\dfrac{4\alpha \delta(r_{+},q)}{r_{+}^{2}}\right) 
\label{eq12}
\end{align}
where our choice $f_{1} = h_{1}$ implies  $ b_{1}=0 $ (as is clear from \eqref{b3b4} in Appendix A), leading to  $B=1$ in 
\eqref{Bx}.   We see that
if $ \delta(r_{+},q)>0 $ then $ \dfrac{4S}{A}<1 $.

The electric potential is  
\begin{equation}\label{eq25}
\phi=\int_{r_{+}}^{\infty}\dfrac{Bq}{r^{2}}dr=\dfrac{q}{r_{+}}
\end{equation}
 
 We now consider the thermodynamics of these black hole solutions, whose basic equations are 
the first law and Smarr formula 
\begin{equation}\label{eqfirstlaw}
dM=TdS+\phi dq
\end{equation}
\begin{equation}\label{eq26}
M=2TS+q\phi
\end{equation}
where there are no pressure/volume terms since we have set $\Lambda=0$. 
From Eq. \eqref{eq26} we have
\begin{equation}\label{eqmass}
M=\dfrac{(1+\delta(r_{+},q))r_{+}}{2}-\dfrac{2\alpha(1+\delta(r_{+},q))\delta(r_{+},q)}{r_{+}}+\dfrac{q^{2}}{r_{+}}
\end{equation}
yielding the mass parameter as a function of the horizon radius and the charge.  

 We now impose the first law \eqref{eqfirstlaw}, which becomes
\begin{equation}\label{flaw1}
\dfrac{\partial M}{\partial r_{+}}dr_{+}+\dfrac{\partial M}{\partial q}dq=T \dfrac{\partial S}{\partial r_{+}}dr_{+}+T \dfrac{\partial S}{\partial q}dq+\phi   dq
\end{equation}
yielding
\begin{multline}\label{flawrp}
\dfrac{\partial M}{\partial r_{+}}-T\dfrac{\partial S}{\partial r_{+}}=0,\hspace{0.5cm}\Longrightarrow\\\left[\dfrac{r_{+}}{2}-\dfrac{\alpha}{r_{+}}\right]\dfrac{\partial \delta(r_{+},q)}{\partial r_{+}}+\dfrac{\alpha \delta(r_{+},q)}{r_{+}}\left[-\dfrac{3\partial \delta(r_{+},q)}{\partial r_{+}}+\dfrac{2}{r_{+}} + \dfrac{2\delta(r_{+},q)}{r_{+}}\right]-\dfrac{q^{2}}{r_{+}^{2}}=0  
\end{multline}
and
\begin{multline}\label{flawq}
\dfrac{\partial M}{\partial q}-T\dfrac{\partial S}{\partial q}-\phi=0,\hspace{0.5cm}\Longrightarrow
\left[\dfrac{r_{+}}{2}-\dfrac{\alpha}{r_{+}}\right]\dfrac{\partial \delta(r_{+},q)}{\partial q}-\dfrac{3\alpha}{r_{+}}\delta(r_{+},q)\dfrac{\partial \delta(r_{+},q)}{\partial q}+\dfrac{q}{r_{+}}=0.
\end{multline}
as differential equations that must be satisified by $\delta(r_{+},q)$.

\begin{figure}[H]\hspace{0.4cm}
\centering
\subfigure[$K=0.9$]{\includegraphics[width=0.4\columnwidth]{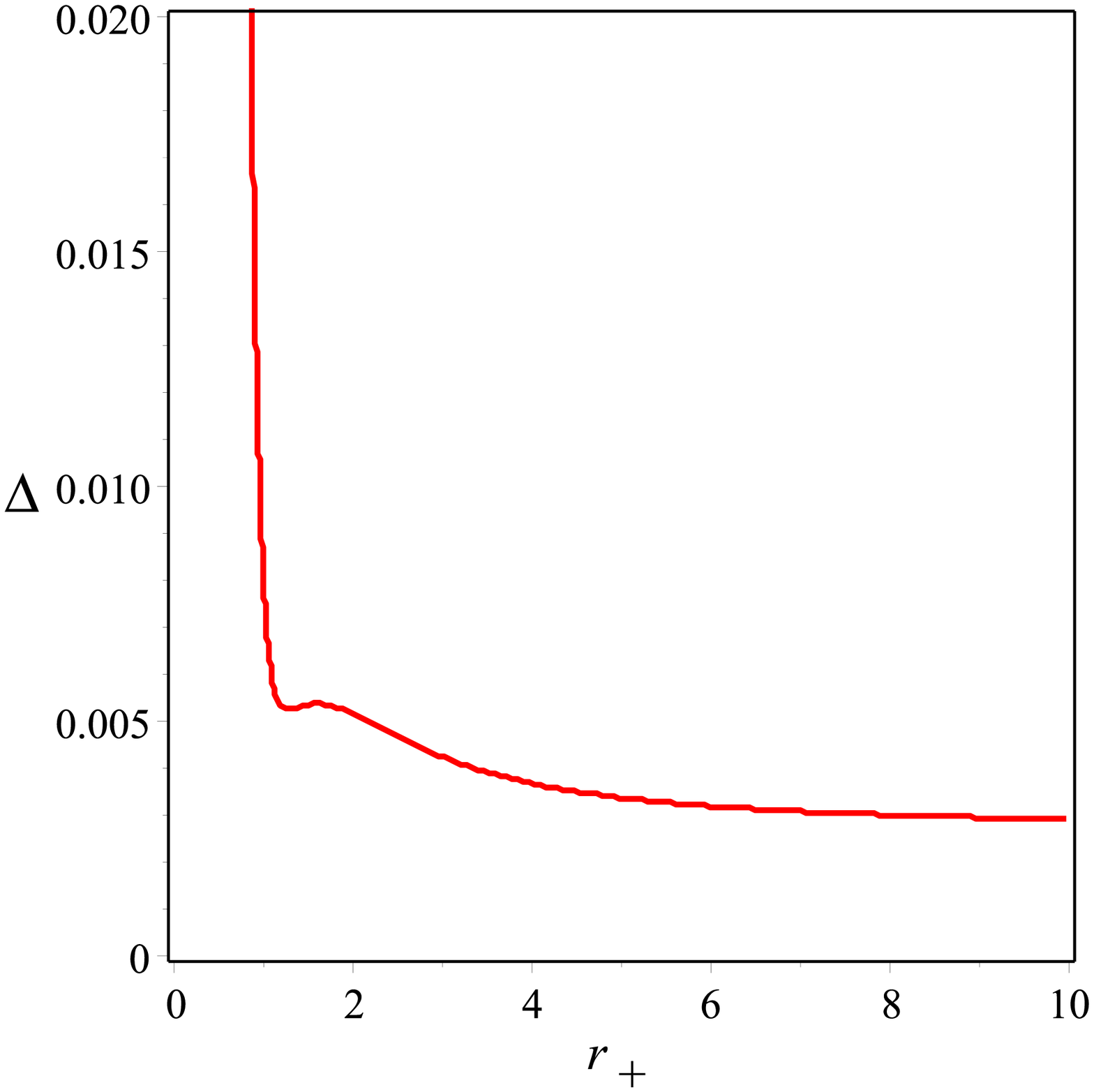}}
\subfigure[$ K=0.9$]{\includegraphics[width=0.4\columnwidth]{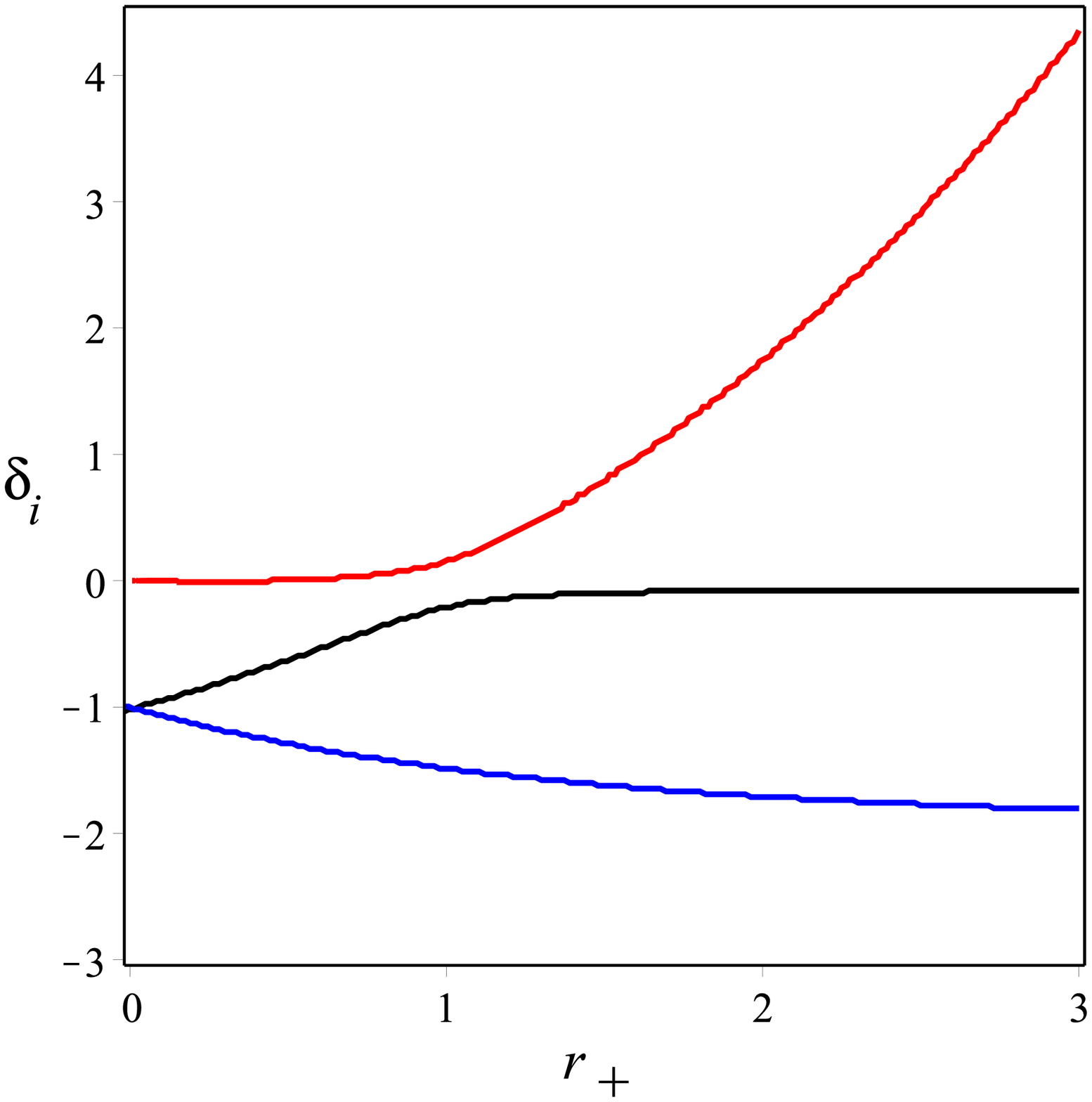}}\\
\subfigure[$K=1$]{\includegraphics[width=0.4\columnwidth]{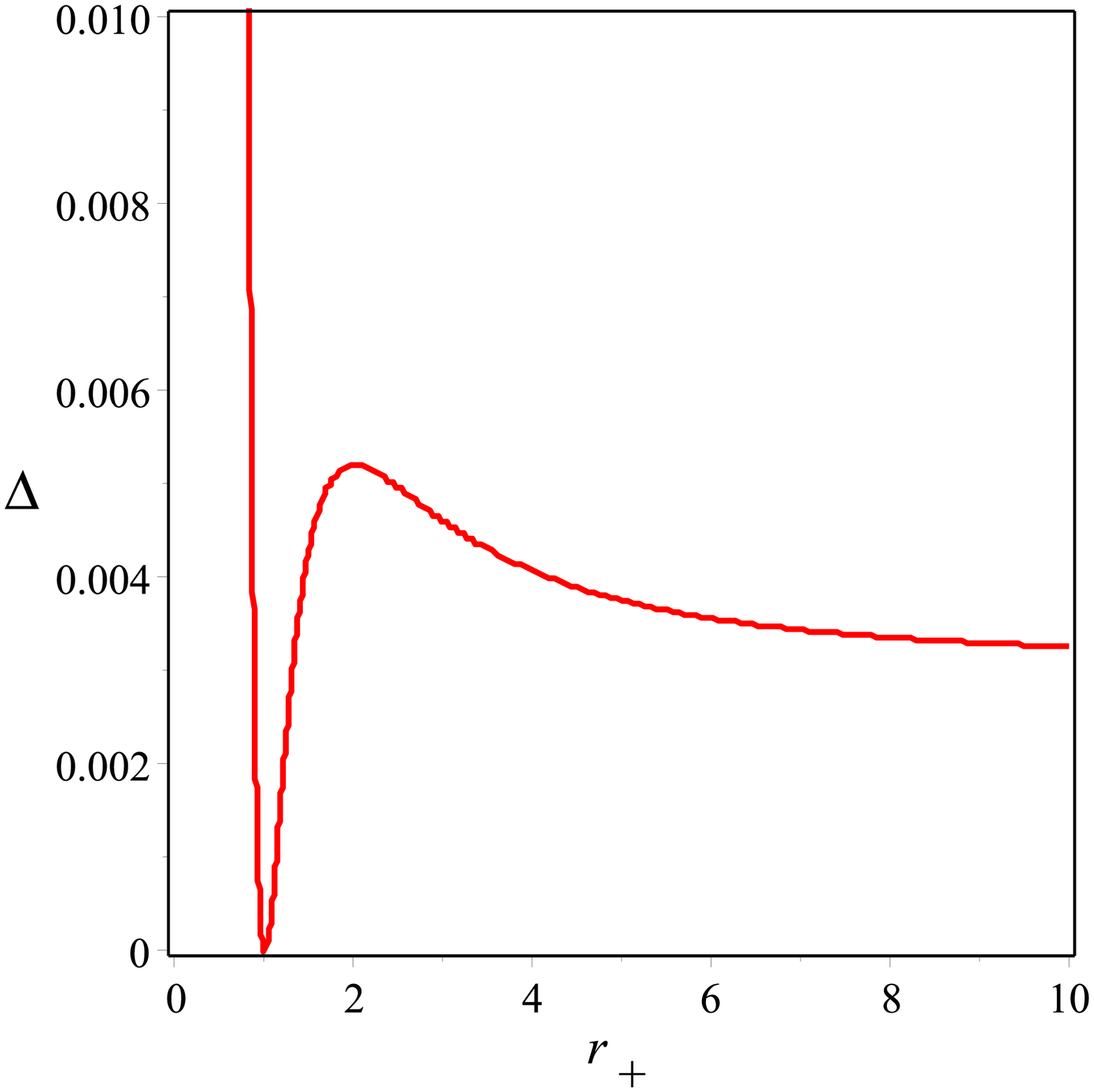}}
\subfigure[$ K=1$]{\includegraphics[width=0.4\columnwidth]{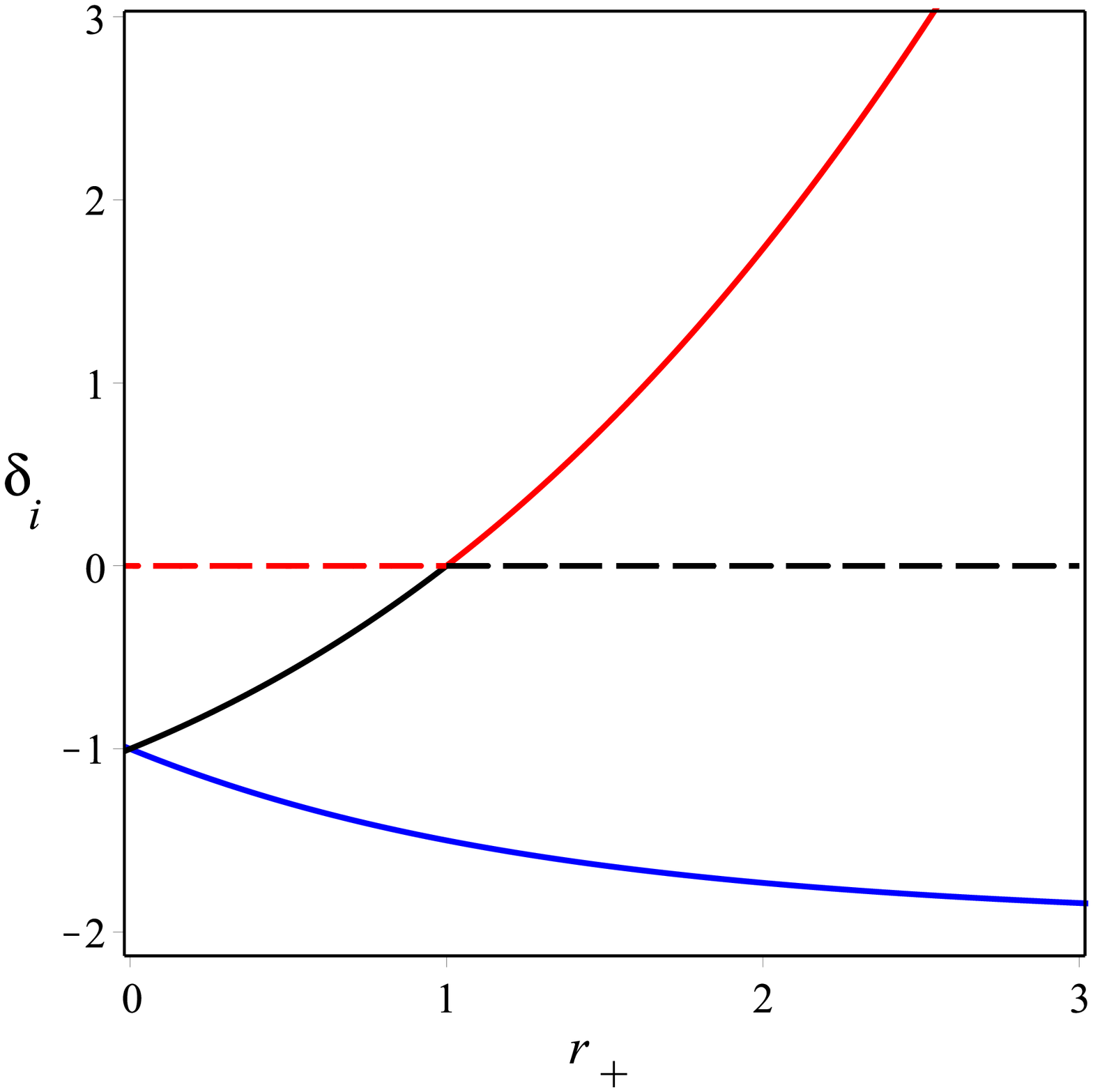}}\\
\subfigure[$K=1.1$]{\includegraphics[width=0.4\columnwidth]{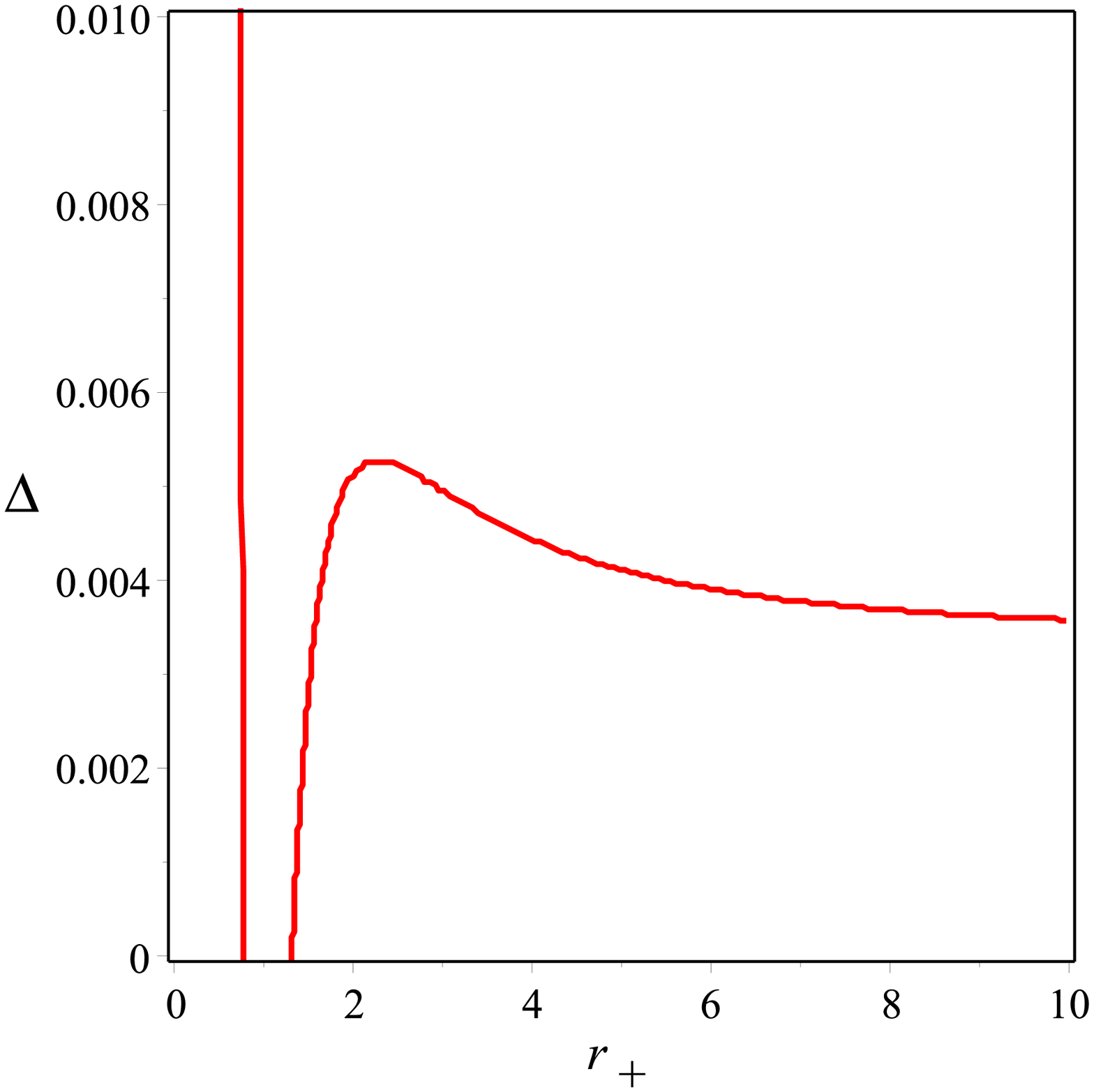}}
\subfigure[$K=1.1$]{\includegraphics[width=0.4\columnwidth]{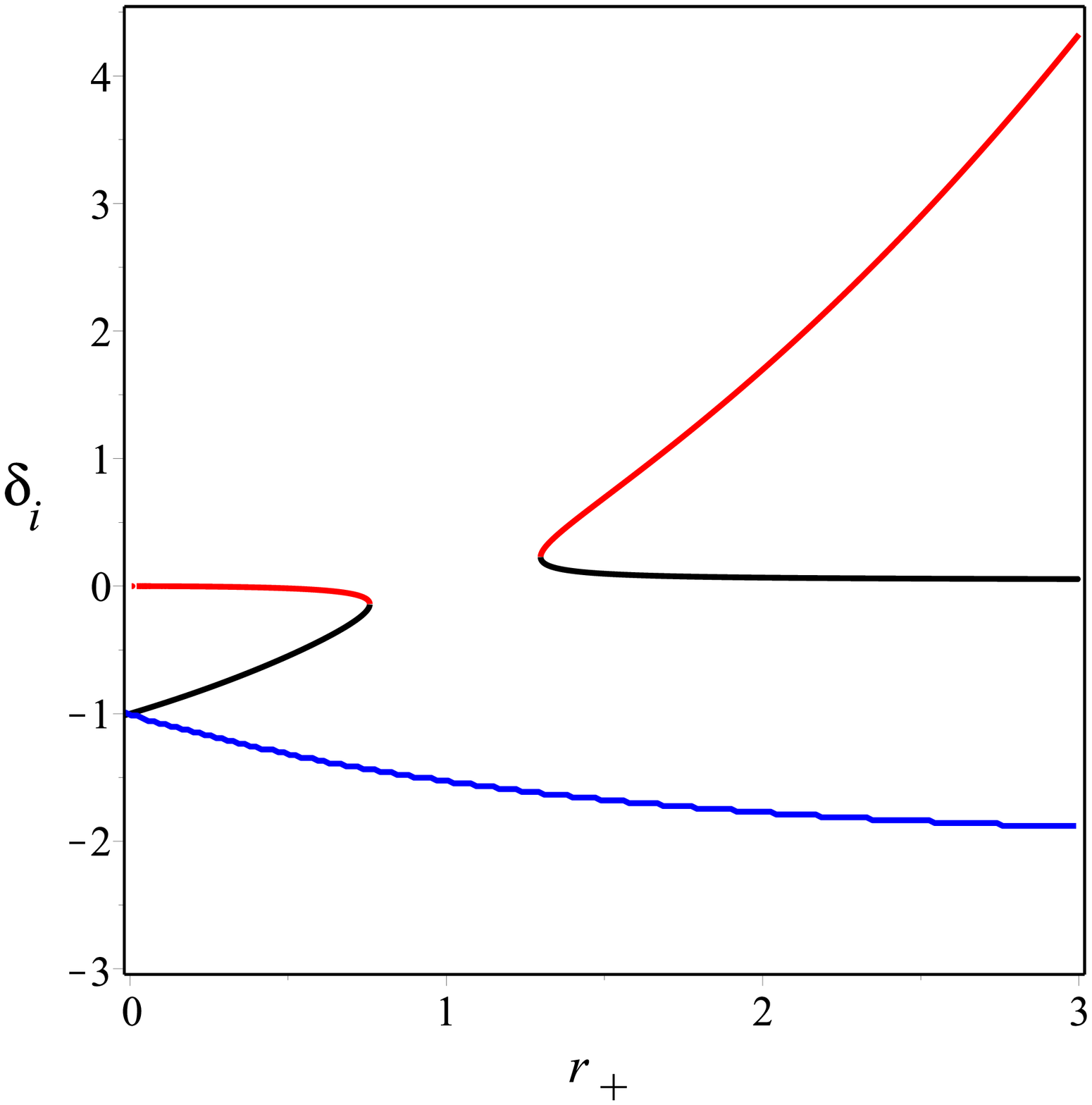}}
\caption{Plots of $ \Delta $ in terms of $r_{+}$  (left) and the behavior of roots ($\delta_{1}$ (red line), $\delta_{2}$ (blue line), $\delta_{3}$ (black line)) in terms of $r_{+}$ (right) for $\alpha=0.5$ and different values of $K$.} 
\label{Delta-delta}
\end{figure}

Consider first the neutral case $q=0$.  Equation (\ref{flawrp}) yields 
\begin{equation}\label{eq36delta}
-\dfrac{(r_{+}^2-2\alpha)\delta(r_{+})(1+\delta(r_{+}))^2}{3r_{+}^2}+\dfrac{\delta(r_{+})^{3}}{3}+\dfrac{\delta(r_{+})^2}{2}+\frac{K-1}{6}= 0.
\end{equation}
As can be seen, the equation (\ref{eq36delta}) is cubic and there is at least one root for $\delta$ depends to the sign of the discriminant $\Delta$. The discriminant for cubic equation (\ref{eq36delta}) is
\begin{equation}\label{eqdiscrim}
\Delta=\dfrac{K(r_{+}^{6}+12\alpha r_{+}^{4}+\alpha^{2}(48K-108)r_{+}^2+64\alpha^{3})}{324 r_{+}^{6}}.
\end{equation}
for positive values of $\alpha$ and $K$, for large and small $r_{+}$ the discriminant is positive which means there are three roots. There will be an intermediate range of $r_{+}$ where equation (\ref{eqdiscrim}) depending on the values of $K$ and $\alpha$ could go negative.
as an algebraic equation for $\delta(r_+)$,  with $K$ a constant of integration and \eqref{eq36delta} trivially satisfied.  
Solving this yields
\begin{align}
\delta_{1}(r_+)  &= A+\dfrac{(r_{+}^2+4\alpha)^{2}}{144\alpha^{2}A}+\dfrac{r_{+}^2-8\alpha}{12\alpha}\\
\delta_{2}(r_+)  &=\dfrac{-(1-\sqrt{3}i)A}{2}+\dfrac{r_{+}^{2}-8\alpha}{12\alpha}-\dfrac{(r_{+}^{2}+4\alpha)^2(1+\sqrt{3}i)}{288\alpha^2 A}\\
\delta_{3}(r_+)  &=\dfrac{-(1+\sqrt{3}i)A}{2}+\dfrac{r_{+}^{2}-8\alpha}{12\alpha}+\dfrac{(r_{+}^{2}+4\alpha)^2(\sqrt{3}i-1)}{288\alpha^2 A}
\end{align} 
where
\begin{align}
&A =\nonumber\\
&\dfrac{\left(r_{+}^6+12\alpha r_{+}^{4}+64\alpha^3 -24( 9K-2) \alpha^2 r_{+}^2+12\sqrt{3}\alpha r_{+} 
\sqrt{K\left( 12 \alpha^2 r_{+}^2 (9 K-4)  -(r_{+}^6+12\alpha r_{+}^{4}+64\alpha^3)\right)}
\right)^{\frac{1}{3}} }{12\alpha}.
\end{align}
For different values of $K$, the discriminant and the roots have been shown in Fig. \ref{Delta-delta}. As can be seen, from Fig. \ref{Delta-delta} for $K\leq 1$, $\Delta$  for all values of $r_{+}$ is positive. This shows there are three real solution for $\delta$ which have been depicted in Fig. \ref{Delta-delta}(b, d, f). For $K=1.1$, $\Delta$ in small $r_{+}$ becomes negative which leads to a solution for $\delta$ as shown in Fig. \ref{Delta-delta}f.  
In Fig. \ref{Delta-delta}d, $\delta$ for the dashed line curves vanishes which is corresponds to the Schwarzschild-like behaviour. Any deviation from this line (solid lines) corresponds to the non- Schwarzschild-like behaviour.  \\

\begin{figure}[H]
\centering
\subfigure[$K=0.9$]{
 \includegraphics[width=0.3\columnwidth]{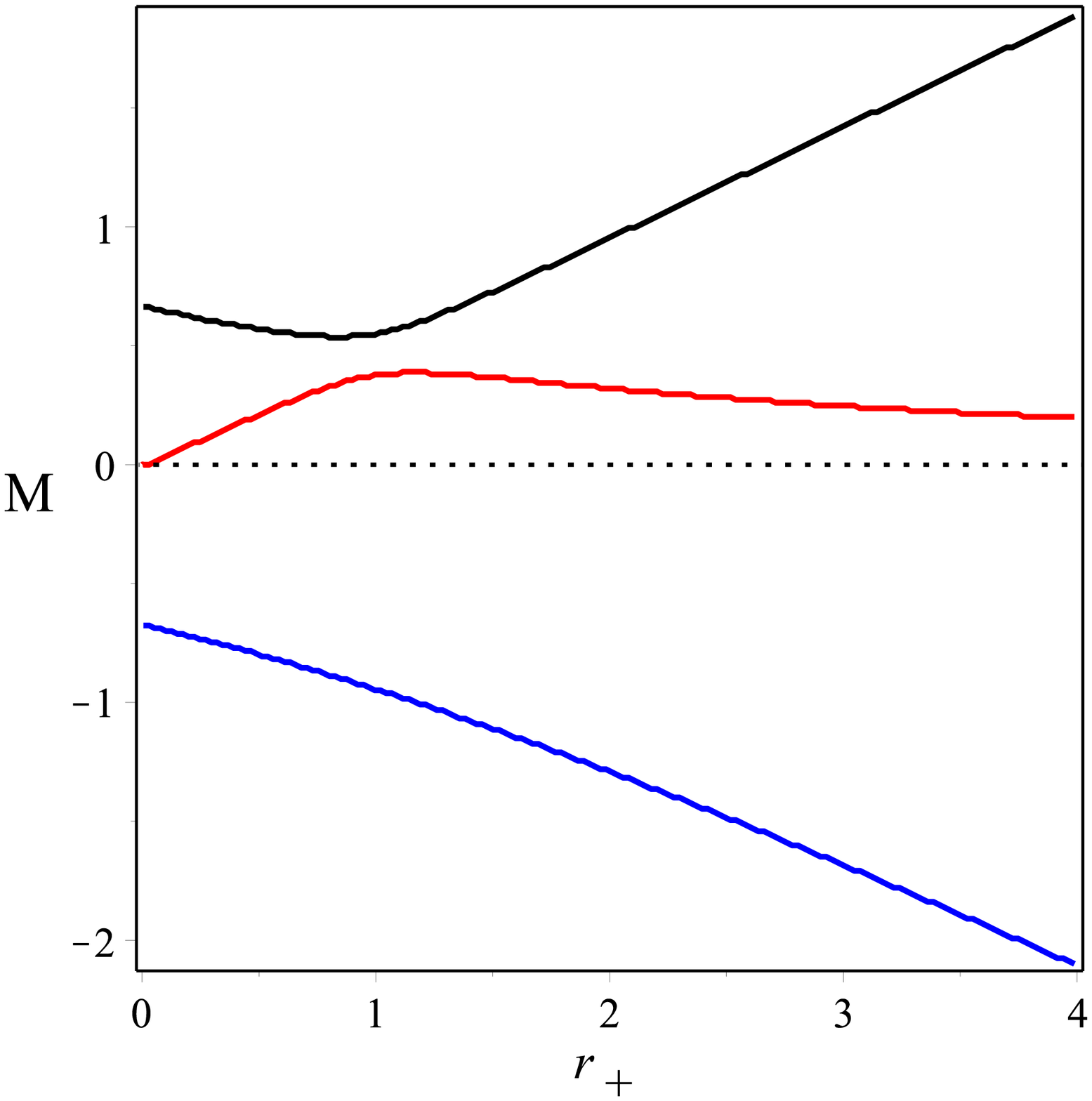}
 \label{fignT}
 }
 \subfigure[$ K=0.9$]
 {
 \includegraphics[width=0.3\columnwidth]{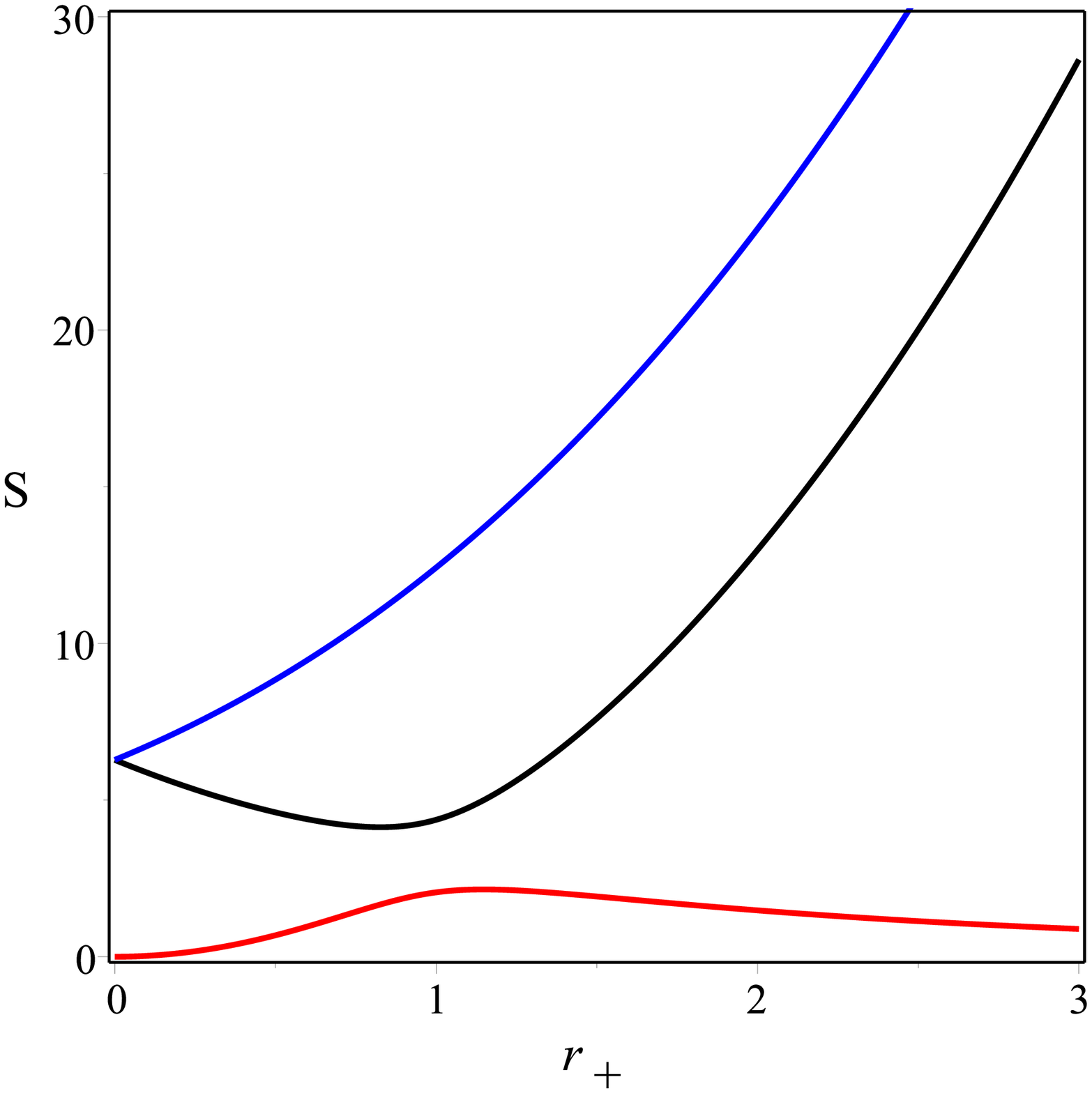}
 \label{fignM}
 }
 \subfigure[$ K=0.9$]
 {
 \includegraphics[width=0.3\columnwidth]{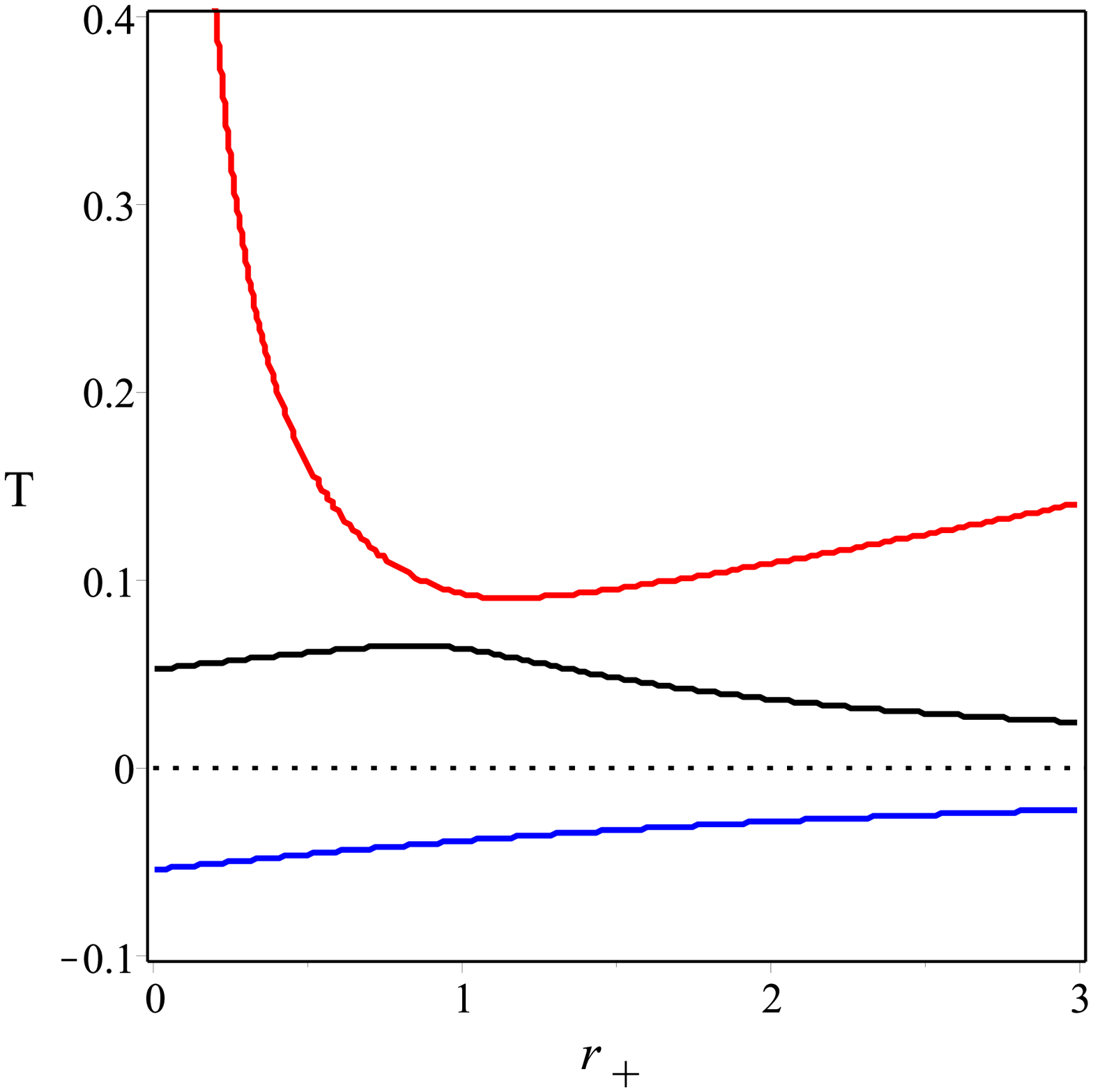}
 \label{fignS}
 }
\subfigure[$ K=1$]{
 \includegraphics[width=0.3\columnwidth]{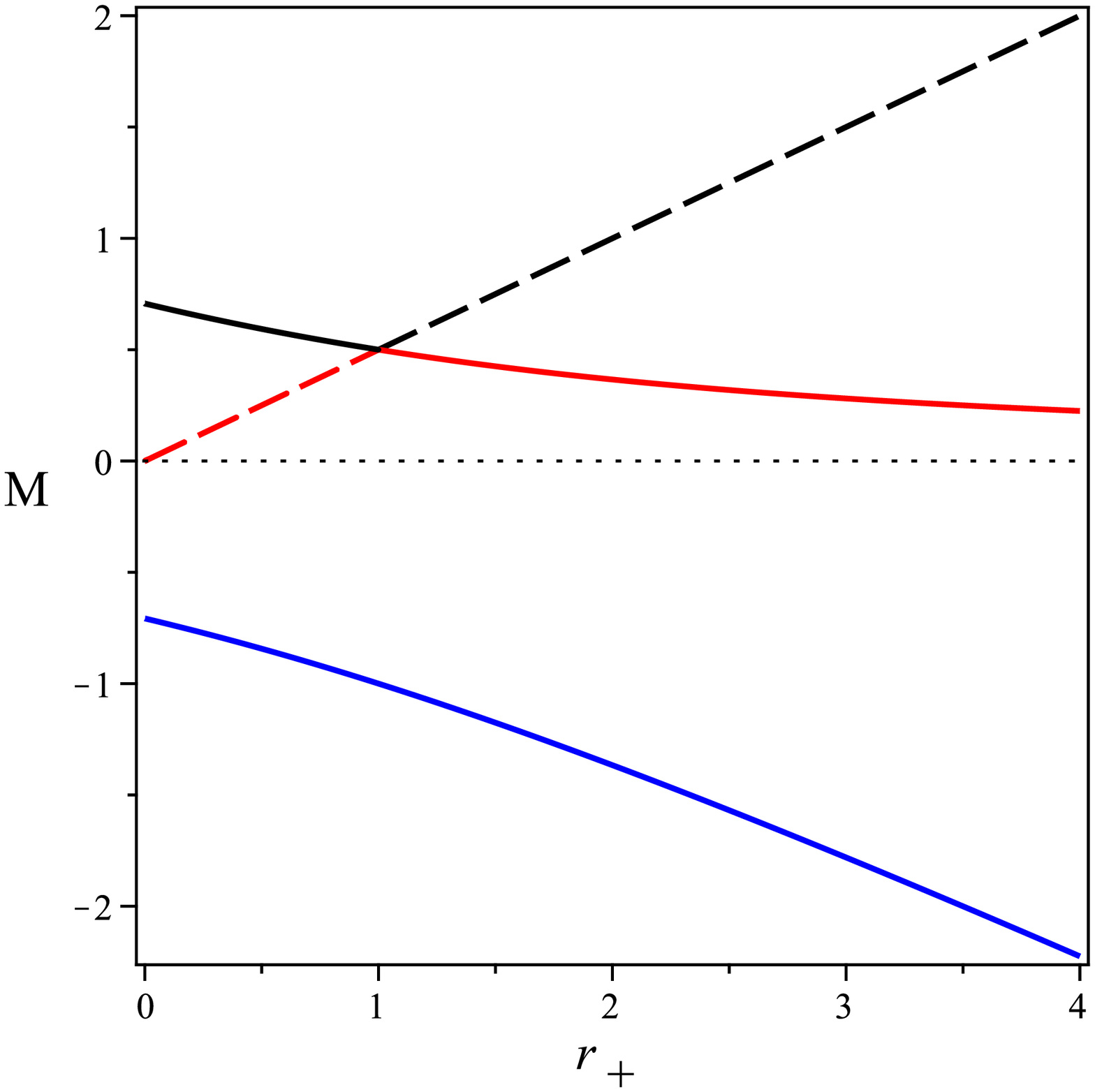}
 \label{fignT}
 }
 \subfigure[$ K=1$]
 {
 \includegraphics[width=0.3\columnwidth]{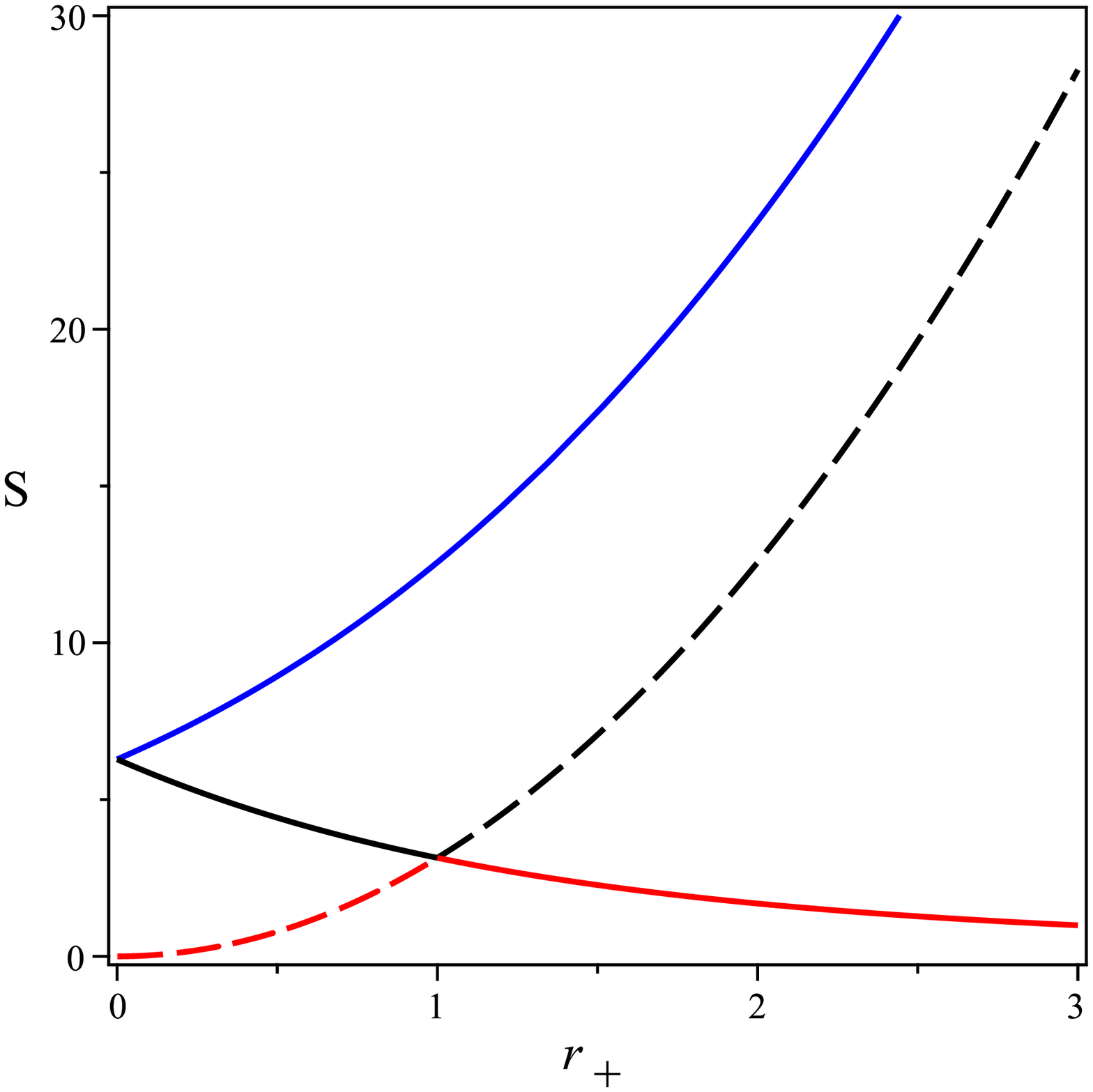}
 \label{fignM}
 }
 \subfigure[$ K=1$]
 {
 \includegraphics[width=0.3\columnwidth]{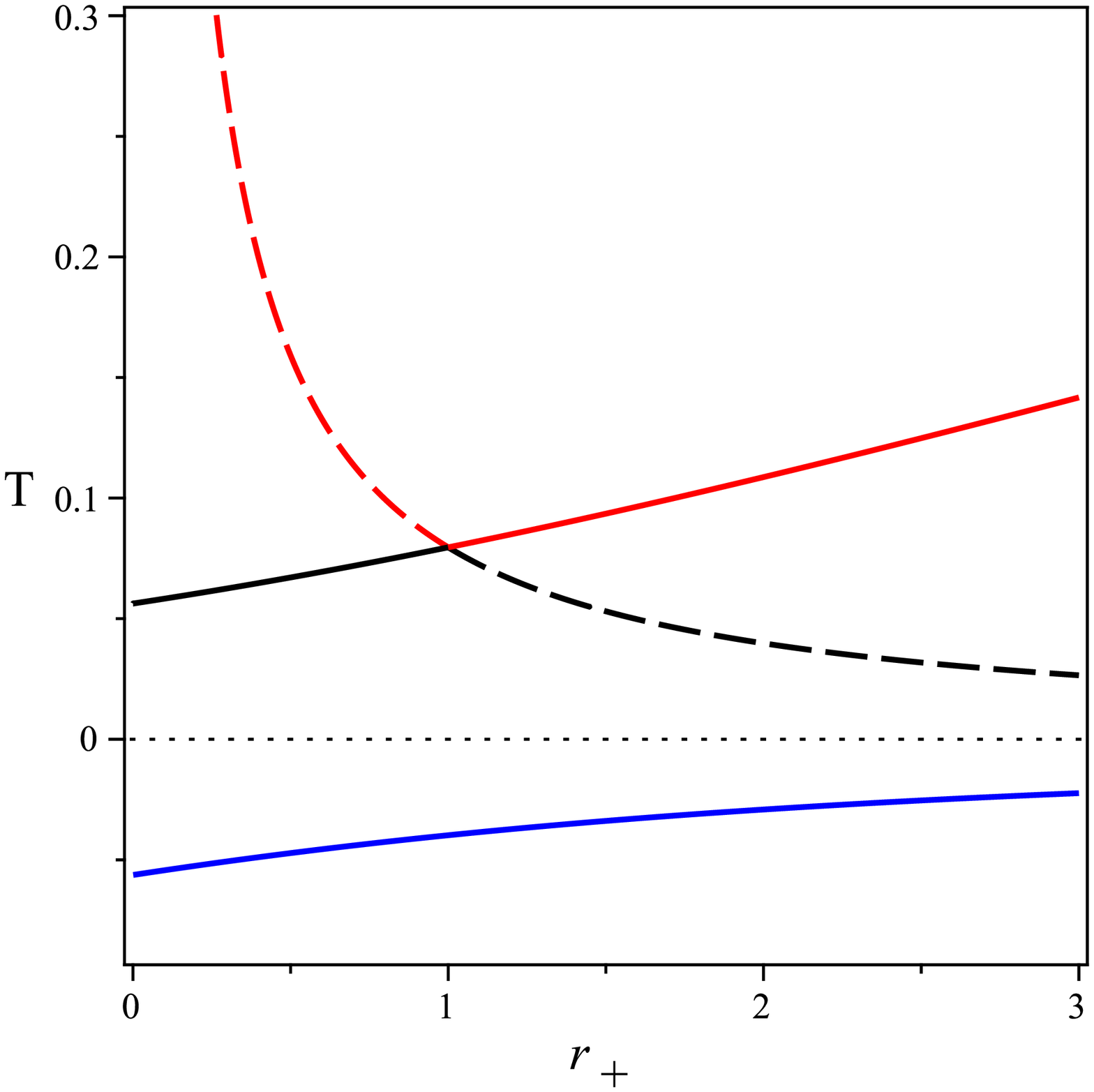}
 \label{fignS}
 }
 \subfigure[$K=1.1$]{
 \includegraphics[width=0.3\columnwidth]{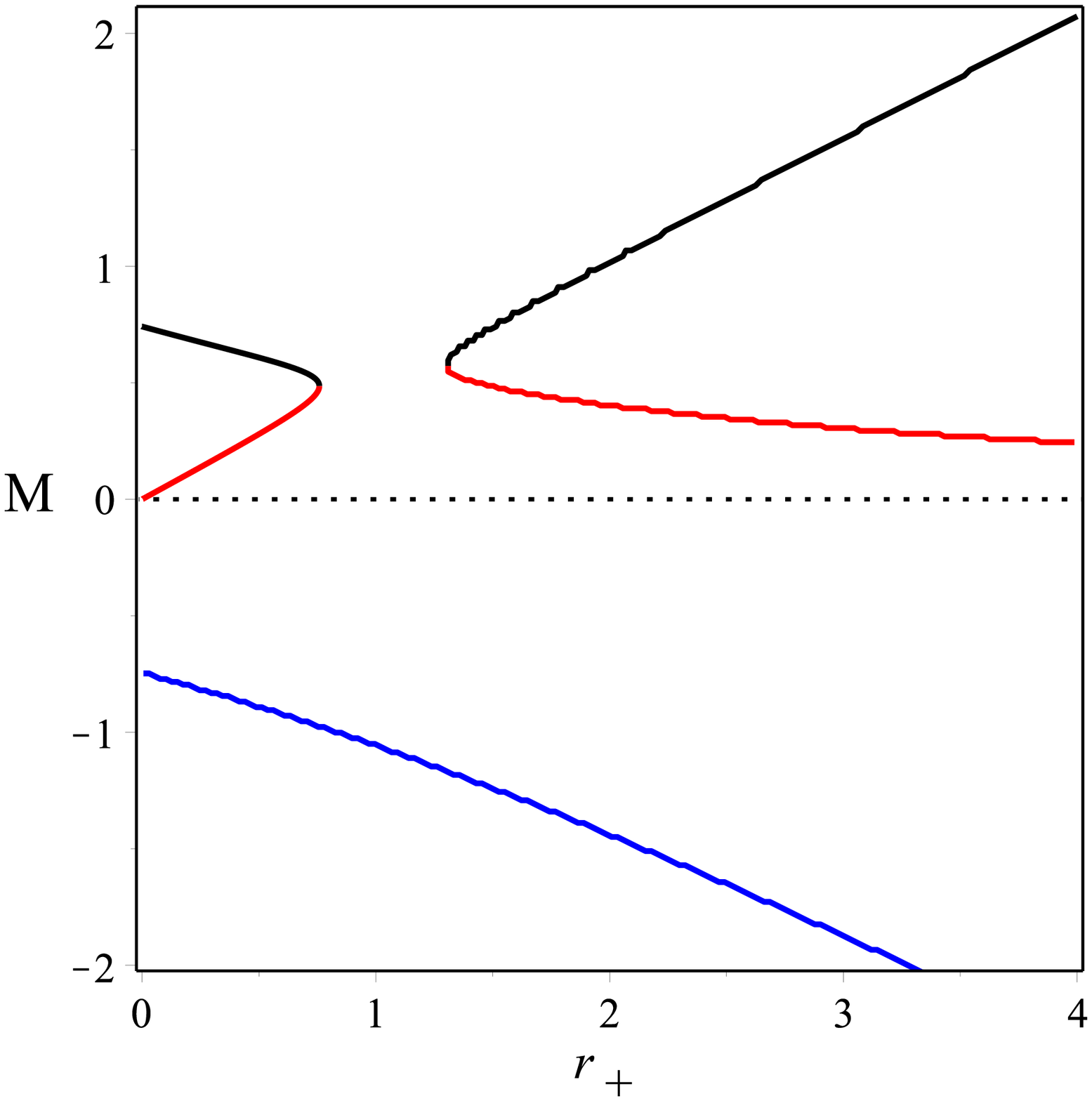}
 \label{fignT}
 }
 \subfigure[$ K=1.1$]
 {
 \includegraphics[width=0.3\columnwidth]{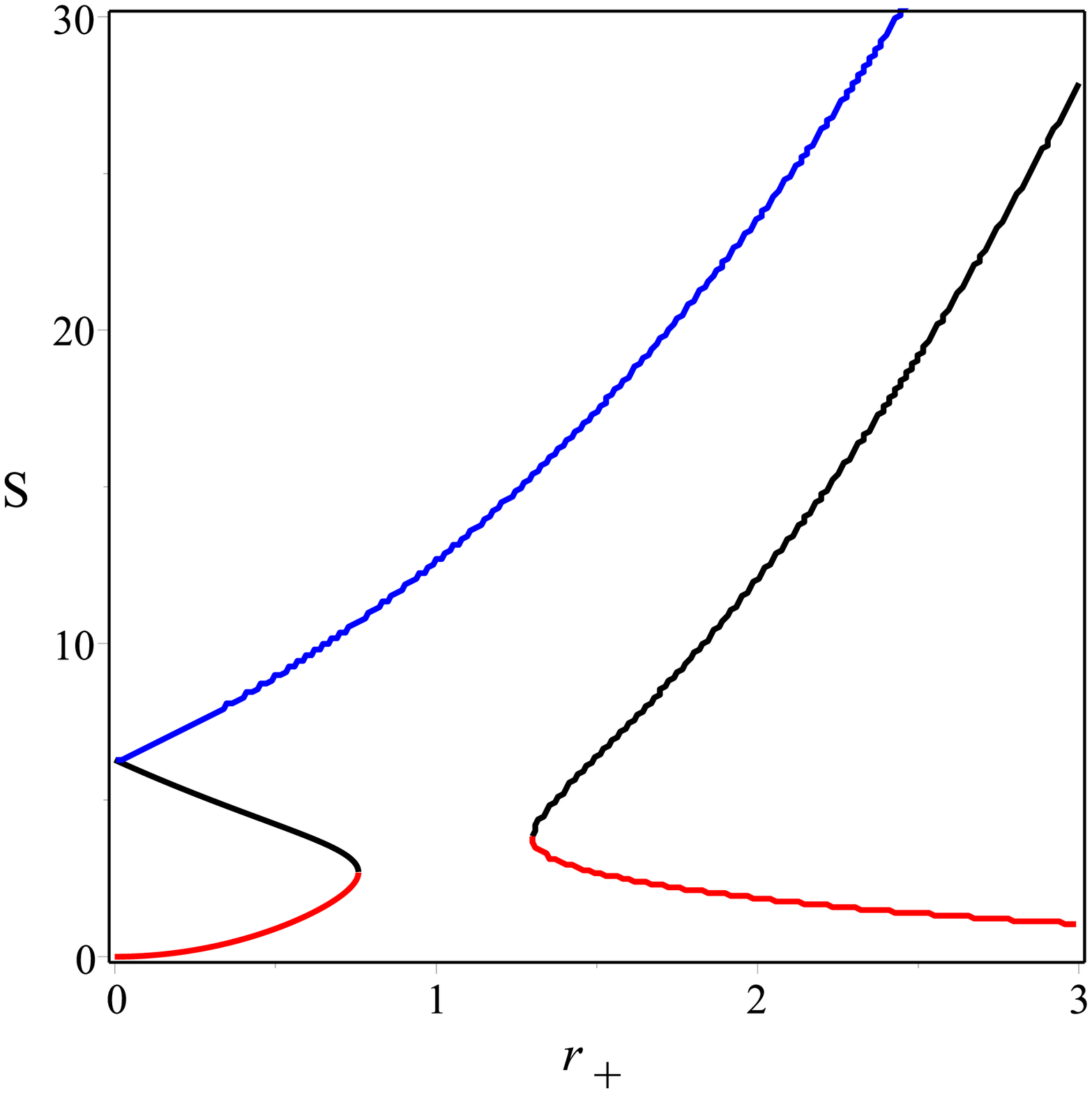}
 \label{fignM}
 }
 \subfigure[$ K=1.1$]
 {
 \includegraphics[width=0.3\columnwidth]{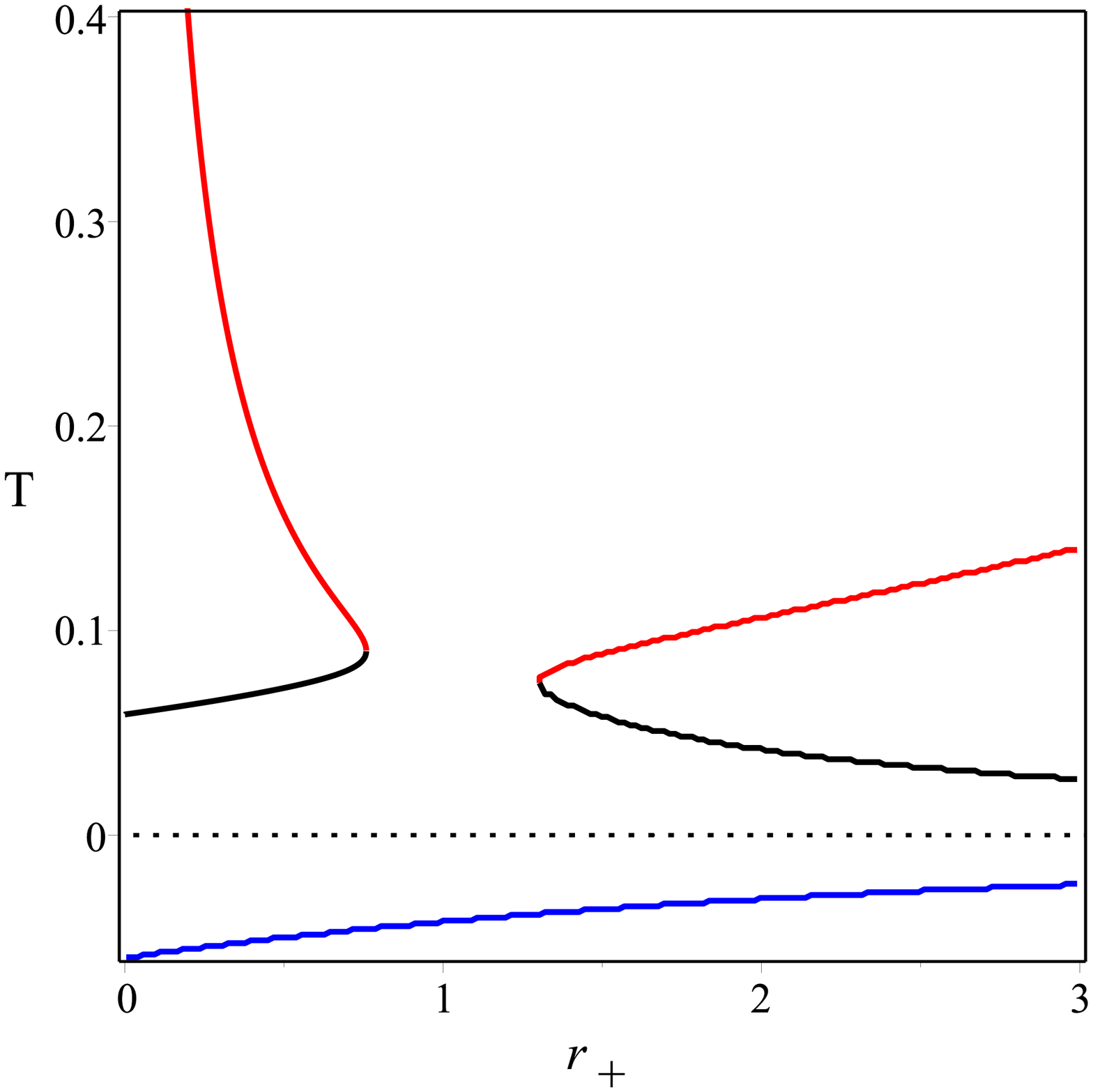}
 \label{fignS}
 }
 \caption{Plots of $ M $, $ S $ and $T$ in terms of $r_{+}$ for $\alpha=0.5$ and different values of $K$. 
In the middle figures the dashed line curves indicate  Schwarzschild-like behaviour and solid line curves non-Schwarzschild-like behaviour. The colors correspond to the colors in figure (\ref{Delta-delta}). }
\label{figmst}
\end{figure}

By inserting $\delta$ into  (\ref{eq20}), (\ref{eq12}) and (\ref{eqmass}), we obtain similar behavior for the temperature, entropy and mass of the black hole. 
In Fig.~\ref{figmst}  we illustrate the behavior of $M$, $S$ and $T$ in terms of $r_{+}$ for different values of $K$.
For $K=1$, the dashed and solid lines correspond to the Schwarzschild and non-Schwarzschild-like behaviour, respectivly. For this case the black solid line correspond to the cold non-Schwarzschild black hole while the red solid line correspond to the hot non-Schwarzschild black holes \cite{Bonanno:2019rsq}.

\begin{figure}[H]\hspace{0.4cm}
\centering
\subfigure[$K=0.9$]{\includegraphics[width=0.4\columnwidth]{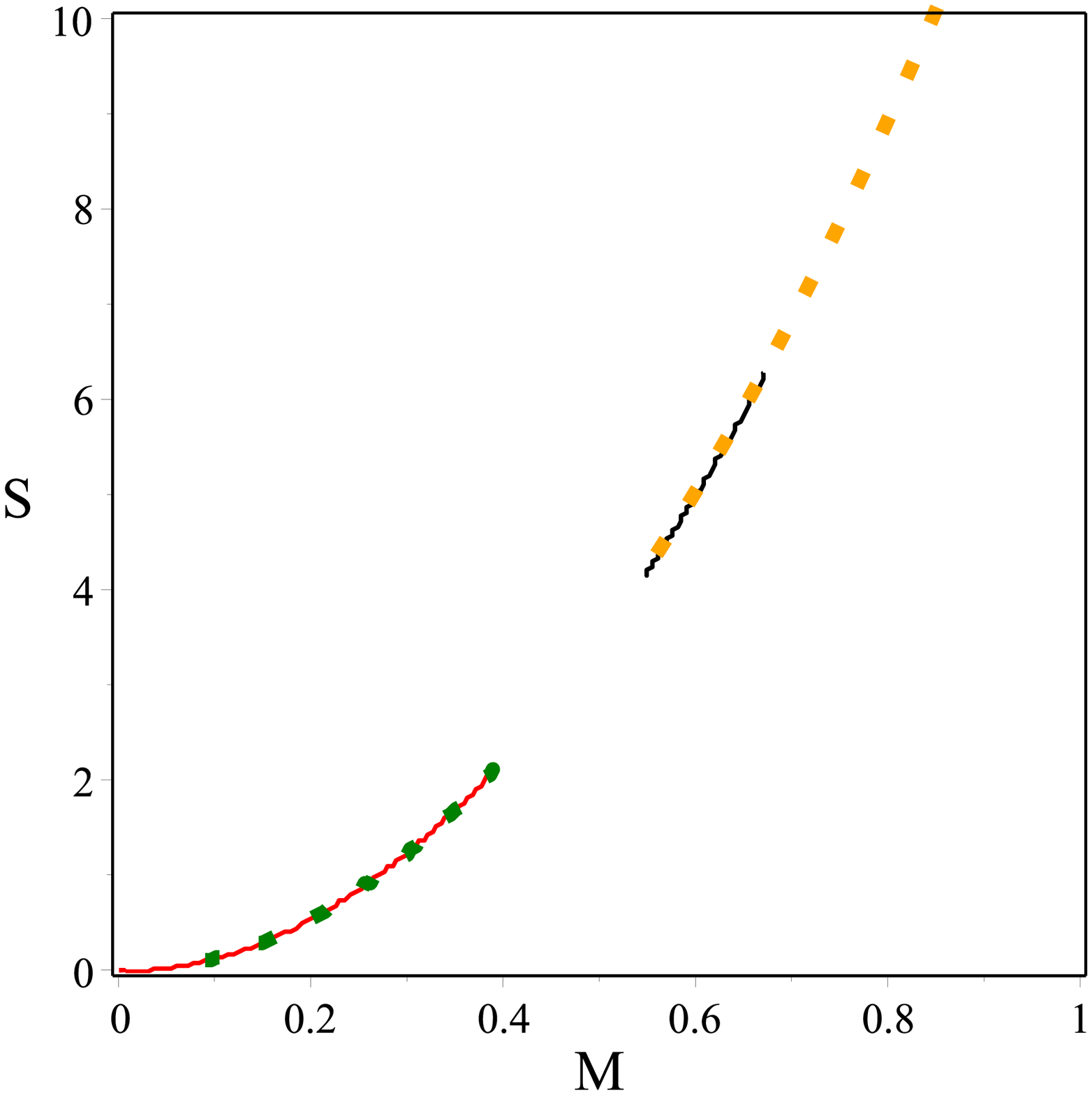}}
\subfigure[$K=0.9$]{\includegraphics[width=0.4\columnwidth]{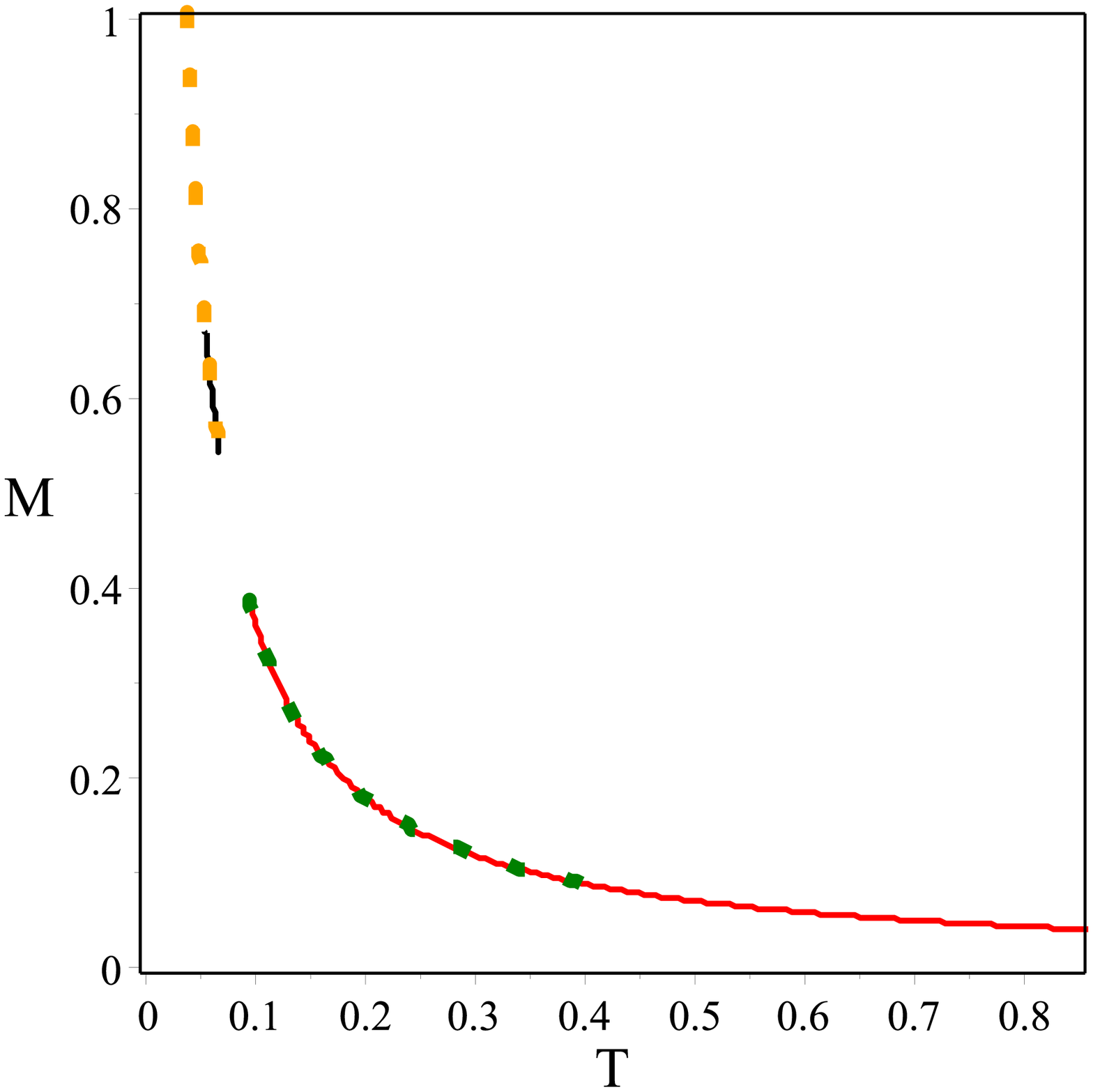}}
\subfigure[$K=1$]{\includegraphics[width=0.4\columnwidth]{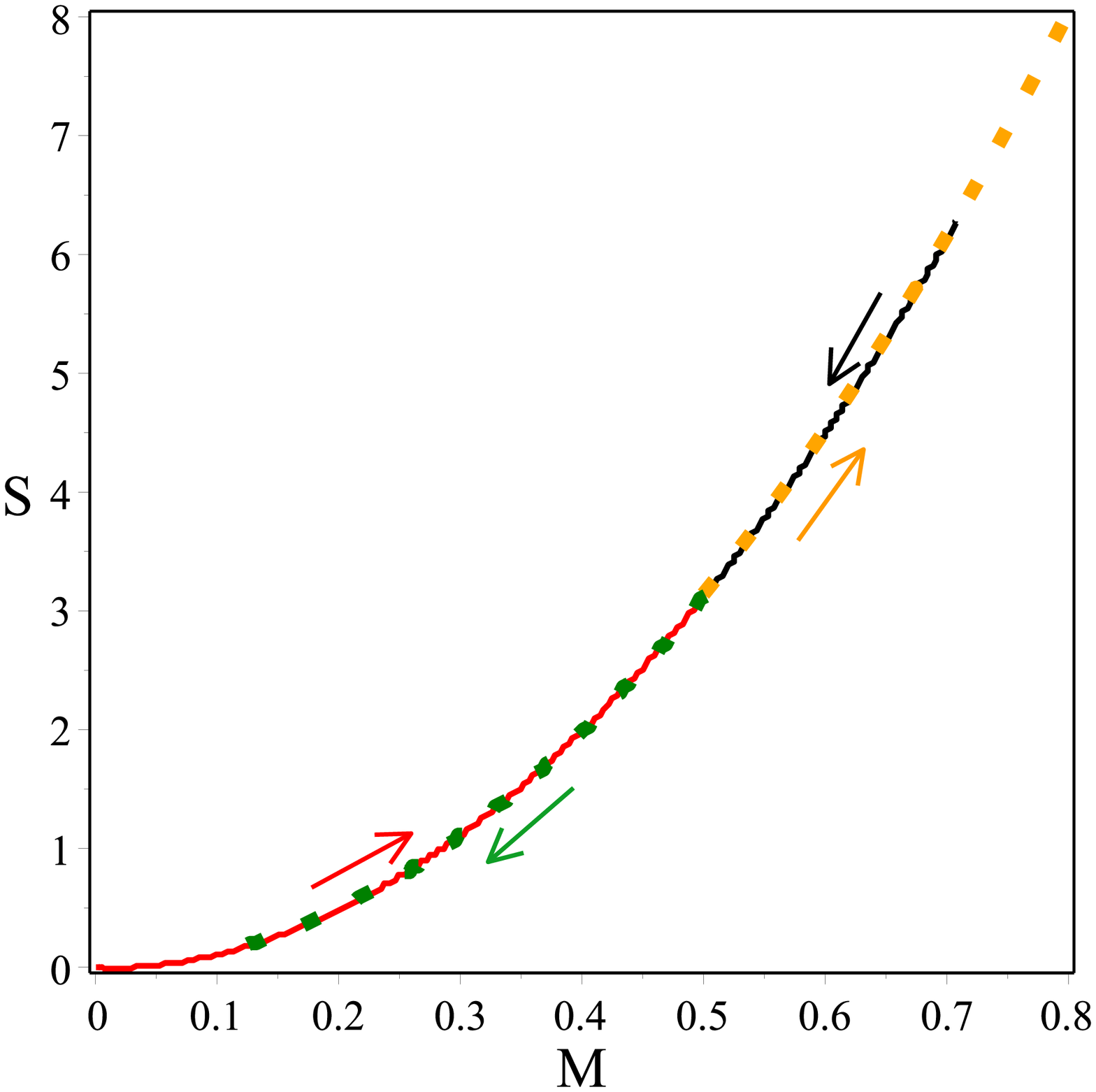}}
\subfigure[$K=1$]{\includegraphics[width=0.4\columnwidth]{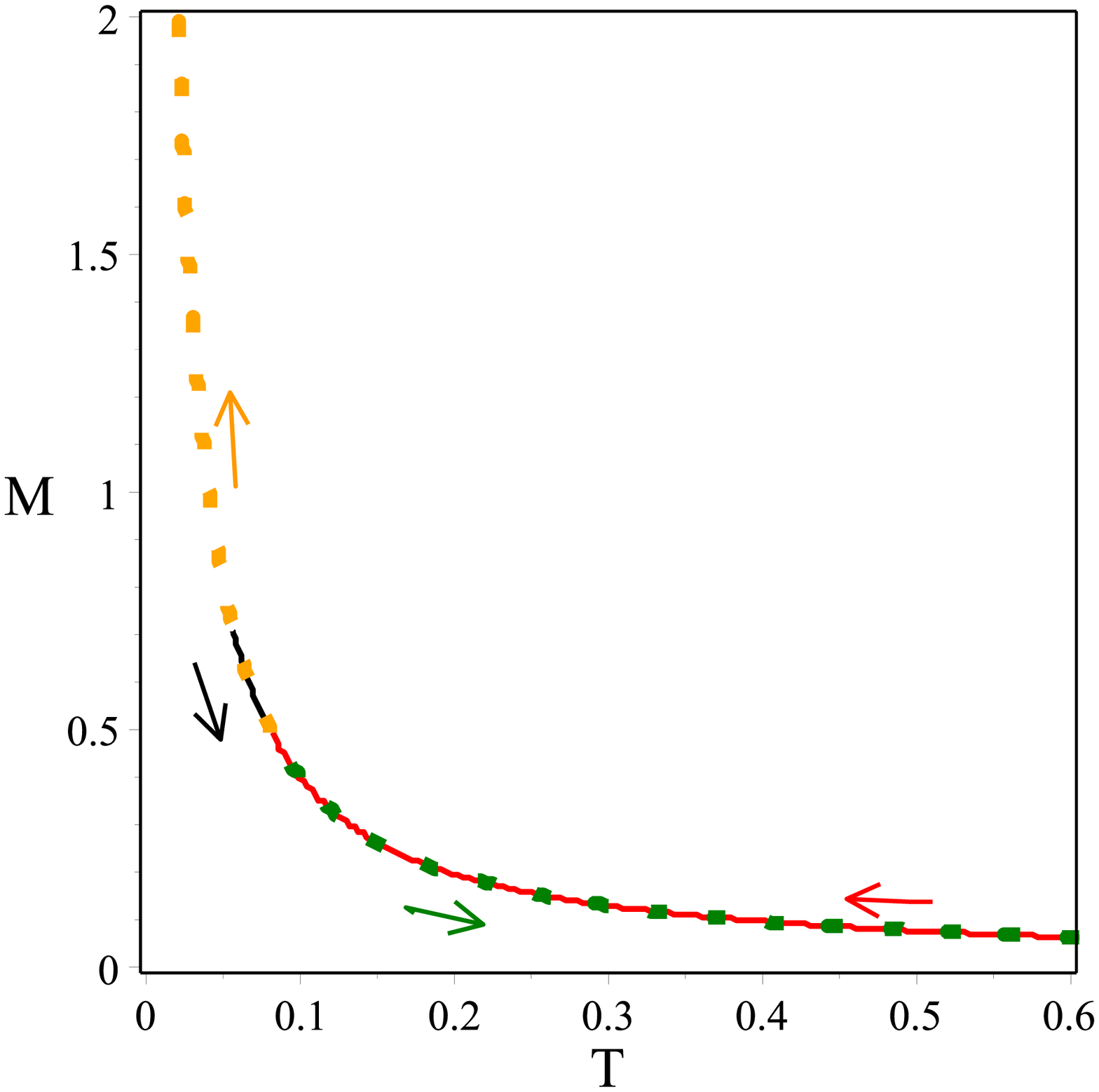}}
\subfigure[$K=1.1$]{\includegraphics[width=0.4\columnwidth]{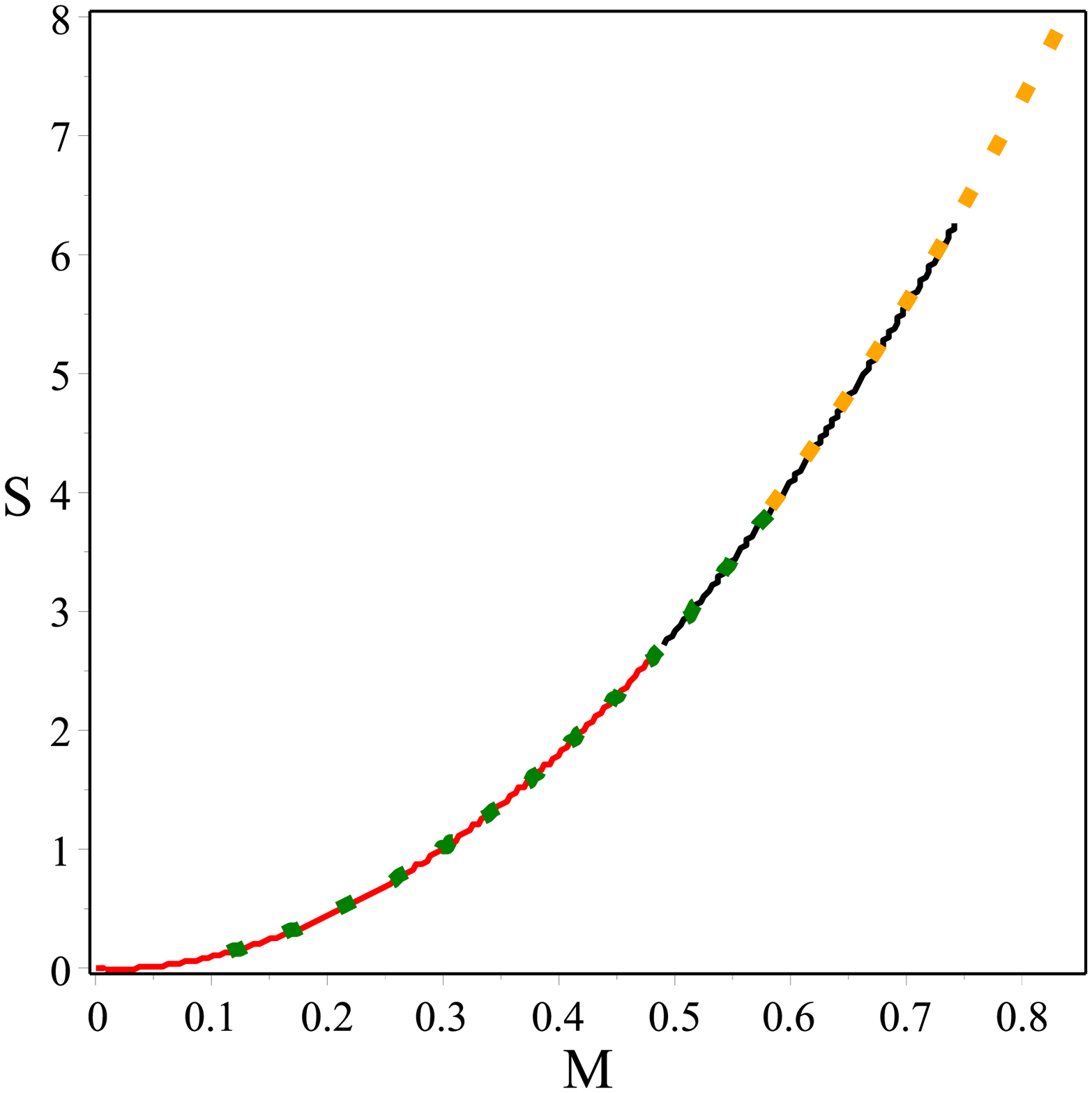}}
\subfigure[$K=1.1$]{\includegraphics[width=0.4\columnwidth]{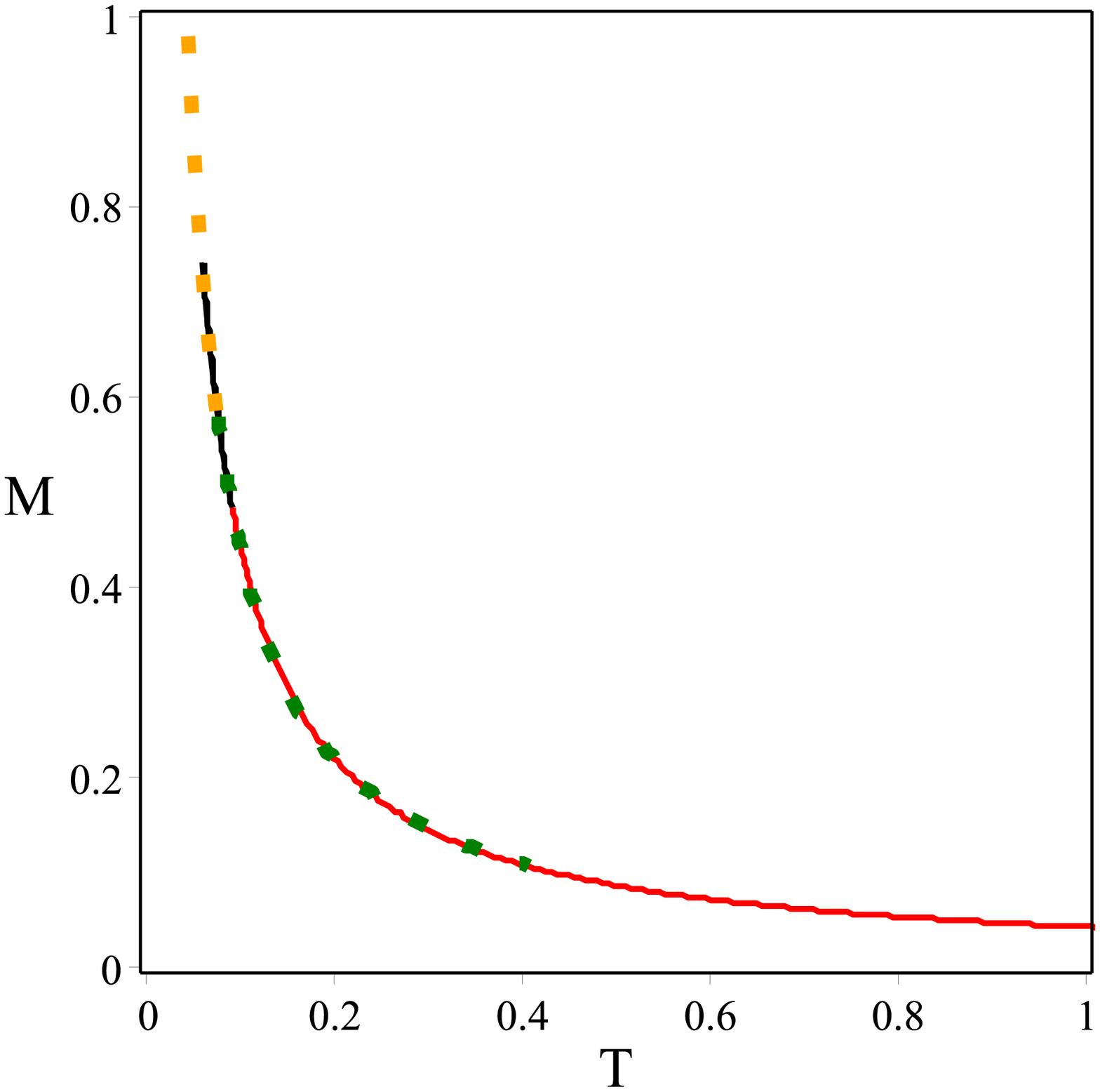}}
\caption{Plots of $ S $ in terms of $M$ (left), and of $ M $ in terms of $T$ (right) for $ \alpha=0.5$ and different values of $K$. The direction of the arrows shows the direction of increasing $r_{+}$. Schwarzschild-like behaviour is shown in red and orange.} 
\label{Ss-Mmplot}
\end{figure}

In Fig. \ref{Ss-Mmplot}, we will depict the behaviour of $S$ as a function of $M$ and $M$ as a function of $T$ for the two solutions of $\delta$ and different values of $K$. The third solution of $\delta $ doesn't have physical meaning (blue curve). This leads to the negative entropy and mass which we did not show in Fig. \ref{Ss-Mmplot}. 
For instance, in Fig. \ref{Ss-Mmplot}(c, d) we observe Schwarzschild-like behaviour that starts with the red curve and then smoothly is connected to the orange dotted line. While non-Schwarzschild-like behaviour first follows the black line and then continues with the green dotted line.
For  $q \neq 0$ we cannot obtain an analytic solution.  However we can obtain a solution to leading
order in $\alpha$ and $q^2$. Expanding 
$\delta(r_{+},q)$ in powers of $\alpha$ we find
 \be\label{deltexp}
\delta(r_+,q) =  -\frac{q^2}{r^2_+} +\alpha\left(C_1  - \frac{2q^2}{r^4_+} \right)+ \cdots
\ee
yielding
\begin{equation}\label{massequ}
M=M_{0}+\alpha M_{1}+\cdots=\dfrac{r_{+}}{2}+\dfrac{q^{2}}{2r_{+}}+\left( \dfrac{q^{2}}{r_{+}^{3}}+\dfrac{C_{1} r_{+}}{2}\right) \alpha+\cdots
\end{equation} 
where $C_1$ is an arbitrary constant of integration with dimension of $1/[length]^{2}$. We plot in Fig.~\ref{plotM1} the quantity $M_1 = (M - M_0)/\alpha$   for different values of $C_1$. Obtaining the higher order terms will necessitate obtaining corrections to the
potential \ref{eq25}.
 \begin{center}
\begin{figure}[H] \hspace{4cm}\includegraphics[width=8.cm]{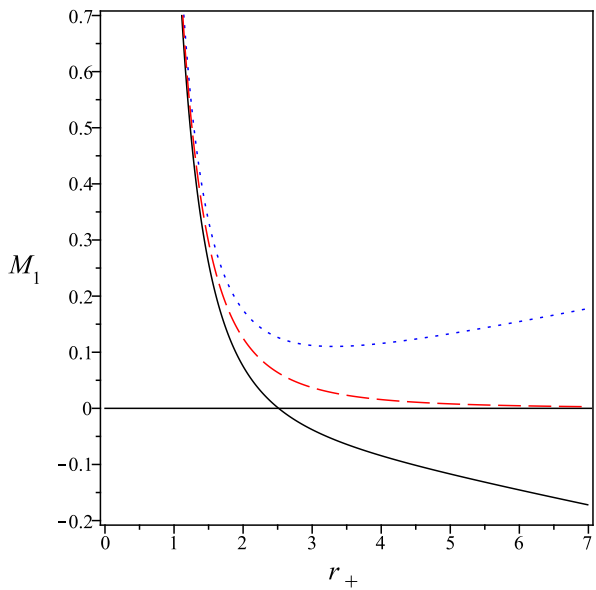}\vspace{0.1cm}\caption{\label{mrplot} \small
Plots of $ M_{1} $ in terms of $r_{+}$ for $ q=1, C_{1}=-0.05$ (\textcolor{black}{solid line}), $ C_{1}=0$ (\textcolor{red}{red dashed line}), $C_{1}=0.05$ (\textcolor{blue}{  blue dotted line}).}
\end{figure}
\label{plotM1}
\end{center}

\section{Particle Orbits in the Schwarzschild-Like Solutions}\label{sectwo}

In this section we examine the behaviour of  time-like and null geodesics for these black hole solutions.

Consider the general form of the spherically symmetric line element:
\begin{equation}\label{eq32}
ds^{2}=-h(r)dt^{2}+\dfrac{dr^{2}}{f(r)}+r^{2}d\theta^{2}+r^{2}\sin^{2}(\theta)d\phi^{2}.
\end{equation}
Since the metric is independent of $ t $ and $ \phi $, there are two conserved quantities:
\begin{equation}
E=h \dot{t}\hspace{0.5cm},\hspace{0.5cm}L=r^{2}\dot{\phi}.
\end{equation}
Without loss of generality we consider the metric 
 on the equatorial plane ($ \theta=\dfrac{\pi}{2} $), and  obtain the geodesic equation 
 \begin{equation}\label{geod}
\dfrac{1}{2}\dfrac{h}{f}\dot{r}^{2}+\dfrac{1}{2}\left[ h(r)\left( \dfrac{l^{2}}{r^{2}}+1\right)\right]=\dfrac{E^{2}}{2}
\end{equation}
for massive neutral particles. The second term is effective potential that we want to study. 
Note that the kinetic energy term has a non-canonical normalization 
\cite{1990AN....311..271F}.

We shall work in the asymptotic regime, where the Yukawa terms can be neglected: for $r=10M,\alpha=0.5M^2,q=0.5M$   the respective ratios of the coefficients of the $C_{2}$ term in (\ref{Yuk})  
to the second and third   terms in \eqref{eqf11} are $2.5 \times 10^{-4}$ and $2 \times 10^{-2}$. 
Eq. (\ref{eq17}) then becomes
\begin{equation}\label{eqapproximat}
h(r)=1-\dfrac{2M}{r}+\dfrac{q^{2}}{r^{2}}-\left[ \dfrac{128M(1+\delta)}{r^{3}}-\dfrac{16 q^{2}(1+\delta)}{r^{3}r_{+}}\right]\alpha, \hspace{0.5cm} B(x)=1 
\end{equation}
yielding $f(r)=h(r)$.  
So, the effective potential is 
\begin{equation}\label{Veff38}
V_{eff} =\dfrac{1}{2}\left[ h(r)\left( \dfrac{l^{2}}{r^{2}}+1\right)\right]=\dfrac{1}{2}\left( 1-\dfrac{2M}{r}+\dfrac{q^{2}}{r^{2}}-\left[ \dfrac{128M(1+\delta)}{r^{3}}-\dfrac{16 q^{2}(1+\delta)}{r^{3}r_{+}}\right]\alpha \right) \left( \dfrac{l^{2}}{r^{2}}+1\right)
\end{equation} 
valid for large $r$ and small $\alpha$,
by using the continued fraction expansion up to order 2.  This expansion is not valid to
higher orders in the continued fraction expansion, which exhibits terms inversely proportional to $\alpha$.

\begin{figure}[H]\hspace{0.4cm}
\centering
\subfigure[]{\includegraphics[width=0.48\columnwidth]{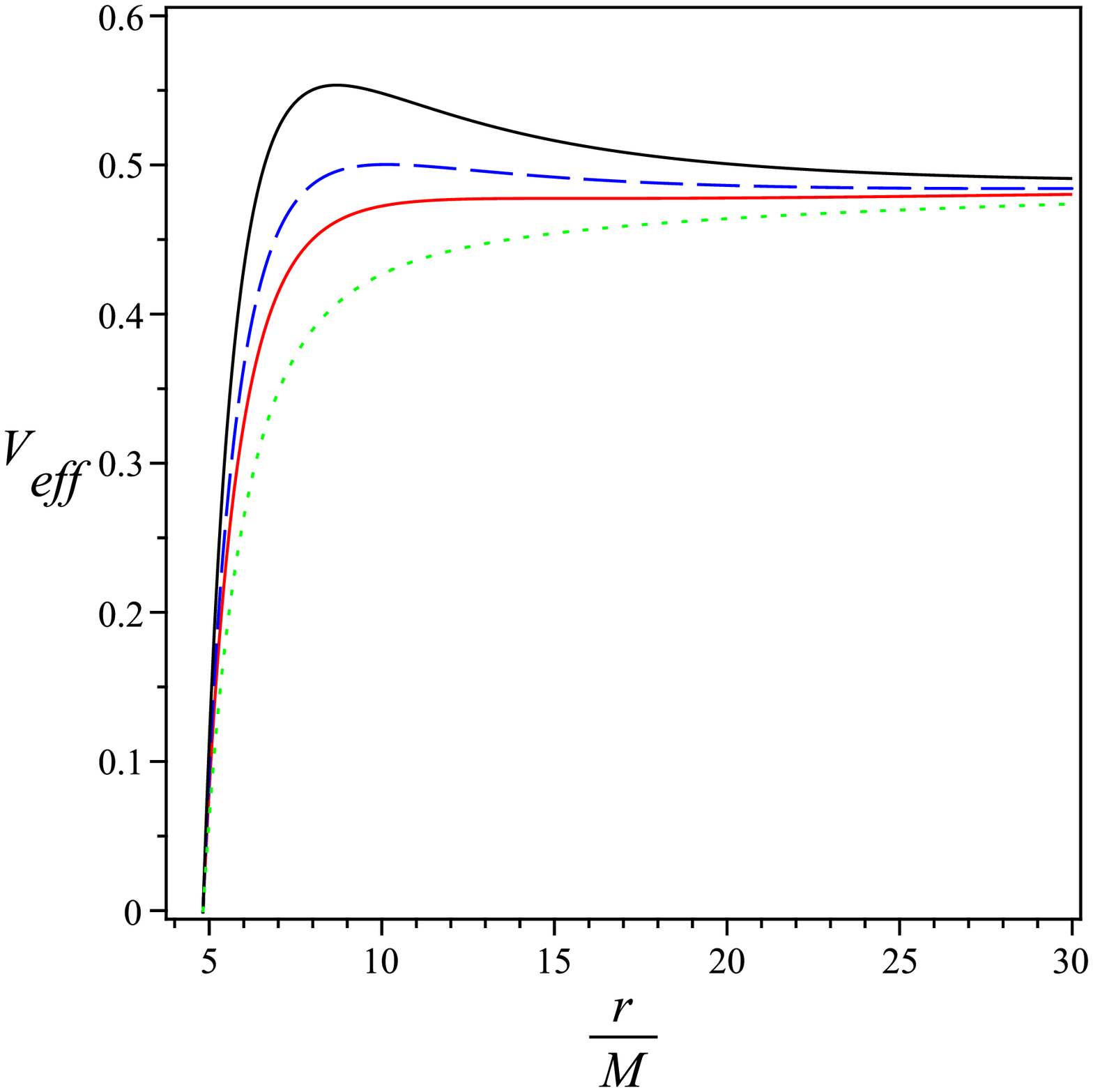}}
\subfigure[]{\includegraphics[width=0.48\columnwidth]{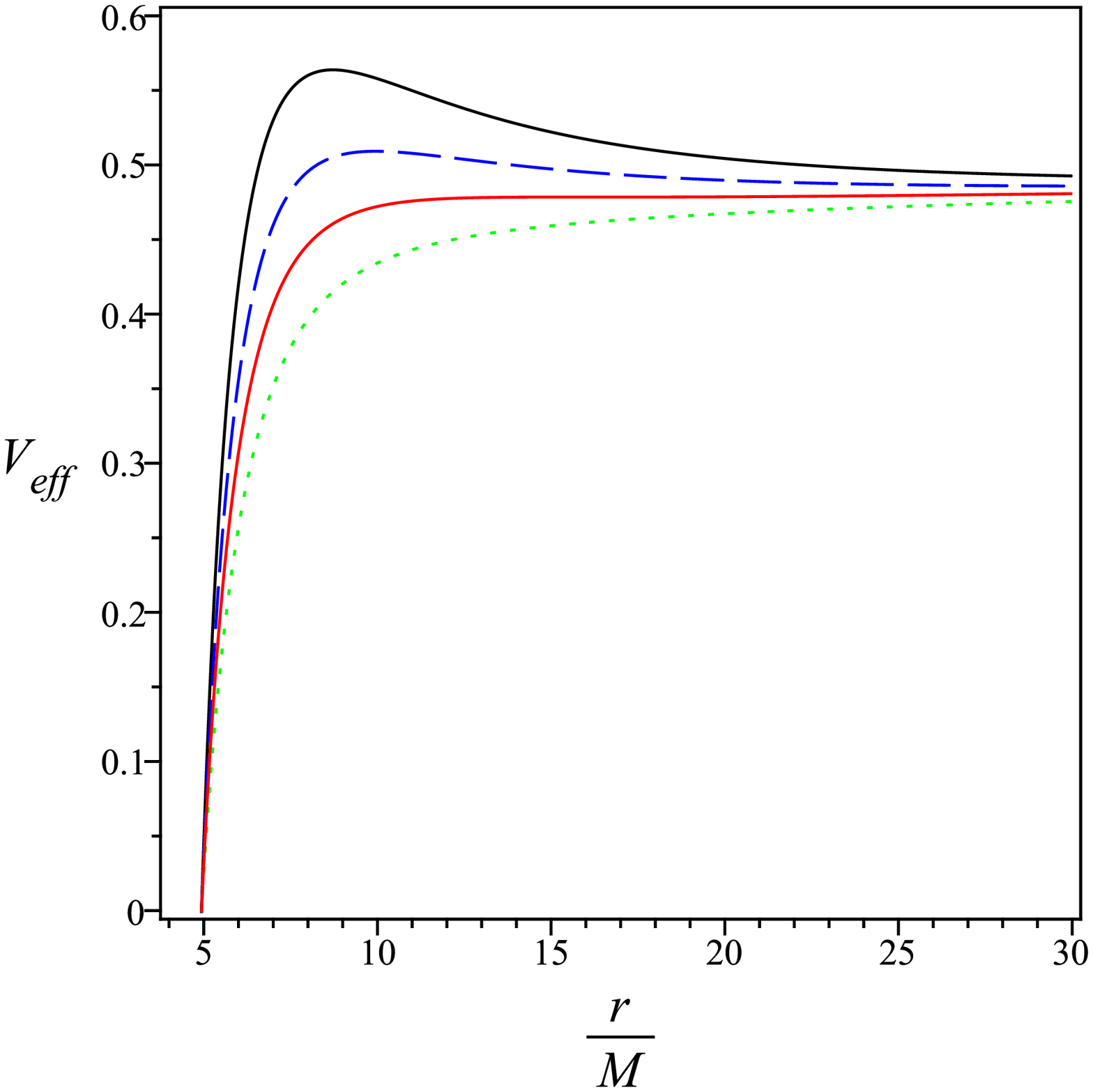}}
\caption{Plots of $ V_{eff} $ in terms of $\frac{r}{M}$ for $\alpha=0.5M^{2},k=1,\Lambda=0, q=0.5M, r_{+}=2M, f_{1}=\frac{0.53}{M} $ and $ \frac{L}{M}=4,5.33,6,7 $ from (\textcolor{green}{  green dotted line}) to (\textcolor{black}{ black solid line})(left), and for $\alpha=0.5M^{2},k=1,\Lambda=0, q=2M, r_{+}=2M, f_{1}=\frac{0.95}{M} $ and $ \frac{L}{M}=4,5.11,6,7 $ from (\textcolor{green}{ green dashed line}) to (\textcolor{black}{  black solid line})(right).
} 
\label{veff0}
\end{figure}

In Fig. \ref{veff0}, we plot the approximation to $ V_{eff} $ given in \eqref{Veff38}.  We have set
  $ \alpha=0.5M^{2} $ and considered different values of $ L $. For large $ L $, there are two extreme points in $V_{eff}$ . The maximum (minimum) point is related to unstable (stable) circular orbits for massive particles. 

\begin{figure}[H]\hspace{0.4cm}
\centering
\subfigure[]{\includegraphics[width=0.48\columnwidth]{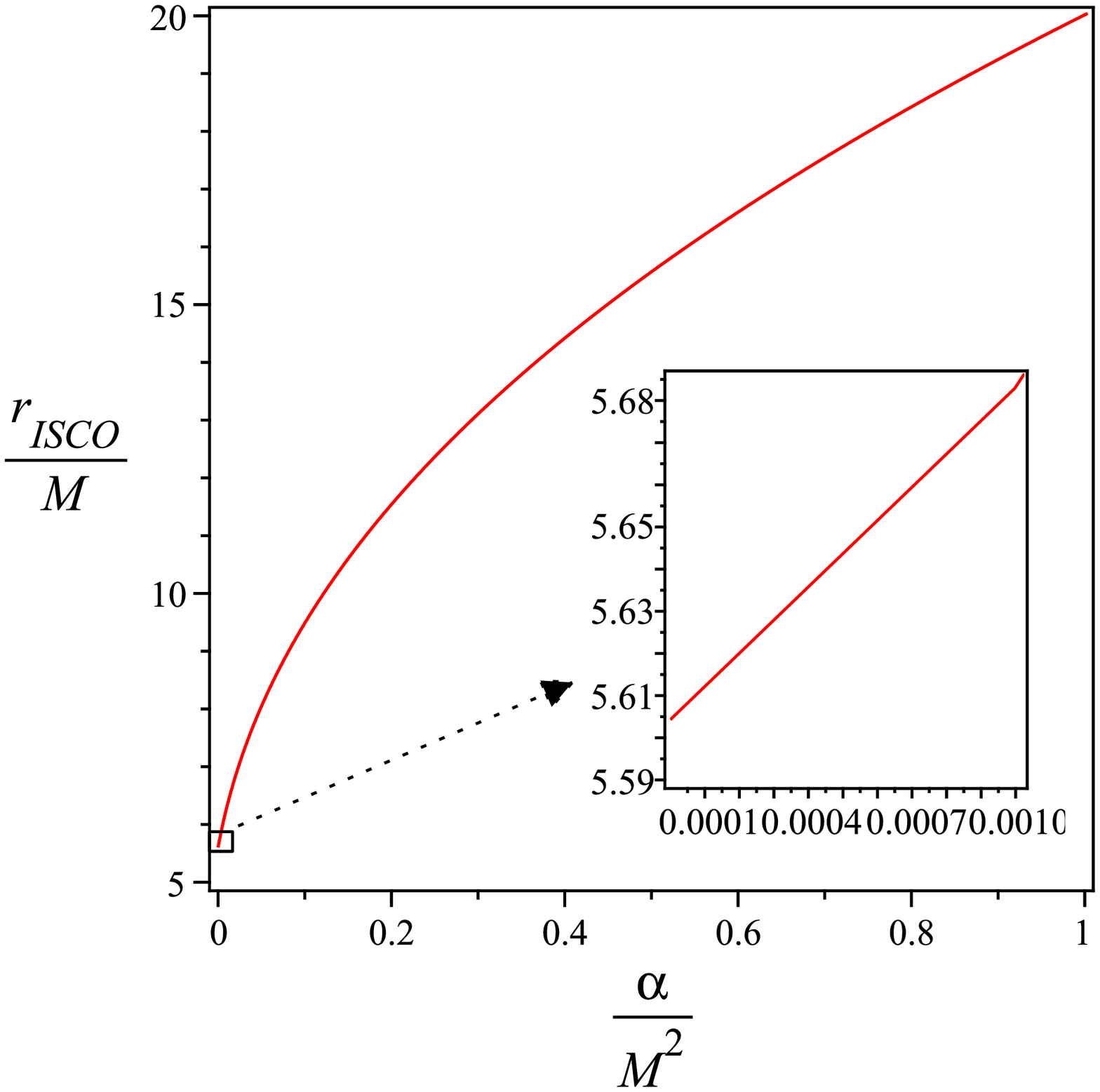}}
\subfigure[]{\includegraphics[width=0.48\columnwidth]{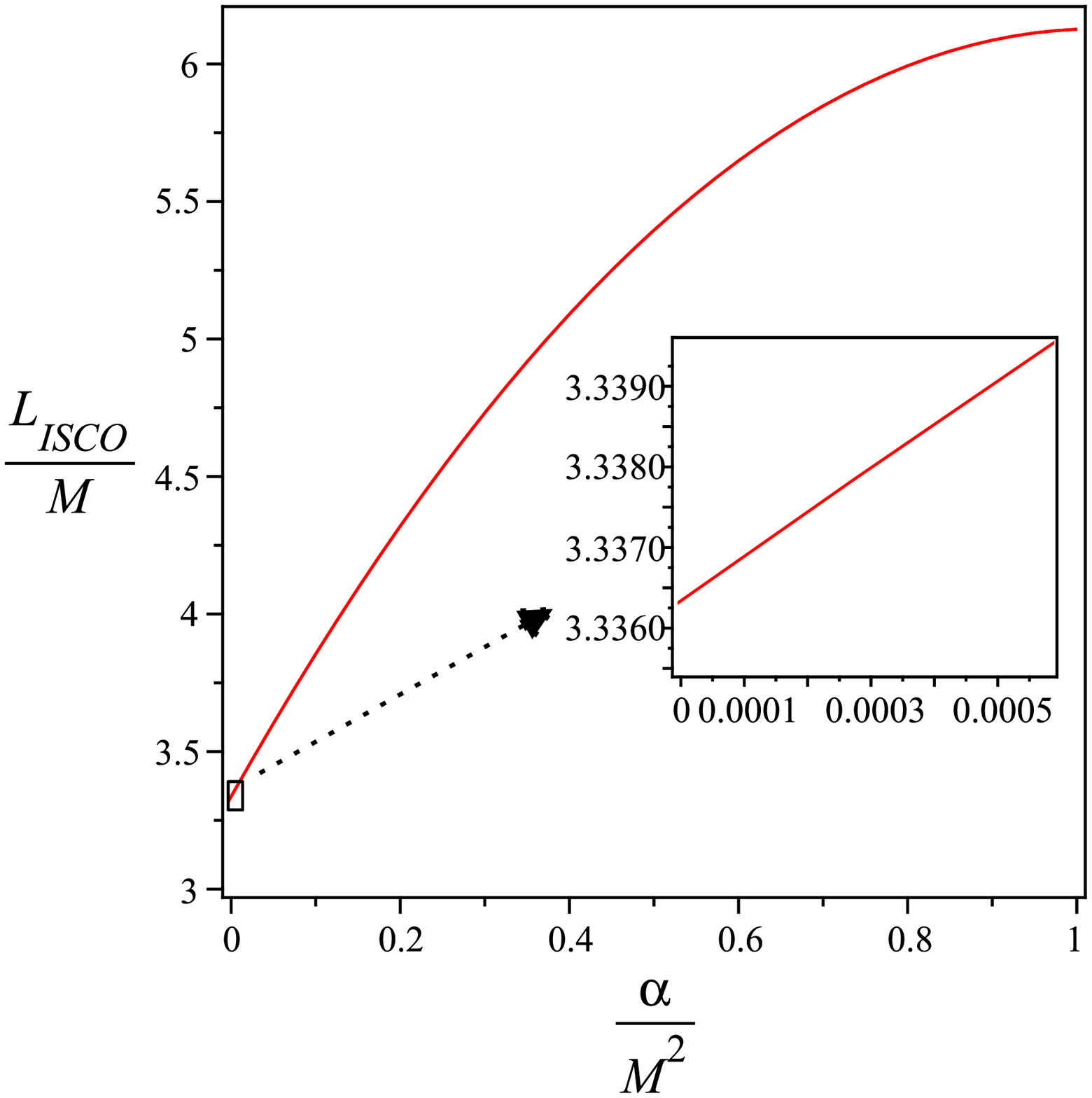}}
\caption{The innermost stable circular orbit (left) and the angular momentum of inflection point (right) in terms of the coupling constant of theory ($ \frac{\alpha}{M^2} $) for $k=1,\Lambda=0, q=0.5M, r_{+}=2M, f_{1}=\frac{0.53}{M} $.} 
\label{figla}
\end{figure}

 We can find the ISCO (Innermost Stable Circular Orbit) of the potential in \eqref{Veff38} by computing the point of inflection of the effective potential.  We find, for example, that for $ \alpha=0.5M^{2}, q=0.5M, r_{+}=2M, f_{1}=\frac{0.53}{M} $,
$ r_{ISCO}=15.57M $ and $ |L_{ISCO}|=5.33M $ and it is different from the respective
Schwarzschild values of  $6M$ and $2\sqrt{3} M$ (Fig. \ref{veff0}a). However  Fig. \ref{figla} indicates that when $\alpha$ goes to zero, $ r_{ISCO} $ and $|L_{ISCO}|$ go to $5.6M$ and $3.33M$, respectively.  
 We can also consider a case that would be above extremality in Einstein gravity:
 $ \alpha=0.5M^{2}, q=2M, r_{+}= 2M, f_{1}=\frac{0.95}{M} $,  for which we obtain $ r_{ISCO}=16.07M $ and $ |L_{ISCO}|=5.11 M $ (Fig. \ref{veff0}b). Note that $q=2M$ is not beyond the extremal value for the parameters. Because $q=\sqrt{2} r_{+}$ is the extremal case while here $q=r_{+}$. 
 
\begin{figure}[H]\hspace{0.4cm}
\centering
\subfigure[]{\includegraphics[width=0.48\columnwidth]{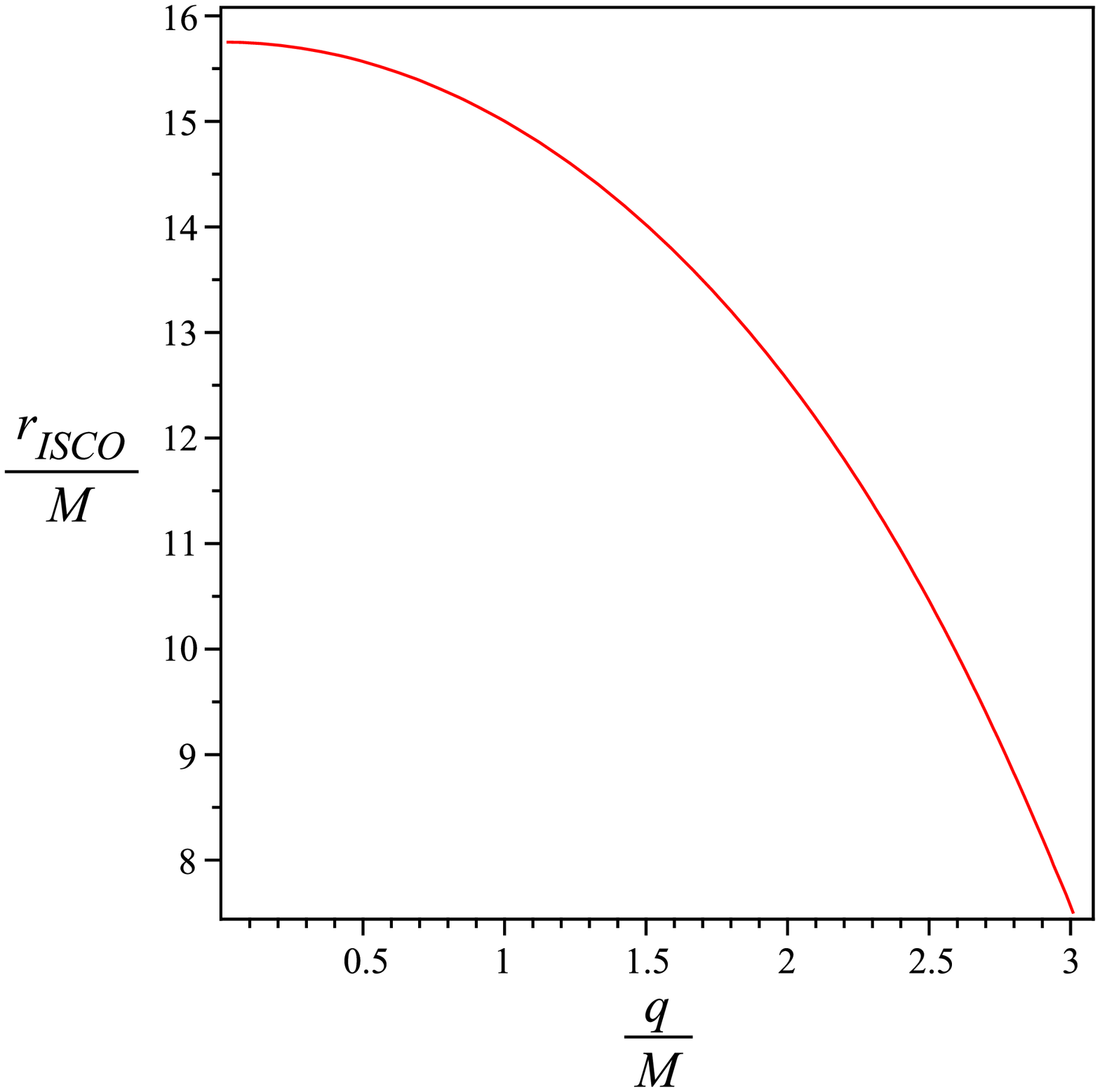}}
\subfigure[]{\includegraphics[width=0.48\columnwidth]{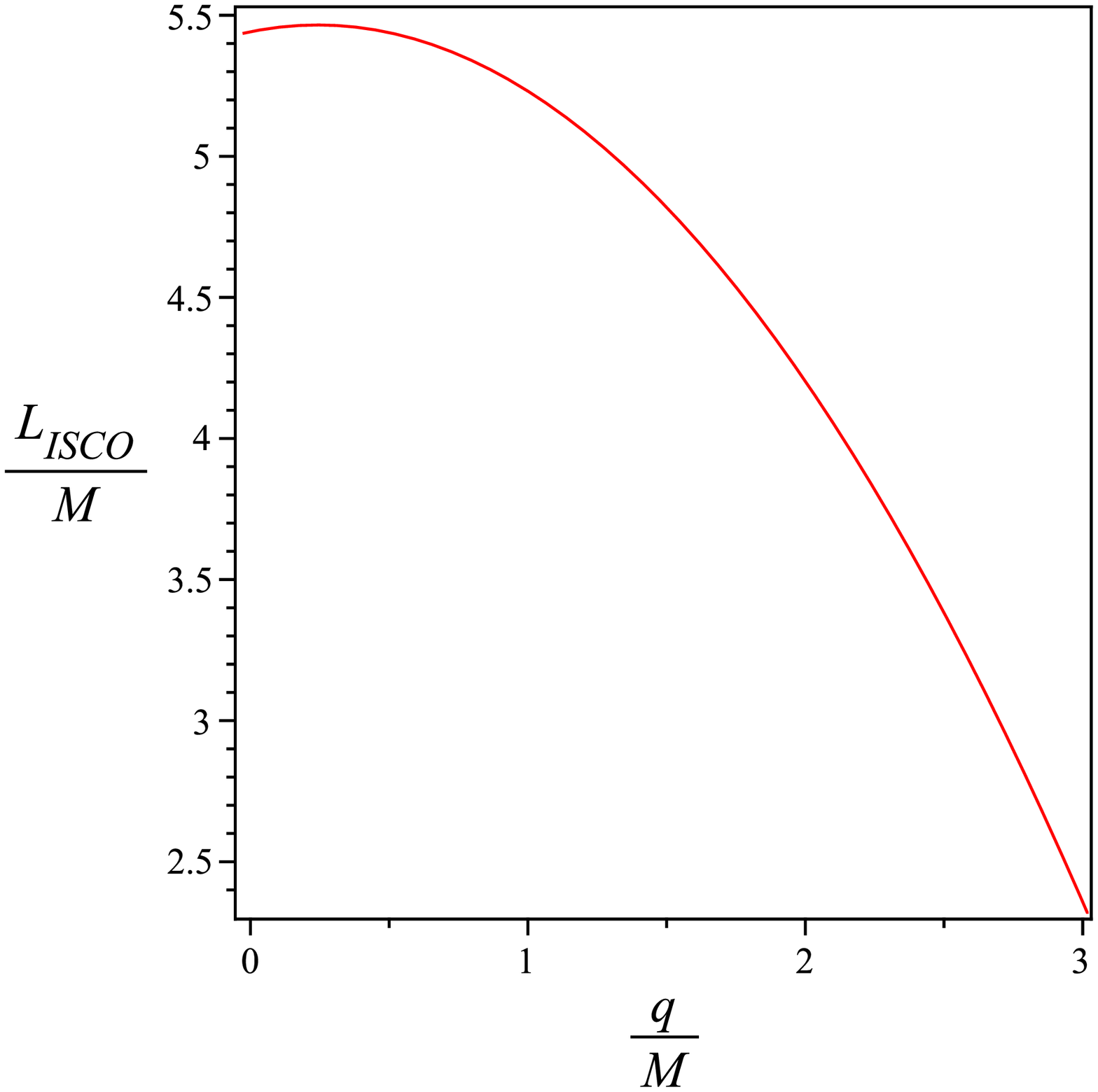}}
\caption{The innermost stable circular orbit (left) and the angular momentum of inflection point (right) in terms of electric charge ($ \frac{q}{M} $) for $k=1,\Lambda=0, \alpha=0.5M^2, r_{+}=2M, f_{1}=\frac{0.53}{M} $.} 
\label{figlb}
\end{figure}

  We find that  $ r_{ISCO} $ and $ L_{ISCO} $ both increase for increasing $ \alpha $, shown in Fig. \ref{figla}.  However, 
as $q$ increases we find that $r_{ISCO}$ and $L_{ISCO}$ decrease.  We illustrate this behaviour  in Fig. \ref{figlb}. 
Also, 

We now turn to a consideration of the behaviour of null geodesics in the vicinity of the black hole. The deflection of the photon as it moves from infinity to $r_m$ and off to infinity for the metric \eqref{metform} can be expressed as
\begin{equation}\label{eq37}
\delta \varphi=\int_{r_{m}}^{\infty}\dfrac{2dr}{\sqrt{\dfrac{f}{h}\dfrac{r^{4}}{b^{2}}-f r^{2}}}-\pi=I-\pi ,
\end{equation}
where $ b=\sqrt{\dfrac{r_{m}^{2}}{h(r_{m})}} $ is the impact parameter of the null ray and $ r_{m} $ is coordinate
distance of closest approach. Here $\pi$ is the change in the angle $ \varphi $ for straight line motion and is therefore subtracted out.
In the asymptotic regime
(up to 2nd order ($ a_{2} $ and $ b_{2} $) in the continued fraction expansion) with $q=0$ we have 
\begin{equation}\label{appmetric}
f(r)\approx h(r) \approx 1-\dfrac{2 M}{r}-\dfrac{128 M (1+\delta) \alpha}{r^{3}}, \hspace{1cm}B\approx 1
\end{equation}
We now calculate the integral in (\ref{eq37}) using (\ref{appmetric}). Writing the term in the denominator
of \eqref{eq37} as  $h(r) r^2 \left(r^2/b^2 h(r) -1\right)$, we have
\begin{multline}
\frac{h(r_{m})}{h(r)}\dfrac{r^{2}}{r_{m}^{2}}-1=\left[ \dfrac{1-\dfrac{2M}{r_{m}}-\dfrac{128M\alpha (1+\delta)}{r_{m}^{3}}}{1-\dfrac{2M}{r}-\dfrac{128M\alpha (1+\delta)}{r^{3}}}\right]\left(\dfrac{r^{2}}{r_{m}^{2}} \right)-1=\\
\left(\dfrac{r}{r_{m}}\right)^{2}\left[1+2M\left(\dfrac{1}{r}-\dfrac{1}{r_{m}} \right)+128M\alpha (1+\delta)\left( \dfrac{1}{r^{3}}-\dfrac{1}{r_{m}^{3}}\right)  \right]-1=\\
\left(\dfrac{r^{2}}{r_{m}^{2}}-1\right)\left[1-\dfrac{2Mr}{r_{m}(r+r_{m})}-\dfrac{128M\alpha (1+\delta)}{r r_{m}^{2}}\left(1+\dfrac{r^{2}}{r_{m}(r+r_{m})}\right) \right]   
\label{expand}   
\end{multline}
for $M \ll r$.  The integrand becomes
\begin{multline}
\int_{r_{m}}^{\infty}\dfrac{1}{\sqrt{\left(\dfrac{1}{r_{m}^{2}}-\dfrac{1}{r^{2}}\right)} }\left[1+\dfrac{M}{r}\left(1+\dfrac{r^{2}}{r_{m}(r+r_{m})}\right)+\dfrac{64M\alpha (1+\delta)}{r^{3}}\left(1+\dfrac{r^{2}}{r_{m}^{2}}\left(1+\dfrac{r^{2}}{r_{m}(r+r_{m})}\right)\right)\right]\dfrac{dr}{r^{2}} 
\label{eq42}
\end{multline}
upon expanding in powers of $M/r$, $M/r_m$, and $\alpha/M^2$.

After making the substitution $\sin(\theta)= \frac{r_{m}}{r}$ the integral becomes
\begin{multline}
\int_{0}^{\frac{\pi}{2}}d\theta \left[1+\dfrac{M}{r_{m}}\left(\sin(\theta)+\dfrac{1}{1+\sin(\theta)}\right) +\dfrac{64M\alpha (1+\delta)}{r_{m}^{3}}\left(\sin^{3}(\theta)+\sin(\theta)+\dfrac{1}{1+\sin(\theta)}\right) \right]\\
=\dfrac{\pi}{2}+\dfrac{2M}{r_{m}}+\dfrac{512M\alpha (1+\delta)}{3r_{m}^{3}} 
\end{multline}
The deflection is as follows
\begin{equation}\label{deflect}
\Delta \varphi=\dfrac{4M}{r_{m}}+\dfrac{1024M\alpha(1+\delta)}{3r_{m}^{3}}
\end{equation}
valid for large $r$ ($r\rightarrow \infty$) and small $\alpha$ ($ \alpha \rightarrow 0 $),   
 where $\delta$ is given in \eqref{eq20}.
So, above is a simple modification of the standard Einstein result of $4M/r_{m}$. The constant $\alpha$ must be small enough such that the extra term is negligible compared to $4M/r_{m}$ on solar system distance scales.

\begin{center}
\begin{figure}[H] \hspace{4cm}\includegraphics[width=8.cm]{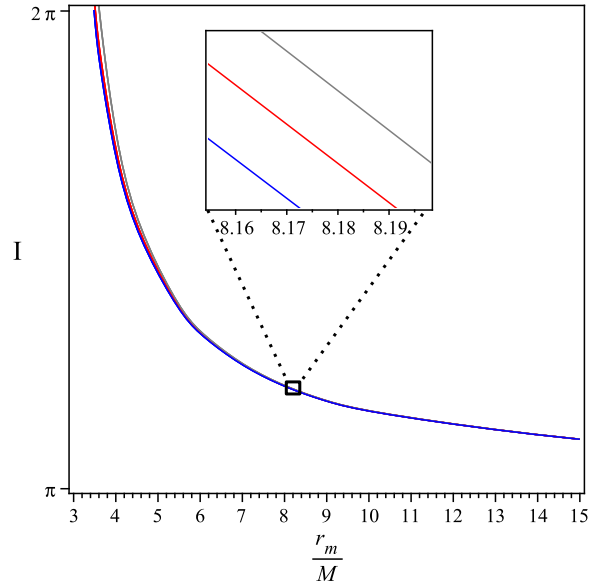}\vspace{0.1cm}\caption{\label{Irpp}
\small
Plots of deflection angle in terms of $r_{m}$ for $k=1,\Lambda=0, q=0.5M, r_{+}=2M, f_{1}=\frac{0.53}{M} $ and $ \alpha=0.05M^{2},0.5M^{2},M^{2} $ from (\textcolor{gray}{ gray dashed line}) to (\textcolor{blue}{ blue solid line}).}
\end{figure}
\end{center}

In Fig. \ref{Irpp}, we plot  the deflection angle in terms of closest distance to the black hole. As $ r_{m} $ increases the deflection angle decreases and goes to $ \pi $. 
However we see that as $\alpha$ increases the deflection angle diverges at smaller values of $ r_{m}$ relative to the Schwarzschild case, we see that deflection angle diverges at $ r_{m} = 3.5M $ as
$\alpha/M^2$ approaches unity.   The location at
which the deflection angle diverges  is the radius of the photon sphere.

We next consider the shadow of these black holes. In fact, we follow up the null geodesics which satisfy the condition $ R^{''}>0 $, i.e, unstable circular orbits.  The angular radius of the shadow as seen by an
observer at $ r_{0} $ is \cite{Synge} 
\begin{equation}
\sin^{2}(\Gamma)=\dfrac{r_{ph}^{2} h(r_{0})}{r_{0}^{2} h(r_{ph})}
\end{equation}
and using of Eq. (\ref{eqapproximat}) for $ h(r) $ with $ q=0 $ we obtain
\begin{align}
\dfrac{r_{ph}^{2}f(r_{0})}{r_{0}^{2}f(r_{ph})} &=\dfrac{r_{ph}^{2}}{r_{0}^{2}}\left[\dfrac{1-\dfrac{2M}{r_{0}}-\dfrac{128M\alpha(1+\delta)}{r_{0}^{3}}}{1-\dfrac{2M}{r_{ph}}-\dfrac{128M\alpha(1+\delta)}{r_{ph}^{3}}} \right] 
\nonumber\\
&= \dfrac{r_{p}^{2}}{r_{0}^{2}}\left[\left( 1-\dfrac{2M}{r_{0}}-\dfrac{128M\alpha(1+\delta)}{r_{0}^{3}}\right)\left(1
+ \dfrac{2M}{r_{ph}} + \dfrac{128M\alpha(1+\delta)}{r_{ph}^{3}}\right)\right]  \nonumber\\
&= \dfrac{r_{p}^{2}}{r_{0}^{2}}\left[1+2M\left(\dfrac{1}{r_{ph}}-\dfrac{1}{r_{0}}\right)+128M\alpha(1+\delta)\left(\dfrac{1}{r_{ph}^{3}}-\dfrac{1}{r_{0}^{3}}\right) \right]  
\end{align}
yielding in turn
\begin{equation}
\sin(\Gamma)=\dfrac{r_{ph}}{r_{0}}+\dfrac{M(r_{0}-r_{ph})}{r_{0}^{2}}+\dfrac{64M\alpha(1+\delta)(r_{0}^{3}-r_{ph}^{3})}{r_{ph}^{2}r_{0}^{4}}
\end{equation}
where $ r_{ph} $ is the radius of the photon sphere and $ \Gamma $ is the angle subtended by the radius of the shadow as seen by an observer at as seen by an observer at $r_{0}$.\\
In the case of small $ \Gamma $, we have $\sin(\Gamma)\approx \Gamma$, so
\begin{equation}
\Gamma=\Gamma_{Ein}+\Gamma_{Con}
\end{equation}
up to 2nd order   in the continued fraction expansion, i.e. the asymptotic regime.
In above equation on the right hand side, the first two term are the Einstein term and the second term  the correction from  the Quadratic  corrections. 
%If $\alpha$ and $\eta$ go to zero the theory goes back to the Einstein Gravity. 

Finally, we consider Shapiro time-delay to obtain a bound on the coupling constant $\alpha$. The general expression for time delay for the metric (\ref{eq32}) is  
\begin{equation}
t(r_{0},r)=\int_{r_{0}}^{r}\dfrac{dr}{\sqrt{f(r)h(r)\left( 1-\dfrac{r_{0}^{2}}{r^{2}}\dfrac{h(r)}{h(r_{0})}\right) }}
\end{equation}
In order to evaluate the integral, we expand the metric at asymptotic regime 
as in \eqref{appmetric}. Similar manipulations as before yield
\begin{equation}
\dfrac{1}{\sqrt{1-\dfrac{r_{0}^{2}}{r^{2}}}}\left(1+\dfrac{M}{r}\left( 1+\dfrac{r_{0}}{(r+r_{0})}\right)+\dfrac{64M(1+\delta)\alpha}{r^{3}}\left( 2+\dfrac{r^{2}}{(r+r_{0})r_{0}}\right) \right)  
\end{equation}
for the integrand. 
Now, integral is elementary and we find that the time required for light to go from $ r_{0} $ to $ r $ is
\begin{multline}
t(r,r_{0})\approx \sqrt{r^{2}-r_{0}^{2}}+M\ln\left(\dfrac{r+\sqrt{r^{2}-r_{0}^{2}}}{r_{0}}\right)+M\sqrt{\dfrac{r-r_{0}}{r+r_{0}}}+\dfrac{M(1+\delta)\alpha}{r_{0}}\sqrt{\dfrac{r-r_{0}}{r+r_{0}}}\left(\dfrac{192}{r_{0}}+\dfrac{128}{r} \right)  
\end{multline}
Working in the asymptotic regime (to 2nd order in the continued fraction expansion)
 schematically this expression is
 \begin{equation}
t(r,r_{0})=t_{SR}(r,r_{0})+\Delta t_{GR}(r,r_{0})+\Delta t_{EC}(r,r_{0})=t_{SR}(r,r_{0})+\Delta t(r,r_{0})
\end{equation}
 where $ t_{SR}=\sqrt{r^{2}-r_{0}^{2}} $ is the special relativistic contribution of the propagation of light in 
 flat spacetime.
So, the maximum round-trip excess time delay is given by
\begin{multline}
\Delta t(r,r_{0})=2\left[ t(r_{2},r_{0})+t(r_{1},r_{0})-\sqrt{r_{1}^{2}-r_{0}^{2}}-\sqrt{r_{2}^{2}-r_{0}^{2}}\right]=2M\ln\left[\dfrac{(r_{1}+\sqrt{r_{1}^{2}-r_{0}^{2}})(r_{2}+\sqrt{r_{2}^{2}-r_{0}^{2}})}{r_{0}^{2}}\right]\\
+2M\left[\sqrt{\dfrac{r_{1}-r_{0}}{r_{1}+r_{0}}} +\sqrt{\dfrac{r_{2}-r_{0}}{r_{2}+r_{0}}}\right]+\dfrac{2M\alpha (1+\delta)}{r_{0}}\left[ \sqrt{\dfrac{r_{1}-r_{0}}{r_{1}+r_{0}}}\left(\dfrac{192}{r_{0}}+\dfrac{128}{r_{1}}\right)+\sqrt{\dfrac{r_{2}-r_{0}}{r_{2}+r_{0}}}\left( \dfrac{192}{r_{0}}+\dfrac{128}{r_{2}}\right)  \right]   
\end{multline}
in the case of $ r_{1}=r_{2}=r $, this becomes
\begin{equation}\label{eqshap}
\Delta t(r,r_{0})=4M\ln\left(\dfrac{r+\sqrt{r^2-r_0^2}}{r_0}\right)+4M\sqrt{\dfrac{r-r_{0}}{r+r_{0}}}+\dfrac{4\alpha (1+\delta)}{M^2}\sqrt{\dfrac{r-r_{0}}{r+r_{0}}}\left( \dfrac{192 M^3}{r^2_{0}}+\dfrac{128 M^3}{r r_{0}}\right)  
\end{equation}
where we have partitioned the expression into the general relativitistic (GR) and Einstein-Conformal (EC) corrections.  Here  $ r_{0} $ is the distance of closest approach of the radar wave to the center of the Sun, $ r_{1} $ is the distance along the line of light from the Earth to the point of closest approach to the Sun, and $ r_{2} $ represents the distance along the path from this point to the planet, where $ r_{1,2}\gg r_{0}$.
 
Taking the smallest possible value of $r$ as the radius of the sun $r_\odot = 6.957\times 10^8$ m, we see that 
the coefficient of the Einstein conformal correction is about $(M_\odot/r_\odot)^2 \sim 4.5 \times 10^{-12}$, 
where $M=M_{\odot}=1477$ m, implying that $\alpha (1+\delta)/M^2$ need not be extremely small.
Deviations of time delay from the prediction of general relativity have been constrained to be less than 0.000012 \cite{Hen}, and so the last term in (\ref{eqshap}) must be no larger than this value \cite{Edery}- \cite{Hen}. Using solar system data (where in units of metres, $r_{1}=r_{2}=10^{11}$ m,  $r_{0}=r_{\odot}=10^{8}$ m), we obtain the constraint 
\begin{equation}\label{eqsh}
\frac{\alpha (1+\delta)}{M_{\odot}^{2}} <  0.05. 
\end{equation}
We note by comparison recent work 
\cite{Ruggiero:2020yoq} making use of exoplanet data to constrain modifications of the form
$\tilde{\alpha}^{(N)}/r^N$  to the effective gravitational potential in the weak-field limit.  For the theory we are considering, $N=3$, but $\tilde{\alpha}^{(N)} = \alpha (1+\delta) M$ from \eqref{Veff38}, so the appropriate parameter to compare to is
 $\tilde{\alpha}^{(2)}$ because of the mass parameter. Inserting units into the bound
in \eqref{eqsh}, we find
\begin{equation}\label{eqsh1}
 \alpha (1+\delta)  <  0.05  \left( \frac{G M_{\odot} }{c^2}\right)^2 c^2  \sim 9.8 \times 10^{21}\; \textrm{m}^4/\textrm{s}^2
\end{equation}
comparable to the limit $\tilde{\alpha}^{(2)} < 10^{22}$ m$^4$/s$^{2}$ obtained from exoplanet data.

\section{Conclusion}\label{con}

We have obtained an analytic approximation to a charged black hole solutions in Einstein Quadratic gravity
by making use of a continued fraction expansion. The key advantage to this approach is that the continued fraction can be used in place of an exact solution, allowing one to study problems that are difficult to address by numeric methods.  We have studied thermodynamics of the black hole in the absence of cosmological constant. Working to leading order in $\alpha$ and $q^2$, we have shown the first law and Smarr formula is satisfied.   

We also investigated phenomenological consequences of the $q=0$ solution.  We found that   for a given value of the mass, the ISCO for a massive test body, as well as its angular momentum at that location grows as the parameter $\alpha$ increases. 

We note that our approximations need to be taken with care.  As for the $q=0$ case \cite{Lu:2015cqa}, the near-horizon expansions \eqref{eq9} and \eqref{eq10} of the metric functions do not have a sensible small-$\alpha$ limit.  Beyond 2nd order in the continued fraction expansion the same thing happens.  
This means that the small-$\alpha$ expansions must be understood as asymptotic expansions, and should
not be taken to apply in the strong-field limit.  

This raises the question as to whether or not the solution presented in section 2 is an appropriate generalization of the Schwarzschild solution.  We present in Appendix B an  alternate  near horizon solution with a   well-defined
$ \alpha\to 0 $ limit, analogous to that obtained in Einstein Cubic gravity \cite{Hen}.  Exploring the physical consequences of this solution remains an interesting subject for future study.

\section*{Acknowledgements}

This work was supported in part by the Natural Sciences and Engineering Research Council of Canada.

\appendix

\section{Explicit Terms in the Continued Fraction Approximation}

We present terms up to order 4 in the continued fraction approximation \eqref{cfrac}:
\begin{align}
&\epsilon=-\dfrac{F_{1}}{r_{+}}-1,\,\,\,\, a_{1}=-1-a_{0}+2\epsilon+r_{+}h_{1},\,\,\,\, a_{2}=-{\dfrac {4a_{1}-5\epsilon+1+3 a_{0}+ h_{2}{{r_{+}}}^{2}}{{ a_{1}}}}
\nonumber \\
&a_{3}=-\dfrac{1}{{a_{1}}{a_{2}}}[-{h_{3}}{{r_{+}}}^{3}+{a_{1}}{{a_{2}}}^{2}+5{a_{1}}{a_{2}}+6{a_{0}}+10{a_{1}}-9\epsilon+1]  
\label{cfrac-a}\\
&a_{4}=-\dfrac{{h_{4}}{{r_{+}}}^{4}+{a_{1}}{{a_{2}}}^{3}+2{a_{1}}{{a_{2}}}^{2}{a_{3}}+{a_{1}}{a_{2}}{{a_{3}}}^{2}+6{a_{1}}{{a_{2}}}^{2}+6{a_{1}}{a_{2}}{a_{3}}+15{a_{1}}{a_{2}}+10{a_{0}}+20{a_{1}} -14\epsilon+1}{{a_{1}}{a_{2}}{a_{3}}} \nonumber 
\end{align}
and 
\begin{eqnarray}\label{b3b4}
&&b^\pm_{1}=-1\pm\sqrt{\dfrac{h_{1}}{f_{1}}},\,\,\,\,\,\,\,\,\,  b_{2}=-\dfrac{{r_{+}}\left( f_{{1}}{h_{2}}-{f_{2}}h_{{1}} \right)(b_{1}+1)+4{b_{1}}f_{{1}}h_{{1}}}{2{{b_{1}}}{f_{{1}}}{h_{{1}}}},\,\,\nonumber\\
&&b_{3}=-\dfrac{1}{{{f_{1}}}^{2}{{h_{1}}}^{2}}[-4{{f_{1}}}^{2}{h_{1}}{h_{3}}{{r_{+}}}^{2}{b_{1}}-4{{f_{1}}}^{2}{h_{1}}{h_{3}}{{r_{+}}}^{2}+{{f_{1}}}^{2}{{h_{2}}}^{2}{{r_{+}}}^{2}{b_{1}}+{{f_{1}}}^{2}{{h_{2}}}^{2}{{r_{+}}}^{2}+2{f_{1}}{f_{2}}{h_{1}}{h_{2}}{{r_{+}}}^{2}{b_{1}}\nonumber\\
&&+2{f_{1}}{f_{2}}{h_{1}}{h_{2}}{{r_{+}}}^{2}
+4{f_{1}}{f_{3}}{{h_{1}}}^{2}{{r_{+}}}^{2}{b_{1}}+4{f_{1}}{f_{3}}{{h_{1}}}^{2}{{r_{+}}}^{2}-3{{f_{2}}}^{2}{{h_{1}}}^{2}{{r_{+}}}^{2}{b_{1}}-3{{f_{2}}}^{2}{{h_{1}}}^{2}{{r_{+}}}^{2}\nonumber\\
&&+8{b_{1}}{{b_{2}}}^{2}{{f_{1}}}^{2}{{h_{1}}}^{2}+24{b_{1}}{b_{2}}{{f_{1}}}^{2}{{h_{1}}}^{2}+24{b_{1}}{{f_{1}}}^{2}{{h_{1}}}^{2}] \label{b5}\\
&&b_{4}=-\dfrac{1}{16{b_{1}}{b_{2}}{b_{3}}{{f_{1}}}^{3}{{h_{1}}}^{3}}[96{b_{1}}{b_{2}}{{f_{1}}}^{3}{{h_{1}}}^{3}+64{b_{1}}{{f_{1}}}^{3}{{h_{1}}}^{3}+64{b_{1}}{{b_{2}}}^{2}{
{f_{1}}}^{3}{{h_{1}}}^{3}+32{b_{1}}{{b_{2}}}^{2}{b_{3}}{{f_{1}}}^{3}{{h_{1}}}^{3}+\nonumber\\
&&16{b_{1}}{b_{2}}{{b_{3}}}^{2}{{f_{1}}}^{3}{{h_{1}}}^{3}+64{b_{1}}{b_{2}}{b_{3}}{{f_{1}}
}^{3}{{h_{1}}}^{3}+16{b_{1}}{{b_{2}}}^{3}{{f_{1}}}^{3}{{h_{1}
}}^{3}+{{f_{1}}}^{3}{{h_{2}}}^{3}{{r_{+}}}^{3}+{f_{1}}^{3}{{ 
h_{2}}}^{3}{{r_{+}}}^{3}{b_{1}}- 5{{f_{2}}}^{3}{{h_{1}}}^{3}{{r_{+}
}}^{3}\nonumber\\
&&-5{{f_{2}}}^{3}{{h_{1}}}^{3}{{r_{+}}}^{3}{b_{1}}+8{{ 
f_{1}}}^{3}{{h_{1}}}^{2}{h_{4}}{{r_{+}}}^{3}+8{{f_{1}}}^{3}{{ 
h_{1}}}^{2}{h_{4}}{{r_{+}}}^{3}{b_{1}}-4{{f_{1}}}^{3}{h_{1}}{
h_{2}}{h_{3}}{{r_{+}}}^{3}-4{{f_{1}}}^{3}{h_{1}}{h_{2}}
{h_{3}}{{r_{+}}}^{3}{b_{1}}\nonumber\\
&&-4{{f_{1}}}^{2}{f_{2}}{{h_{1}
}}^{2}{h_{3}}{{r_{+}}}^{3}-4{{f_{1}}}^{2}{f_{2}}{{h_{1}}}^{
2}{h_{3}}{{r_{+}}}^{3}{b_{1}}+{{f_{1}}}^{2}{f_{2}}{h_{1}}{
{h_{2}}}^{2}{{r_{+}}}^{3}+{{f_{1}}}^{2}{f_{2}}{h_{1}}{{h_{2}
}}^{2}{{r_{+}}}^{3}{b_{1}}-4{{f_{1}}}^{2}{f_{3}}{{ h_{1}}}^{2}
{h_{2}}{{r_{+}}}^{3}\nonumber\\
&&-4{{f_{1}}}^{2}{f_{3}}{{h_{1}}}^{2}{
h_{2}}{{r_{+}}}^{3}{b_{1}}-8{{f_{1}}}^{2}{f_{4}}{{h_{1}}}^
{3}{{r_{+}}}^{3}-8{{f_{1}}}^{2}{f_{4}}{{h_{1}}}^{3}{{r_{+}}}^
{3}{b_{1}}+3{f_{1}}{{f_{2}}}^{2}{{h_{1}}}^{2}{h_{2}}{{ 
r_{+}}}^{3}+3{f_{1}}{{f_{2}}}^{2}{{h_{1}}}^{2}{h_{2}}{{r_{+}}}
^{3}{b_{1}}\nonumber\\
&&+12{f_{1}}{f_{2}}{f_{3}}{{h_{1}}}^{3}{{r_{+}}
}^{3}+12{f_{1}}{f_{2}}{f_{3}}{{h_{1}}}^{3}{{r_{+}}}^{3}{
b_{1}}]  \nonumber
\end{eqnarray}
The quantities  $f_2$ and $h_2$ are respectively given in \eqref{eq9} and \eqref{eq10}
and
\begin{multline}
f_{3}={\dfrac {-30{q}^{2}{ f_{1}}{ h_{2}}-180\alpha k{{f_{1}}}^{3}+
336\alpha{r_{+}}{{f_{1}}}^{4}+60k{r_{+}}{{f_{1}}}^{2}-30
{{r_{+}}}^{3}{f_{2}}{{f_{1}}}^{2}-120{{r_{+}}}^{3}{{f_{1}}}
^{2}{h_{2}}+120\Lambda{{r_{+}}}^{3}{{f_{1}}}^{2}}{\alpha{{
r_{+}}}^{3}{{f_{1}}}^{3}}}-\\\dfrac{15{q}^{2}{
 f_{2}}{f_{1}}-728\Lambda\alpha{{ r_{+}}}^{2}{{ f_{1}}}^{3}-
204\alpha k{r_{+}}{{f_{1}}}^{2}{h_{2}}-728\Lambda\alpha{
{r_{+}}}^{3}{{f_{1}}}^{2}{ h_{2}}-120{{r_{+}}}^{2}{{ f_{1}}}^{3}+
507\alpha{{r_{+}}}^{2}{{f_{1}}}^{3}{f_{2}}}{\alpha{{
 r_{+}}}^{3}{{ f_{1}}}^{3}}+\\\dfrac{75\alpha{{r_{+}
}}^{3}{{ f_{2}}}^{2}{{f_{1}}}^{2}+90k{{r_{+}}}^{2}{f_{1}}{
h_{2}}+513\alpha{{r_{+}}}^{2}{{f_{1}}}^{3}{ h_{2}}+90\Lambda{{
r_{+}}}^{4}{f_{1}}{h_{2}}-54\alpha{{r_{+}}}^{3}{{f_{1}}}^{2
}{{ h_{2}}}^{2}-120\alpha k{r_{+}}{f_{2}}{{f_{1}}}^{2}}{\alpha{{
r_{+}}}^{3}{{f_{1}}}^{3}}-\\\dfrac{320
\Lambda \alpha{{ r_{+}}}^{3}{f_{2}}{{f_{1}}}^{2}+320{\Lambda}
^{2}\alpha{{r_{+}}}^{3}{{f_{1}}}^{2}+160\Lambda \alpha k{ 
r_{+}}{{f_{1}}}^{2}+240\Lambda \alpha k{{r_{+}}}^{2}{f_{1}}{
h_{2}}+240{\Lambda}^{2}\alpha{{r_{+}}}^{4}{f_{1}}{h_{2}}+303
\alpha{{r_{+}}}^{3}{{ f_{1}}}^{2}{f_{2}}{h_{2}}}{\alpha{{
 r_{+}}}^{3}{{ f_{1}}}^{3}}\nonumber
\end{multline}
\begin{multline}
h_{3}={\dfrac {-6{q}^{2}{f_{1}}{h_{2}}-36\alpha k{
{f_{1}}}^{3}-192\alpha{r_{+}}{{f_{1}}}^{4}+12 k{r_{+}}{{
f_{1}}}^{2}-6{{r_{+}}}^{3}{ f_{2}}{{f_{1}}}^{2}-24{{r_{+}}}^
{3}{{f_{1}}}^{2}{h_{2}}+24\Lambda{{r_{+}}}^{3}{{f_{1}}}^{2}}{216\alpha{{
r_{+}}}^{3}{{f_{1}}}^{3}}}-\\\dfrac{3
{q}^{2}{f_{2}}{f_{1}}+200\Lambda\alpha {{r_{+}}}^{2}{{ 
f_{1}}}^{3}+132\alpha k{r_{+}}{{f_{1}}}^{2}{ h_{2}}+200\Lambda
\alpha{{r_{+}}}^{3}{{f_{1}}}^{2}{ h_{2}}-24{{r_{+}}}^{2}{{ 
f_{1}}}^{3}-201\alpha{{r_{+}}}^{2}{{f_{1}}}^{3}{f_{2}}+15\alpha
{{r_{+}}}^{3}{{f_{2}}}^{2}{{f_{1}}}^{2}}{216\alpha{{
r_{+}}}^{3}{{f_{1}}}^{3}}+\\\dfrac{18 k{{r_{+}}}^{2}{ 
f_{1}}{ h_{2}}-459\alpha{{r_{+}}}^{2}{{f_{1}}}^{3}{h_{2}}+18
\Lambda{{r_{+}}}^{4}{f_{1}}{h_{2}}-54\alpha{{r_{+}}}^{3}{{
f_{1}}}^{2}{{ h_{2}}}^{2}-24\alpha k{ r_{+}}{f_{2}}{{ f_{1}}}
^{2}-64\Lambda\alpha{{ r_{+}}}^{3}{f_{2}}{{f_{1}}}^{2}}{216\alpha{{
r_{+}}}^{3}{{f_{1}}}^{3}}+\\\dfrac{64{
\Lambda}^{2}\alpha{{r_{+}}}^{3}{{f_{1}}}^{2}+32\Lambda\alpha 
k{r_{+}}{{f_{1}}}^{2}+48\Lambda\alpha k{{r_{+}}}^{2}{f_{1}}
{ h_{2}}+48{\Lambda}^{2}\alpha{{r_{+}}}^{4}{f_{1}}{h_{2}}-
69\alpha{{ r_{+}}}^{3}{{ f_{1}}}^{2}{f_{2}}{ h_{2}}}{216\alpha{{
r_{+}}}^{3}{{f_{1}}}^{3}}\nonumber
\end{multline}

\section{An alternate Near-horizon Solution}

An alternate  near horizon solution with a   well-defined
$ \alpha\to 0 $ limit can be obtained by taking 
 $ r_+$ and $ h_{2}$ to be the undetermined constants of integration.  This yields 
\begin{eqnarray}
&&f_{1}=\dfrac{1}{96 r_{+}{}^{2}(r_{+} h_{2}+2 h_{1})\alpha}[80 \Lambda \alpha r_{+}^{3}h_{1}+48\alpha k r_{+}h_{1}-6 r_{+}^{3} h_{1} -3q^{2}+(3072 \Lambda^{2} \alpha^{2}r_{+}^{7}h_{1}h_{2} \nonumber  \\ 
&&+12544 \Lambda^{2}\alpha^{2}r_{+}^{6}h_{1}^{2}+3072 \Lambda \alpha^{2}k r_{+}^{5}h_{1}h_{2}+1152\Lambda \alpha r_{+}^{7}h_{1}h_{2}+13824 \Lambda \alpha^{2}k r_{+}^{4}h_{1}^{2}+1344\Lambda \alpha r_{+}^{6} h_{1}^{2} \nonumber \\
&&+1152\alpha k r_{+}^{5}h_{1}h_{2}+13824\Lambda \alpha^{2}k r_{+}^{4}h_{1}^{2}+1344\Lambda \alpha r_{+}^{4}h_{1}^{2}+1152\alpha k r_{+}^{5}h_{1}h_{2}-480\Lambda \alpha q^{2}r_{+}^{3}h_{1}+ \nonumber\\  
&&2304\alpha k^2 r_{+}^{2}h_{1}^{2}+1728\alpha k r_{+}^{4}h_{1}^{2}+36r_{+}^{4}h_{1}^{2}-288\alpha k q^{2}r_{+}h_{1}+36q^{2}r_{+}^{3}h_{1}+9q^{4})^{\frac{1}{2}}]
\label{f1-2nd}
\end{eqnarray}
and
\begin{equation}
f_{2}=\dfrac{1}{h_{1}r_{+}^{2}}[-3f_{1}h_{2}r_{+}^{2}-8f_{1}h_{1}r_{+}+4kh_{1}+8\Lambda h_{1}r_{+}^{2}]
\label{f2-2nd}
\end{equation}
\begin{multline}
f_{3}=\dfrac{1}{r_{+}^{2}h_{1}}[8\Lambda r_{+}^{2}h_{1}h_{2}-5r_{+}^{2}f_{1}h_{1}h_{3}-r_{+}^{2}f_{1}h_{2}^{2}-3r_{+}^{2}f_{2}h_{1}h_{2}+8\Lambda r_{+}h_{1}^{2}-13r_{+}f_{1}h_{1}h_{2}-7r_{+}f_{2}h_{1}{}^{2}+\\ 
4kh_{1}h_{2}-6f_{1}h_{1}{}^{2}]\nonumber
\end{multline}
\begin{multline}
h_{3}=\dfrac{1}{216 h_{1} \alpha r_{+}^{3}f_{1}^{2}}[48 \Lambda^{2}\alpha r_{+}^{4}h_{1}h_{2}+64\Lambda^{2}\alpha r_{+}^{3}h_{1}+200\Lambda \alpha r_{+}^{3}f_{1}h_{1}h_{2}-64\Lambda \alpha r_{+}^{3}f_{2}f_{1}^{2}-54\alpha r_{+}^{3}f_{1}^{2}h_{2}^{2}\\-69\alpha r_{+}^{3}f_{1}f_{2}h_{1}h_{2}+15\alpha r_{+}^{3}f_{2}^{2}h_{1}^{2}+48\Lambda \alpha k r_{+}^{2}h_{1}h_{2}+200\Lambda \alpha r_{+}^{2}f_{1}h_{1}^{2}
+18\Lambda r_{+}^{4}h_{1}h_{2}-459\alpha r_{+}^{2}f_{1}^{2}h_{1}h_{2}\\-201\alpha r_{+}^{2}f_{1}f_{2}h_{1}^{2}+32\Lambda \alpha k r_{+}h_{1}^{2}+24\Lambda r_{+}^{3}h_{1}^{2}+132\alpha k r_{+}f_{1}h_{1}h_{2}
-24\alpha k r_{+}f_{2}h_{1}^{2}-192\alpha r_{+}f_{1}^{2}h_{1}^{2}\\-24r_{+}^{3}f_{1}h_{1}h_{2}-6r_{+}^{3}f_{2}h_{1}^{2}-36\alpha k f_{1} h_{1}^{2}+18 k r_{+}^{2}h_{1}h_{2}-24r_{+}^{2}f_{1}h_{1}^{2}+12k r_{+}h_{1}^{2}-6q^{2}f_{1}h_{2}-3q^{2}f_{2}h_{1}
]\nonumber
\end{multline}
The small $\alpha$ limit is  
\begin{equation}
f_{1}=\dfrac{2r_{+}^{3}h_{1}(\Lambda r_{+}^{2}+k)}{2r_{+}^{3}h_{1}+q^{2}}+\mathcal{O}(\alpha)
\end{equation} 
\begin{equation}
f_{2}=\dfrac{2(-3\Lambda r_{+}^{6} h_{2}-3k r_{+}^{4}h_{2}+4\Lambda q^{2}r_{+}^{2}-4kr_{+}^{3}h_{1}+2kq^{2})}{r_{+}^{2}(2r_{+}^{3}h_{1}+q^{2})}+\mathcal{O}(\alpha)
\end{equation}
with more complicated expressions for $f_3$ and $h_3$ that we shall not write down here.
As we are interested in the charged generalization of the black hole solutions obtained in \eqref{eq1}, we shall postpone investigation of this alternate solution for future study.


\begin{thebibliography}{99}
  
  \bibitem{tHooft:1974toh} 
  G.~'t Hooft and M.~J.~G.~Veltman,
  %``One loop divergencies in the theory of gravitation,''
  Ann.\ Inst.\ H.\ Poincare Phys.\ Theor.\ A {\bf 20}, 69 (1974).
  %1034 citations counted in INSPIRE as of 08 Oct 2019
  
 \bibitem{Stelle:1976gc} 
  K.~S.~Stelle,
  %``Renormalization of Higher Derivative Quantum Gravity,''
  Phys.\ Rev.\ D {\bf 16}, 953 (1977).
  doi:10.1103/PhysRevD.16.953
  %%CITATION = doi:10.1103/PhysRevD.16.953;%%
  %1734 citations counted in INSPIRE as of 08 Oct 2019 
  
  \bibitem{Smilga:2013vba} 
  A.~V.~Smilga,
  %``Supersymmetric field theory with benign ghosts,''
  J.\ Phys.\ A {\bf 47}, no. 5, 052001 (2014)
  doi:10.1088/1751-8113/47/5/052001
  [arXiv:1306.6066 [hep-th]].
  %%CITATION = doi:10.1088/1751-8113/47/5/052001;%%
  %8 citations counted in INSPIRE as of 08 Oct 2019
  
\bibitem{Smilga:2004cy} 
  A.~V.~Smilga,
  %``Benign versus malicious ghosts in higher-derivative theories,''
  Nucl.\ Phys.\ B {\bf 706}, 598 (2005)
  doi:10.1016/j.nuclphysb.2004.10.037
  [hep-th/0407231].
  %%CITATION = doi:10.1016/j.nuclphysb.2004.10.037;%%
  %90 citations counted in INSPIRE as of 08 Oct 2019 
 
 \bibitem{Lu:2015cqa} 
  H.~Lu, A.~Perkins, C.~N.~Pope and K.~S.~Stelle,
  %``Black Holes in Higher-Derivative Gravity,''
  Phys.\ Rev.\ Lett.\  {\bf 114}, no. 17, 171601 (2015)
  doi:10.1103/PhysRevLett.114.171601
  [arXiv:1502.01028 [hep-th]].
  %%CITATION = doi:10.1103/PhysRevLett.114.171601;%%
  %98 citations counted in INSPIRE as of 05 Oct 2019

\bibitem{Lin:2016jjl} 
  K.~Lin, A.~B.~Pavan, G.~Flores-Hidalgo and E.~Abdalla,
  %``New Electrically Charged Black Hole in Higher Derivative Gravity,''
  Braz.\ J.\ Phys.\  {\bf 47}, no. 4, 419 (2017)
  doi:10.1007/s13538-017-0505-0
  [arXiv:1605.04562 [gr-qc]].
  %%CITATION = doi:10.1007/s13538-017-0505-0;%%
  %3 citations counted in INSPIRE as of 05 Oct 2019
 
   
  \bibitem{Rezzolla:2014mua} 
  L.~Rezzolla and A.~Zhidenko,
  %``New parametrization for spherically symmetric black holes in metric theories of gravity,''
  Phys.\ Rev.\ D {\bf 90}, no. 8, 084009 (2014)
  doi:10.1103/PhysRevD.90.084009
  [arXiv:1407.3086 [gr-qc]].
  %%CITATION = doi:10.1103/PhysRevD.90.084009;%%
  %42 citations counted in INSPIRE as of 08 Oct 2019

\bibitem{Leaver:1985ax} 
  E.~W.~Leaver,
  %``An Analytic representation for the quasi normal modes of Kerr black holes,''
  Proc.\ Roy.\ Soc.\ Lond.\ A {\bf 402}, 285 (1985).
  doi:10.1098/rspa.1985.0119
  %%CITATION = doi:10.1098/rspa.1985.0119;%%
  %607 citations counted in INSPIRE as of 08 Oct 2019  
  
  \bibitem{Deser} 
  S.~Deser and B.~Tekin,
  %``Gravitational energy in quadratic curvature gravities,''
  Phys.\ Rev.\ Lett.\  {\bf 89}, 101101 (2002)
  doi:10.1103/PhysRevLett.89.101101
  [hep-th/0205318].
  %%CITATION = doi:10.1103/PhysRevLett.89.101101;%%
  %237 citations counted in INSPIRE as of 08 Oct 2019
  
 
\bibitem{Altas:2018pkl} 
  E.~Altas and B.~Tekin,
  %``New approach to conserved charges of generic gravity in AdS spacetimes,''
  Phys.\ Rev.\ D {\bf 99}, no. 4, 044016 (2019)
  doi:10.1103/PhysRevD.99.044016
  [arXiv:1811.11525 [hep-th]].
  %%CITATION = doi:10.1103/PhysRevD.99.044016;%%
  %6 citations counted in INSPIRE as of 19 Jan 2020 
  
  
\bibitem{Rez} 
  L.~Rezzolla and A.~Zhidenko,
  %``New parametrization for spherically symmetric black holes in metric theories of gravity,''
  Phys.\ Rev.\ D {\bf 90}, no. 8, 084009 (2014)
  doi:10.1103/PhysRevD.90.084009
  [arXiv:1407.3086 [gr-qc]].
  %%CITATION = doi:10.1103/PhysRevD.90.084009;%%
  %42 citations counted in INSPIRE as of 08 Oct 2019

\bibitem{Edery} 
  A.~Edery and M.~B.~Paranjape,
  %``Classical tests for Weyl gravity: Deflection of light and radar echo delay,''
  Phys.\ Rev.\ D {\bf 58}, 024011 (1998)
  doi:10.1103/PhysRevD.58.024011
  [astro-ph/9708233].
  %%CITATION = doi:10.1103/PhysRevD.58.024011;%%
  %74 citations counted in INSPIRE as of 08 Oct 2019

\bibitem{Asada} 
  H.~Asada,
  %``Gravitational time delays along multiple light paths as a probe of physics beyond Einstein gravity,''
  Phys.\ Lett.\ B {\bf 661}, 78 (2008)
  doi:10.1016/j.physletb.2008.02.006
  [arXiv:0710.0477 [gr-qc]].
  %%CITATION = doi:10.1016/j.physletb.2008.02.006;%%
  %13 citations counted in INSPIRE as of 08 Oct 2019

\bibitem{Will} 
  C.~M.~Will,
  %``The Confrontation between General Relativity and Experiment,''
  Living Rev.\ Rel.\  {\bf 17}, 4 (2014)
  doi:10.12942/lrr-2014-4
  [arXiv:1403.7377 [gr-qc]].
  %%CITATION = doi:10.12942/lrr-2014-4;%%
  %950 citations counted in INSPIRE as of 08 Oct 2019

\bibitem{Hen} 
  R.~A.~Hennigar, M.~B.~J.~Poshteh and R.~B.~Mann,
  %``Shadows, Signals, and Stability in Einsteinian Cubic Gravity,''
  Phys.\ Rev.\ D {\bf 97}, no. 6, 064041 (2018)
  doi:10.1103/PhysRevD.97.064041
  [arXiv:1801.03223 [gr-qc]].
  %%CITATION = doi:10.1103/PhysRevD.97.064041;%%
  %17 citations counted in INSPIRE as of 08 Oct 2019

\bibitem{Wald1} 
  R.~M.~Wald,
  %``Black hole entropy is the Noether charge,''
  Phys.\ Rev.\ D {\bf 48}, no. 8, R3427 (1993)
  doi:10.1103/PhysRevD.48.R3427
  [gr-qc/9307038].
  %%CITATION = doi:10.1103/PhysRevD.48.R3427;%%
  %1533 citations counted in INSPIRE as of 08 Oct 2019

\bibitem{Wald2} 
  V.~Iyer and R.~M.~Wald,
  %``Some properties of Noether charge and a proposal for dynamical black hole entropy,''
  Phys.\ Rev.\ D {\bf 50}, 846 (1994)
  doi:10.1103/PhysRevD.50.846
  [gr-qc/9403028].
  %%CITATION = doi:10.1103/PhysRevD.50.846;%%
  %1293 citations counted in INSPIRE as of 08 Oct 2019

\bibitem{Lu:2015psa} 
  H.~Lü, A.~Perkins, C.~N.~Pope and K.~S.~Stelle,
  %``Spherically Symmetric Solutions in Higher-Derivative Gravity,''
  Phys.\ Rev.\ D {\bf 92}, no. 12, 124019 (2015)
  doi:10.1103/PhysRevD.92.124019
  [arXiv:1508.00010 [hep-th]].
  %%CITATION = doi:10.1103/PhysRevD.92.124019;%%
  %66 citations counted in INSPIRE as of 08 Oct 2019

\bibitem{1990AN....311..271F}
H. Fuchs 
      % ``Deviation of circular geodesics in static spherically symmetric space-times"
Astron. Nach. {\bf 311}, 271 (1990).


\bibitem{Cramer:1994qj} 
  J.~G.~Cramer, R.~L.~Forward, M.~S.~Morris, M.~Visser, G.~Benford and G.~A.~Landis,
  %``Natural wormholes as gravitational lenses,''
  Phys.\ Rev.\ D {\bf 51}, 3117 (1995)
  doi:10.1103/PhysRevD.51.3117
  [astro-ph/9409051].
  %%CITATION = doi:10.1103/PhysRevD.51.3117;%%
  %124 citations counted in INSPIRE as of 14 Jan 2019

\bibitem{Torres:1998xd} 
  D.~F.~Torres, G.~E.~Romero and L.~A.~Anchordoqui,
  %``Might some gamma-ray bursts be an observable signature of natural wormholes?,''
  Phys.\ Rev.\ D {\bf 58}, 123001 (1998)
  doi:10.1103/PhysRevD.58.123001
  [astro-ph/9802106].
  %%CITATION = doi:10.1103/PhysRevD.58.123001;%%
  %53 citations counted in INSPIRE as of 14 Jan 2019
  
\bibitem{Synge}
J. L. Synge, Mon. Not. R. astr. Soc. 131 (1966).  


\bibitem{Ruggiero:2020yoq} 
  M.~L.~Ruggiero and L.~Iorio,
  %``Probing Some Modified Gravity Models with Exoplanets,''
  arXiv:2001.04122 [gr-qc].

\bibitem{Bonanno:2019rsq}
A.~Bonanno and S.~Silveravalle,
%``Characterizing black hole metrics in quadratic gravity,''
Phys. Rev. D \textbf{99}, no.10, 101501 (2019)
doi:10.1103/PhysRevD.99.101501
[arXiv:1903.08759 [gr-qc]].
%5 citations counted in INSPIRE as of 31 May 2020
  
\end{thebibliography}
\end{document}